\documentclass[twocolumn]{aastex62}

\newcommand{\Msun}{{\rm M}_\odot}
\newcommand{\kms}{\textrm{km}\,\textrm{s}^{-1}}

\usepackage{savesym}
\savesymbol{tablenum}
\usepackage{siunitx}
\restoresymbol{SIX}{tablenum}
\usepackage{graphicx,epsf}
\usepackage{subfigure}
\usepackage{graphicx}	
\usepackage{amsmath}	
\usepackage{amssymb}	
\usepackage{units}
\usepackage{newtxtext,newtxmath}
\usepackage{siunitx}

\graphicspath{{./}{figures/}}

\shorttitle{Ca-rich Transient SN~2016hnk}
\shortauthors{Jacobson-Gal\'{a}n et al.}

\begin{document}

\title{Ca hnk: Calcium-rich Transient SN~2016hnk from the Helium Shell Detonation of a Sub-Chandrasekhar White Dwarf}

\correspondingauthor{Wynn Jacobson-Gal\'{a}n (he, him, his)}
\email{wynn@u.northwestern.edu}

\author[0000-0002-3934-2644]{Wynn V. Jacobson-Gal\'{a}n}
\affil{Department of Physics and Astronomy, Northwestern University, 2145 Sheridan Road, Evanston, IL 60208, USA}
\affil{Center for Interdisciplinary Exploration and Research in Astrophysics (CIERA), 1800 Sherman Ave, Evanston, IL 60201, USA}
\affil{Department of Astronomy and Astrophysics, University of California, Santa Cruz, CA 95064,
USA}

\author{Abigail Polin}
\affil{Department of Physics, University of California, Berkeley, CA 94720, USA}
\affil{Department of Astronomy, University of California, Berkeley, CA 94720, USA}
\affil{Lawrence Berkeley National Laboratory, Berkeley, CA 94720, USA}

\author{Ryan J. Foley}
\affil{Department of Astronomy and Astrophysics, University of California, Santa Cruz, CA 95064,
USA}

\author{Georgios~Dimitriadis}
\affil{Department of Astronomy and Astrophysics, University of California, Santa Cruz, CA 95064,
USA}


\author{Charles~D.~Kilpatrick}
\affil{Department of Astronomy and Astrophysics, University of California, Santa Cruz, CA 95064,
USA}

\author{Raffaella Margutti}
\affil{Department of Physics and Astronomy, Northwestern University, 2145 Sheridan Road, Evanston, IL 60208, USA}
\affil{Center for Interdisciplinary Exploration and Research in Astrophysics (CIERA), 1800 Sherman Ave, Evanston, IL 60201, USA}

\author{David~A.~Coulter}
\affil{Department of Astronomy and Astrophysics, University of California, Santa Cruz, CA 95064,
USA}

\author{Saurabh~W.~Jha}
\affil{Department of Physics and Astronomy, Rutgers, the State University of New Jersey, 136 Frelinghuysen Road, Piscataway, NJ 08854 USA}

\author{David~O.~Jones}
\affil{Department of Astronomy and Astrophysics, University of California, Santa Cruz, CA 95064,
USA}

\author{Robert~P.~Kirshner}
\affil{Harvard College Observatory, 60 Garden Street, Cambridge MA 02138}

\author{Yen-Chen Pan}
\affil{Division of Science, National Astronomical Observatory of Japan, 2-21-1 Osawa, Mitaka, Tokyo 181-8588, Japan}
\affil{EACOA Fellow}

\author{Anthony~L.~Piro}
\affil{The Observatories of the Carnegie Institution for Science, 813 Santa Barbara Street, Pasadena, CA 91101, USA}

\author{Armin Rest}
\affil{Space Telescope Science Institute, Baltimore, MD 21218}
\affil{Department of Physics and Astronomy, The Johns Hopkins University, Baltimore, MD 21218}

\author{C\'{e}sar~Rojas-Bravo}
\affil{Department of Astronomy and Astrophysics, University of California, Santa Cruz, CA 95064,
USA}

\begin{abstract}
We present observations and modeling of SN~2016hnk, a Ca-rich supernova (SN) that is consistent with being the result of a He-shell double-detonation explosion of a C/O white dwarf.  We find that SN~2016hnk is intrinsically red relative to typical thermonuclear SNe and has a relatively low peak luminosity ($M_B = -15.4$~mag), setting it apart from low-luminosity Type Ia supernovae (SNe~Ia). SN~2016hnk has a fast-rising light curve that is consistent with other Ca-rich transients ($t_r = 15$~d).  We determine that SN~2016hnk produced $0.03 \pm 0.01 M_{\odot}$ of ${}^{56}\textrm{Ni}$ and $0.9 \pm 0.3 M_{\odot}$ of ejecta.  The photospheric spectra show strong, high-velocity \ion{Ca}{2} absorption and significant line blanketing at $\lambda < 5000$~\AA, making it distinct from typical (SN~2005E-like) Ca-rich SNe.  SN~2016hnk is remarkably similar to SN~2018byg, which was modeled as a He-shell double-detonation explosion.  We demonstrate that the spectra and light curves of SN~2016hnk are well modeled by the detonation of a $0.02 \ \Msun$ helium shell on the surface of a $0.85 \ \Msun$ C/O white dwarf. This analysis highlights the second observed case of a He-shell double-detonation and suggests a specific thermonuclear explosion that is physically distinct from SNe that are defined simply by their low luminosities and strong [Ca II] emission. 


\end{abstract}

\keywords{supernovae:general --- 
supernovae: individual (SN~2016hnk) --- surveys --- white dwarfs}

\section{Introduction} \label{sec:intro}
Type Ia supernovae (SNe Ia) are produced by the thermonuclear explosion of a C/O white dwarf (WD) in a binary system \citep{Hoyle60, Colgate69, Woosley86}. Due to their standardizable light-curve evolution, SNe Ia are used as cosmological probes of the Universe's accelerating expansion \citep{reiss98, perlmutter99}. However, there exists an amalgam of sub-luminous thermonuclear objects that deviate from typical SN Ia characteristics (see \cite{taubenberger17} for a review). Such physically distinct classes of ``WD SNe'' include: 91bg-like \citep{filippenko92AJ}, 02es-like \citep{ganeshalingam12}, ``fast-decliners'' \citep{kasliwal10, perets11, drout13}, 06bt-like \citep{foley10}, SNe Iax \citep{foley13} and Ca-rich transients \citep{filippenko03, perets10}. 
 
``Ca-rich'' SNe are a particularly heterogeneous group of objects that, similar to other thermonuclear objects, are thought to arise from a progenitor system containing a WD. Ca-rich objects are primarily characterized by peak magnitudes of -14 to -16.5, quickly evolving light curves, and strong calcium features in photospheric and nebular phase spectra \citep{taubenberger17}. In nebular spectra, these objects show weak iron line transitions, with [Ca II] $\lambda\lambda7291, 7324$ being the dominant emission feature. Ca-rich transients have detectable [O I] $\lambda\lambda6300, 6364$ emission in nebular spectra, but a defining feature of the class is an integrated [Ca II]/[O I] flux ratio greater than $\sim$2. Furthermore, the majority of these objects exhibit low ejecta and ${}^{56}\textrm{Ni}$ masses of $\lesssim 0.5\Msun$ and $\lesssim 0.1\Msun$, respectively (\citealt{perets10, lunnan17}; however see \citealt{milisavljevic17}). 


Significant variations in the physical characteristics of ``Ca-rich'' transients have resulted in a highly diverse class of objects. For example, there is a substantial spread in [Ca II]/[O I] flux ratios amongst Ca-rich transients: objects such as SN~2003dg, PTF09dav and PTF10iuv have negligible [O I] emission, while SN~2012hn has a comparable oxygen composition to Type IIb/IIP SNe during the nebular phase \citep[e.g.,][]{valenti14}. Additionally, the presence of H$\alpha$ emission in PTF09dav and iPTF15eqv is unlike anything observed within the Ca-rich class \citep{sullivan11,milisavljevic17}. The lack of helium in addition to the probable presence of Sc II, Cr II \& Sr II also makes PTF09dav an outlier within the class. Furthermore, the location of iPTF15eqv in a star-forming, late-type galaxy, plus its large inferred ejecta mass ($\approx 2-4\Msun$), is difficult to reconcile with other homogeneous properties of Ca-rich transients \citep{milisavljevic17}. Lastly, many other thermonuclear objects are discovered with strong Ca II absorption (e.g., SN~2018byg; \citealt{de19}) and even SNe~Iax can be considered ``Ca-rich'' based on their [Ca II]/[O I] flux ratios. However, despite the constraints given by the current sample of Ca-rich objects, a core-collapse SN origin cannot yet be fully ruled out given the diversity in physical characteristics and host environments within this class. 

The significant fraction of objects found in old stellar environments on the outskirts of early-type galaxies supports a WD origin for at least some Ca-rich transients \citep{perets11, kasliwal12}. Parenthetically, \cite{foley15} demonstrate that these SNe tend to be found in dense environments of merged/merging galaxies with a large range of observed velocity offsets that are anti-correlated with the projected offsets. This finding is also supported by \cite{lunnan17} who showed that Ca-rich transients were typically produced in group or cluster environments of early-type elliptical galaxies. Furthermore, non-detections of star forming regions at Ca-rich transient explosion sites indicated that their progenitors were likely not formed at the SN location i.e. in situ \citep{lyman14}. 

A variety of progenitor scenarios involving the thermonuclear explosion of a WD have been proposed to explain the peculiar properties of Ca-rich transients and their environments. Accretion-induced collapse (AIC) of a WD into a neutron star (NS) will produce a sub-luminous explosion. However, the combination of a very rapid rise time ($\sim$1d), high expansion velocities and over-production of intermediate mass elements makes this an unlikely scenario \citep{metzger09}. Similar discrepancies between models and observations also rule out the tidal disruptions of WDs by NSs or stellar-mass black holes (BHs) \citep{metzger12, margalit16, toonen18, zenati2019b, Zenati2019}. Tidal detonations of a WD by interaction with an intermediate-mass black hole (IMBH) does occur in dense stellar systems, but is difficult to reconcile due to the lack of observed IMBHs at the locations of Ca-rich transients \citep{rosswog08, macleod14, sell15}. 


Given the range of observed properties, a highly promising progenitor channel for peculiar thermonuclear SNe such as Ca-rich transients is the detonation of a helium-accreting WD. One such model involves the explosion of a sub-Chandrasekhar mass, C/O WD via an initial detonation of a surface layer of accreted helium \citep{nomoto82a, nomoto82b, Woosley86, woosley94, livne95}. Alternatively, thin helium shell detonations that only partially disrupt the core can result in rapidly evolving, low luminosity explosions often labeled as .Ia SNe \citep{bildsten07, shen10}. However, the rise-times and ejecta masses produced in .Ia explosions are inconsistent with those of Ca-rich objects. 

A helium shell detonation scenario has been employed to model photometric and spectroscopic signatures in SNe Ia \citep{sim10, kromer10, woosley11, polin19, townsley19}. While this model can reproduce the observational properties of normal SNe Ia, there exists a wide parameter space within the double-detonation scenario that can be used to explain atypical objects. Variations on the initial WD mass and chemical composition, as well as helium shell mass, can explain the sub-luminous SNe Ia such as 91bg-like objects \citep{shen18}. Additionally, a helium shell detonation was invoked to explain Ca-rich transient SN~2005E, which showed observational consistency to a low mass WD detonation with a thick helium shell \citep{waldman11}.

Recently, observations of SN~Ia 2018byg (ZTF18aaqeasu) were presented as the first evidence of a helium shell double detonation \citep{de19}. SN~2018byg showed spectroscopic and photometric consistency with a $0.15\Msun$ helium shell detonation on a $0.75\Msun$, sub-Chandrasekhar mass WD in which material was mixed in the outer layers of ejecta. This specific thick shell model was presented by \cite{polin19} who showed that the parameter space of such an explosion scenario is well-matched to observations of SNe~Ia. In Section \ref{sec:models}, we present specific helium shell detonation models and discuss the viability of this explosion mechanism in explaining observations of SN~2016hnk.

In this paper we present observations, analysis and modeling of the peculiar thermonuclear transient SN~2016hnk. Upon discovery, this object was originally classified as a SN~Ia with photometric and spectroscopic similarities to both SN~1991bg and Ca-rich transient PTF09dav. \cite{sell18} identified the object as a candidate Ca-rich gap transient and observed the SN with \textit{Chandra} to test a progenitor scenario involving a tidal detonation of a WD by an intermediate-mass black hole. Finding no evidence of predicted X-ray emission, \cite{sell18} were able to rule out this progenitor scenario. However, the observed X-ray detection limits cannot constrain a WD + NS (or stellar-mass BH) progenitor scenario. \cite{galbany19} (hereafter G19) present multi-wavelength observations of SN~2016hnk and conclude that this object belongs to the 91bg-like subclass. Through NLTE spectral modeling of SN~2016hnk, G19 argue for a progenitor scenario involving a low luminosity detonation of a Chandrasekhar mass WD. We discuss these interpretations of SN~2016hnk throughout the paper as well as present alternative conclusions on the origin and classification of this intriguing object. 

In Section \ref{sec:observation}, we present observations and data reduction of SN~2016hnk. In Section \ref{sec:LC_analysis}, we present photometric properties of SN~2016hnk and discuss how these measurements compare to different classes of transient objects. In Section \ref{sec:spectra_analysis}, we present spectroscopic properties and spectral comparisons. In Section \ref{sec:models}, we compare light curve and spectral models of helium shell detonations to SN~2016hnk. In Section \ref{sec:discussion}, we discuss both the potential progenitor scenarios for SN~2016hnk and how our findings compare to G19.

\newpage
\section{Observations} \label{sec:observation}

\subsection{Detection and Classification}\label{subsec:classification}

SN~2016hnk was discovered by the Asteroid Terrestrial-impact Last Alert System \citep[ATLAS;][]{tonry18} on 2016 October 27 (MJD 57688) using the ACAM1 instrument with a cyan filter \citep{tonry16}. SN~2016hnk has a discovery apparent magnitude of 17.91 and is located at $\alpha = 02^{\textrm{h}}13^{\textrm{m}}16.63^{\textrm{s}}$, $\delta = -07^{\circ}39'40.80^{\prime \prime}$. SN~2016hnk is located 1.4\arcsec\:east and 12.4\arcsec\:north of the nucleus of SBa galaxy MCG -01-06-070. In this paper, we use the reported host-galaxy distance and redshift of 72.9 Mpc and  0.016268, respectively, in standard $\Lambda$CDM cosmology ($H_{0}$ = 70 km s$^{-1}$ Mpc$^{-1}$, $\Omega_M = 0.27$, $\Omega_{\Lambda} = 0.63$). 

SN~2016hnk was first classified as a peculiar SN Ia by \cite{cannizzaro16} who noted its resemblance to SN~1991bg (91bg-like) near maximum light. The initial spectrum was described as being red, with the presence of strong Si II $\lambda 6355$, O I $\lambda 7774$, and NIR Ca II spectral features. \cite{dimitriadis16} made a similar classification and indicated the similarity of SN~2016hnk with SNe 1991bg and 1999by. Finally, the classification by \cite{pan16} illustrated a similarity between SN~2016hnk and Ca-rich transient PTF09dav \citep{sullivan11} given their similar colors, peak absolute magnitude and spectral features. \cite{pan16} note the distinct absorption profiles at  5350, 5540 and 7120 \AA, which \cite{sullivan11} attributes to Sc II and Ti II, respectively. We discuss these specific spectral features in Section \ref{subsec:spectra_compare}. 

\subsection{Early-time Photometry}\label{subsec:photometry}

We imaged SN~2016hnk between 01 November 2016 and 17 February 2017 with the Direct camera on the Swope 1-m telescope at Las Campanas Observatory, Chile.  Observations were performed in Johnson \textit{BV} and Sloan \textit{ugri} filters.  We performed bias-subtraction and flat-fielding, stitching, registration, and photometric calibration using {\tt photpipe} \citep{Rest+05}.  For our photometric calibration, we used stars in the PS1 DR1 catalog \citep{Flewelling+16} transformed from \textit{gri} magnitudes to the \textit{uBVgri} Swope natural system following the Supercal method \citep{Scolnic+15}.  Difference imaging was performed using Swope \textit{BVgri} and \textit{u} templates obtained on 15 November 2017 and 10 December 2017, respectively.  Final photometry was performed in the difference images with DoPhot \citep{Schechter+93}. A Swope \textit{Bri} image of SN~2016hnk, mapped to RGB channels, from 31 October 2016 is shown in Figure \ref{fig:fits_image}. 

Additional imaging of SN~2016hnk was obtained by the Las Cumbres Observatory Global Telescope \citep[LCOGT;][]{Brown13} and the Foundation Supernova Survey with Pan-STARRS \citep{foley18}. We performed a similar photometric reduction on the LCOGT data as the Swope imaging, which included standard subtractions and difference imaging. PS1 images of SN~2016hnk were reduced with the same custom-built pipeline as for the PS1 MDF survey data. The basic data processing is performed by the PS1 IPP \citep{magniter06, magnier13, waters16}. Single epoch images are then processed through a frame-subtraction analysis using \texttt{photpipe}, which determines an appropriate spatially varying convolution kernel using \texttt{HOTPANTS} \citep{becker15}. After the convolutions are performed, the template image is
subtracted from the survey image. Detections of  significant flux excursions in the difference images are found using \texttt{DoPHOT} \citep{1993PASP..105.1342S}.

SN~2016hnk has a reported Milky Way reddening and associated extinction of \textit{E(B-V)} = 0.0224 mag and $A_{V} = 0.069$ mag \citep{schlegel98, schlafly11}, respectively, which we correct for using a standard \cite{fitzpatrick99} reddening law and \textit{$R_V$} = 3.1. We do not correct for host-galaxy contamination given the absence of Na I D absorption in all spectra at the host redshift. We demonstrate the effect of further de-reddening in the color curves shown in Figure \ref{fig:colors} and discuss the intrinsic color of SN~2016hnk in Section \ref{sec:reddening}.

\cite{galbany19} report a host extinction of \textit{E(B-V)} = 0.45 mag using the observed ratio of H$\alpha$ and H$\beta$ fluxes from host-galaxy spectra. However, because the exact vertical location of the source in the galaxy is unknown, Na I D absorption is the most accurate indicator of extinction at the site of the explosion. Thus all multi-color optical photometry of SN~2016hnk in Figure \ref{fig:optical_LC} has only been corrected for MW extinction. All photometric observations are listed in Table \ref{tab:phot_table}.

\subsection{Late-time Keck Imaging}\label{subsec:keck}

\begin{figure*}[t]
\includegraphics[width=\textwidth]{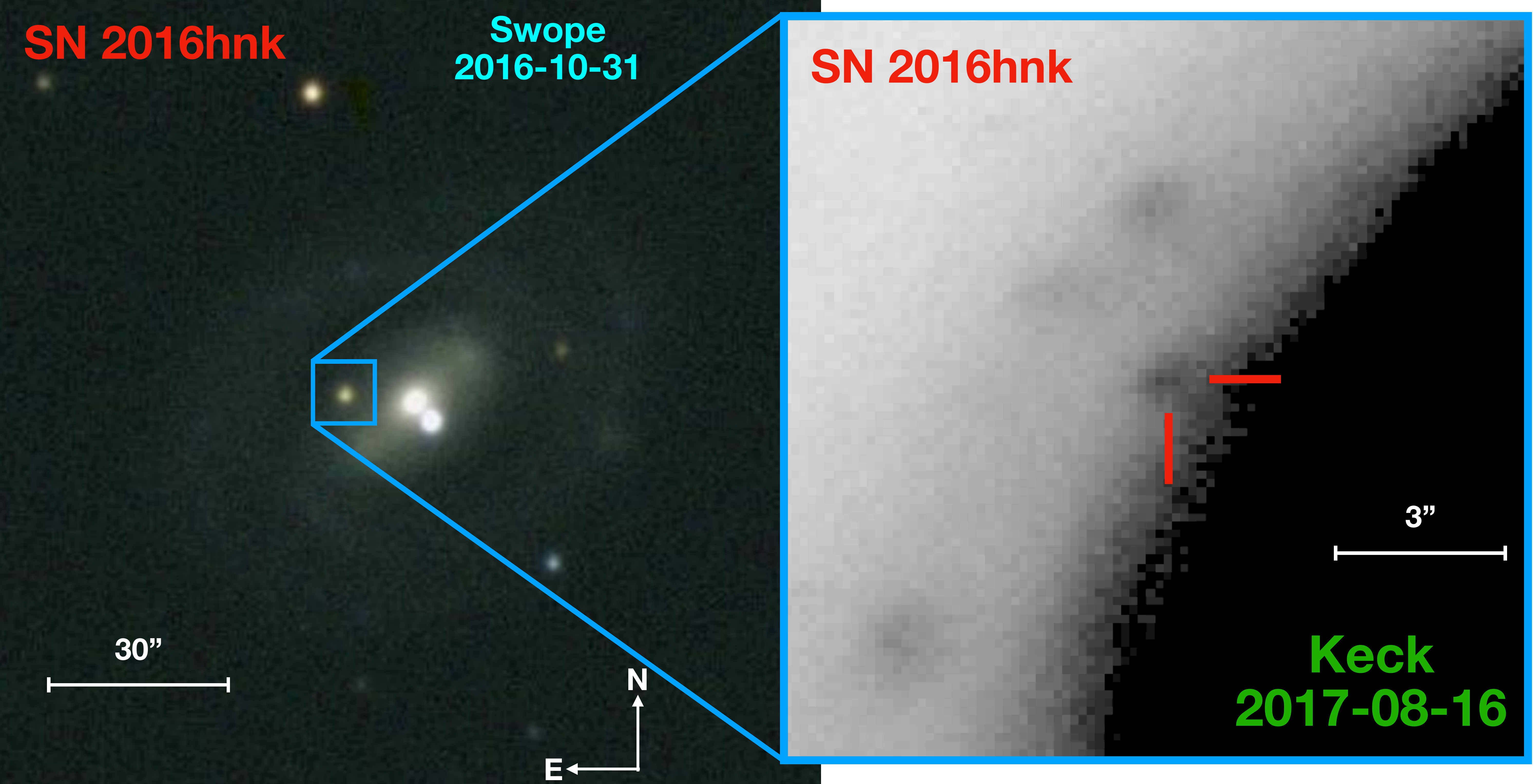}
\caption{\textit{Left:} RGB false-color image of SN~2016hnk taken by the Swope telescope two days after \textit{B}-band maximum. RGB image generated using \textit{r, V} \& \textit{B}-bands. \textit{Right:} \textit{I}-band Keck LRIS image at +291d. Source marked in red. \label{fig:fits_image}}
\end{figure*}

Final photometric observations of SN~2016hnk were obtained using the imaging camera on the Keck telescope Low-Resolution Imaging Spectrometer \citep[LRIS;][]{oke95}. The source was observed on 20 July 2017 in \textit{BR} as well as on 16 August 2017 in \textit{BVRI}.  Observations were performed in the blue and red channels simultaneously with
the B + R filters and V + I filters and the D560 dichroic, and reduced using \texttt{photpipe}. For our photometric calibration, we used secondary calibrators in each image with magnitudes derived from SDSS standard stars transformed to the BVRI system \citep{2011MNRAS.417.2230B,2015ApJS..219...12A}. Keck \textit{BVRI} templates were acquired on 25 September 2019 and image subtraction was performed using \texttt{HOTPANTS}. Final photometry was performed on the difference images using \texttt{photpipe}. The only late-time detection of SN~2016hnk was in $I$-band at a phase of +291d relative to maximum light. The $I$-band detection image is shown in Figure \ref{fig:fits_image} and has an associated apparent magnitude of $23.57 \pm 0.09$ mag. 

\begin{figure*}[t]
\centering
\includegraphics[width=\textwidth]{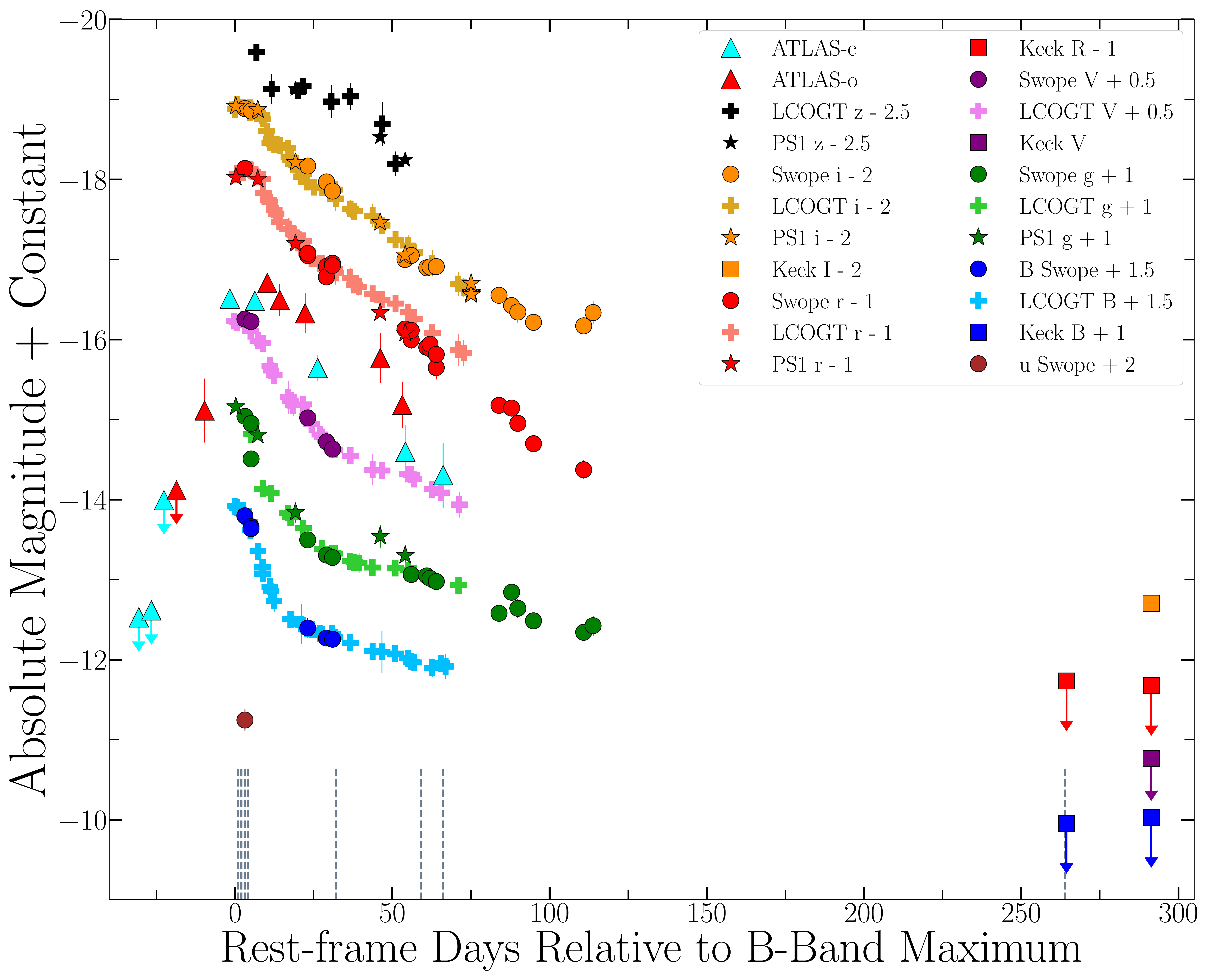}
\caption{Optical light curve of SN~2016hnk with respect to B-band maximum. ATLAS data presented as triangles, Pan-STARRS as stars, Swope as circles, LCOGT as plus signs, and Keck as squares. All Keck LRIS data points are non-detections with the exception of the \textit{I}-band point at +291d shown in orange. The epochs of our spectral observations are marked by grey dashed lines. \label{fig:optical_LC}}
\end{figure*}

All recovered magnitudes in other Keck filters are reported as upper limits as shown in the full optical light curve in Figure \ref{fig:optical_LC} and listed in Table \ref{tab:phot_table2}. 

\subsection{Optical Spectroscopy}\label{subsec:spectra}

\begin{figure*}
\centering
\includegraphics[width=0.85\textwidth]{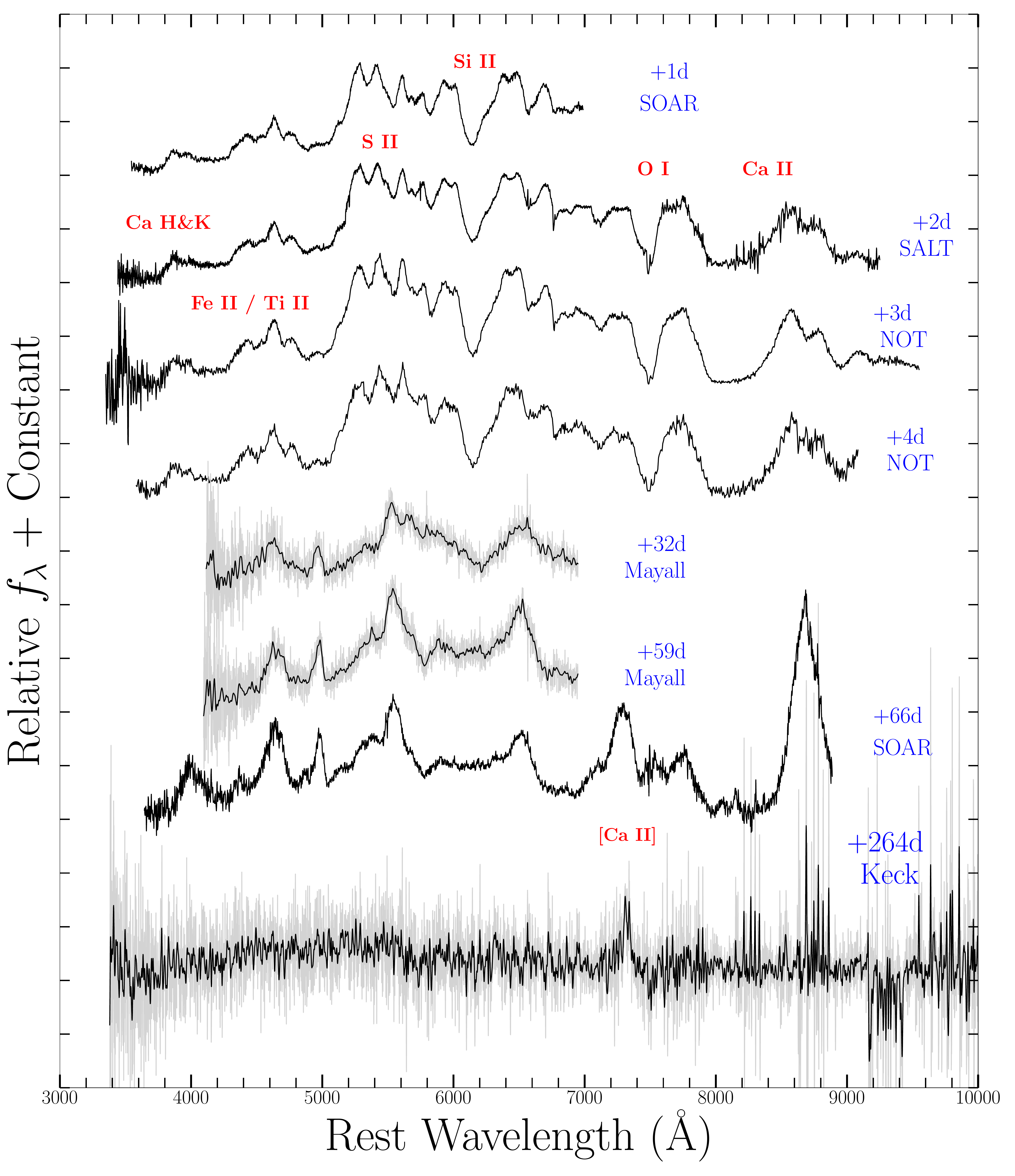}
\caption{Spectral observations of SN~2016hnk with phases (blue) marked with respect to \textit{B}-band maximum. Raw spectra are shown in gray and spectra smoothed with Gaussian-filters in black lines.  \label{fig:spectral_series}}
\end{figure*}

In Figure \ref{fig:spectral_series}, we present optical spectral observations of SN~2016hnk from +1 to +264d relative to \textit{B}-band maximum. We first observed SN~2016hnk on 30 October 2016 with the Goodman High Throughput Spectrograph \citep{clemens04} on the Southern Astrophysical Research Telescope (SOAR). SN~2016hnk was then observed using the Kitt Peak Ohio State Multi-Object Spectrograph \citep[KOSMOS;][]{martini14} on 30 November and 27 December 2016 as well as with SOAR on 03 January 2017. 

For all spectral observations, standard CCD processing and spectrum extraction were accomplished with
IRAF. The data were extracted using the optimal algorithm of
\citet{1986PASP...98..609H}.  Low-order polynomial fits to calibration-lamp
spectra were used to establish the wavelength scale, and small
adjustments derived from night-sky lines in the object frames were
applied.  We employed our own IDL routines to flux calibrate the data
and remove telluric lines using the well-exposed continua of the
spectrophotometric standard stars \citep{1988ApJ...324..411W, 2003PASP..115.1220F}.
Details of our spectroscopic reduction techniques are described in
\citet{2012MNRAS.425.1789S}.

SN~2016hnk was last observed by Keck/LRIS on 20 July 2017. At this time the SN had faded significantly and was undetectable in guide camera images, and a blind offset from a nearby star was necessary to acquire the target.  In the two-dimensional spectrogram, we do not detect continuum emission from the SN, but do see a broad absorption feature consistent with the wavelength of [\ion{Ca}{2}] $\lambda\lambda7291$, 7324.  Because of the bright and spatially varying background caused by the SN proximity to the center of its host galaxy, we chose to perform a two-dimensional background subtraction using a 3rd-order Legendre polynomial in the spatial direction.  Details of this method can be found in \citet{Foley09a}.  While this method produced significantly better results than others attempted, there is still some residual continuum that we believe to be galactic emission.

Additional early-time spectral observations (+2-4d after peak) were retrieved through the
WISeREP archive\footnote{\url{http://wiserep.weizmann.ac.il/}} \citep{Yaron12} and are presented within the complete list of spectral observations in Table \ref{tab:spec_table}.


\section{Light Curve Analysis}\label{sec:LC_analysis}

\subsection{Photometric Properties}\label{subsec:LC_properties}

We fit a low-order polynomial to the SN~2016hnk light curve to find best fit $B$- and $r$-band peak absolute magnitudes of $M_B = -15.40 \pm 0.088$ at MJD $57690.2\pm0.7$ and $M_r = -17.17 \pm 0.04$ at MJD $57693.3\pm0.6$, respectively. We calculate a \cite{phillips93} decline parameter value of $\Delta \rm{m}_{15}(B) = 1.31 \pm 0.085$ mag from our $B$-band light curve fits. All values are in agreement with those presented in G19.  



\begin{figure*}
\centering
\subfigure[]{\includegraphics[width=.38\textwidth]{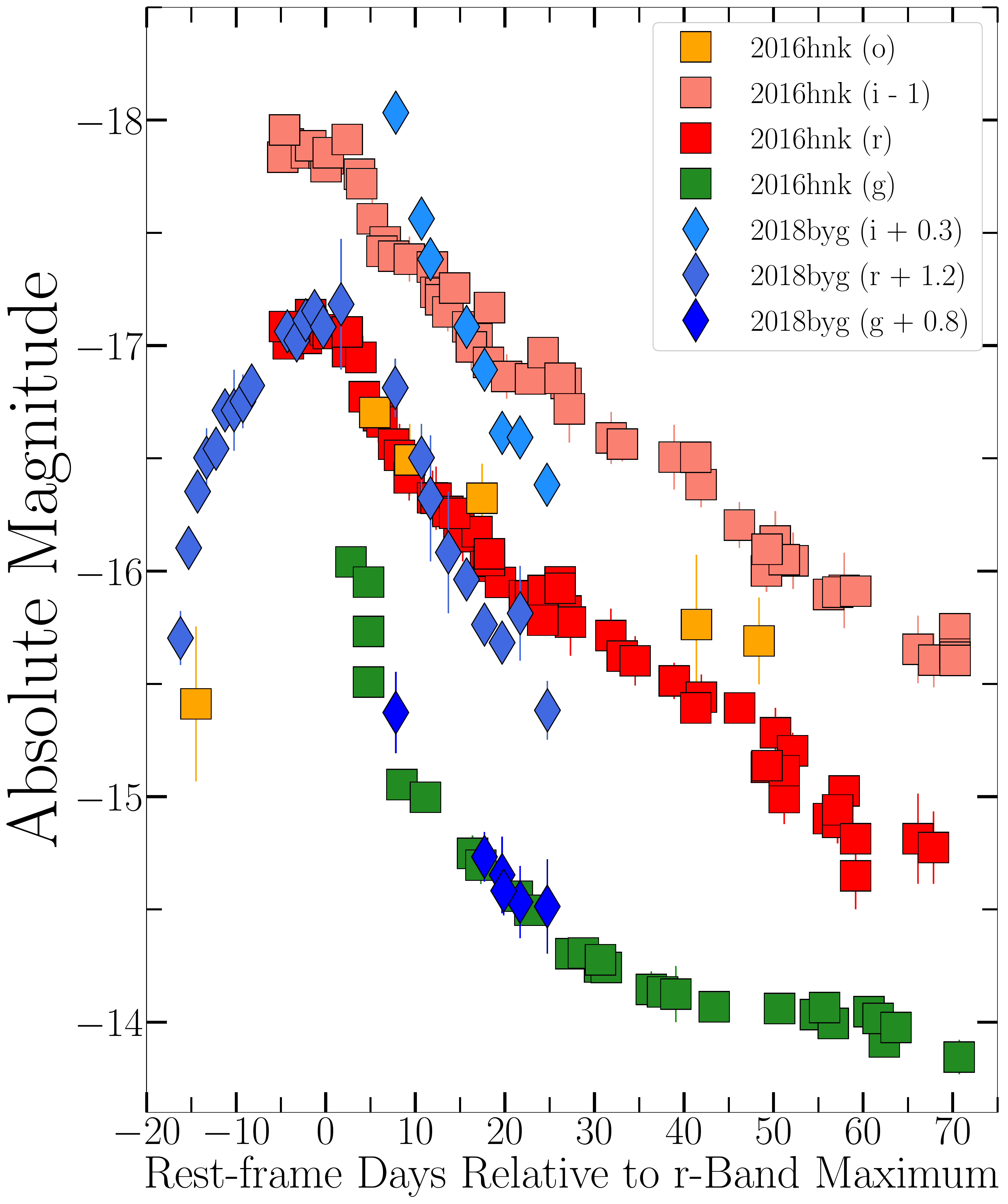}}
\subfigure[]{\includegraphics[width=.6\textwidth]{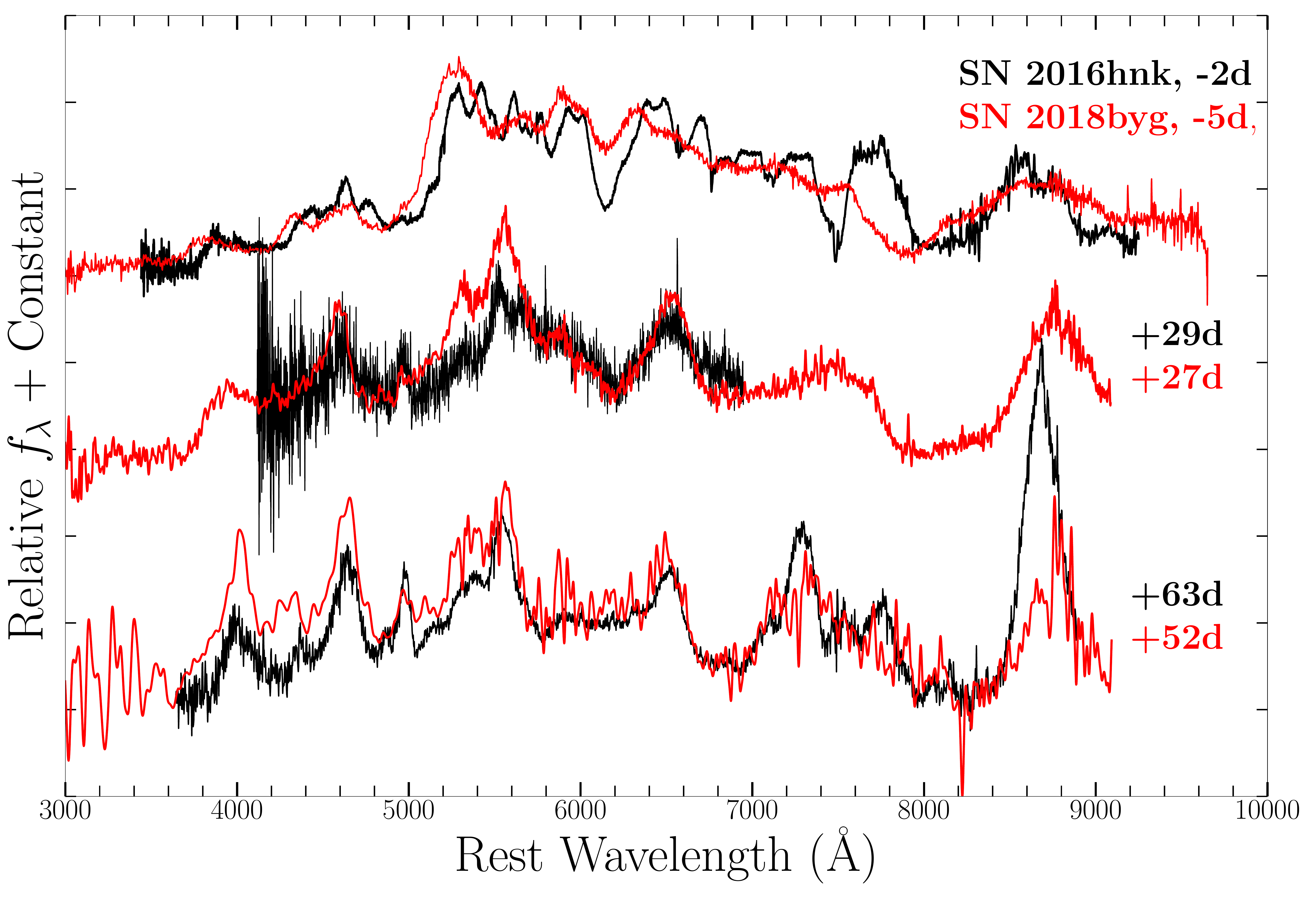}}
\caption{\textit{Left:} Light curve comparison of \textit{r}- and \textit{g}-band photometry for each object relative to \textit{r}-band maximum. SN~2018byg has been shifted in magnitude to match SN~2016hnk. \textit{Right:} Spectral comparison of SNe~2016hnk and 2018byg at multiple epochs. Phases relative to \textit{r}-band maximum.  \label{fig:2018byg_compare}}
\end{figure*}

Based on the ATLAS light curve, the last non-detections in ATLAS cyan and orange filters were at MJD 57667.56 and 57671.55, respectively, with the first detection being in ATLAS-o at MJD 57680.52. We can use these data to constrain the rise time if we match the ATLAS-o observations with all \textit{r}-band data given the similarity in transmission functions of each filter. Consequently, the rise time of SN~2016hnk is in a range of $12.8\textrm{d} \ < \ t_r \ < \ 21.8\textrm{d}$. We constrain the rise-time further by showing that SN~2016hnk has a faster rise than the 17 days quoted for SN~2018byg (Figure \ref{fig:2018byg_compare}a) and a similar rise to PTF11kmb (Figure \ref{fig:LC_compare}b), which has $t_r = 15$ days \citep{lunnan17}. From this comparison, we estimate a rise-time for SN~2016hnk of $t_r = 15 \pm 2$ days. 

We present light curve and spectral comparisons of SNe~2016hnk and 2018byg in Figure \ref{fig:2018byg_compare}. In \ref{fig:2018byg_compare}(a) we show the parallel between \textit{r}- and \textit{g}-band photometry for these objects. Both SNe exhibit a fast rise time of $\lessapprox17$ days, with SN~2016hnk having a slower decline in \textit{r}-band.

\begin{figure}[h]
\centering
\includegraphics[width=0.45\textwidth]{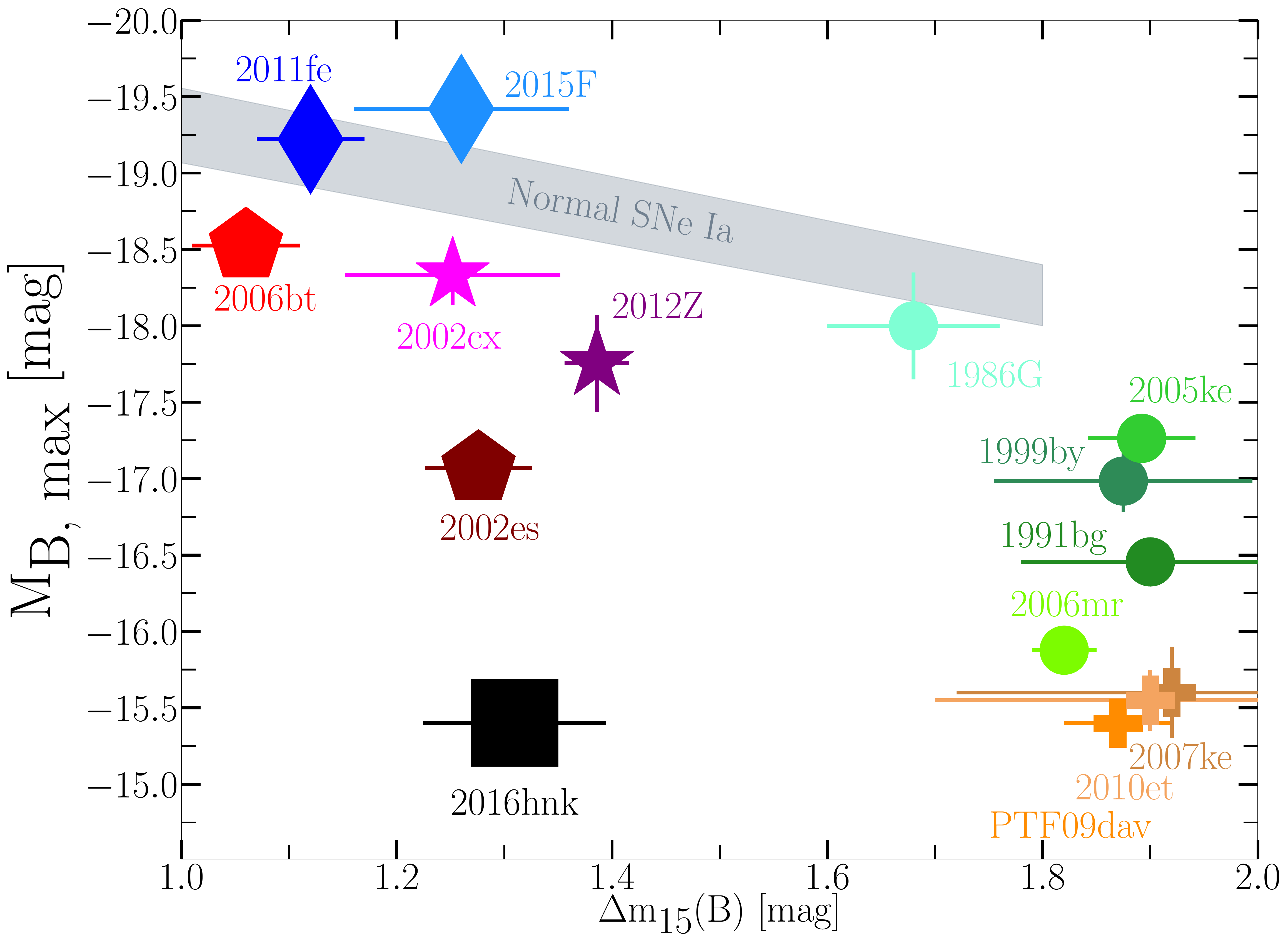}
\caption{$\Delta$m$_{15}$ vs. M$_{\textrm{B, max}}$ for SN~2016hnk (black square), normal SNe Ia (diamonds + grey region), 91bg-like SNe Ia (circles), SNe Iax (stars), 02es-like SNe Ia (pentagons) and Ca-rich objects (plus signs). Some uncertainities on M$_{\textrm{B, max}}$ are smaller than plotted marker size. \label{fig:MB_dm15}}
\end{figure}

\begin{figure*}
\centering
\subfigure[]{\includegraphics[width=.45\textwidth]{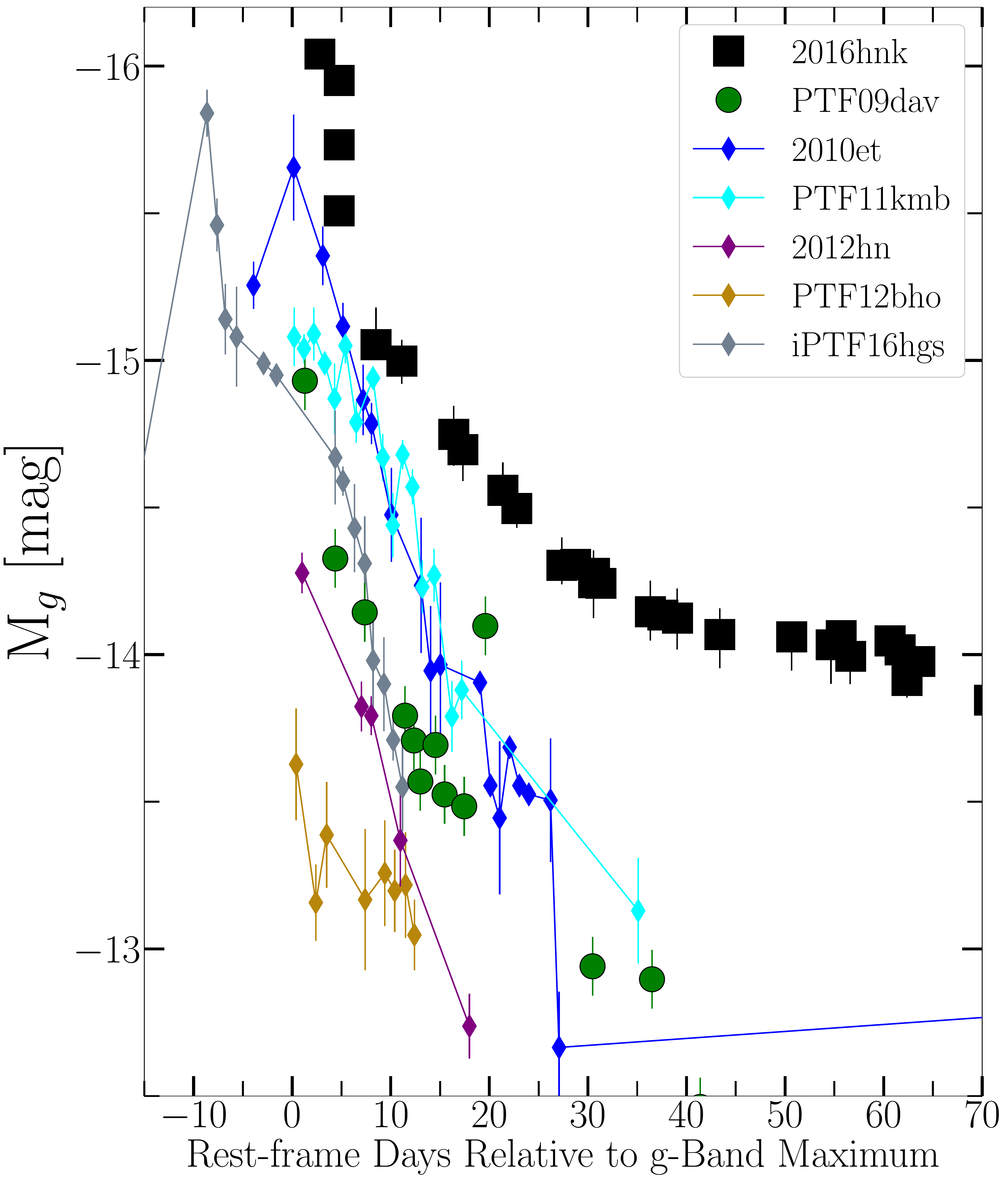}}
\subfigure[]{\includegraphics[width=0.45\textwidth]{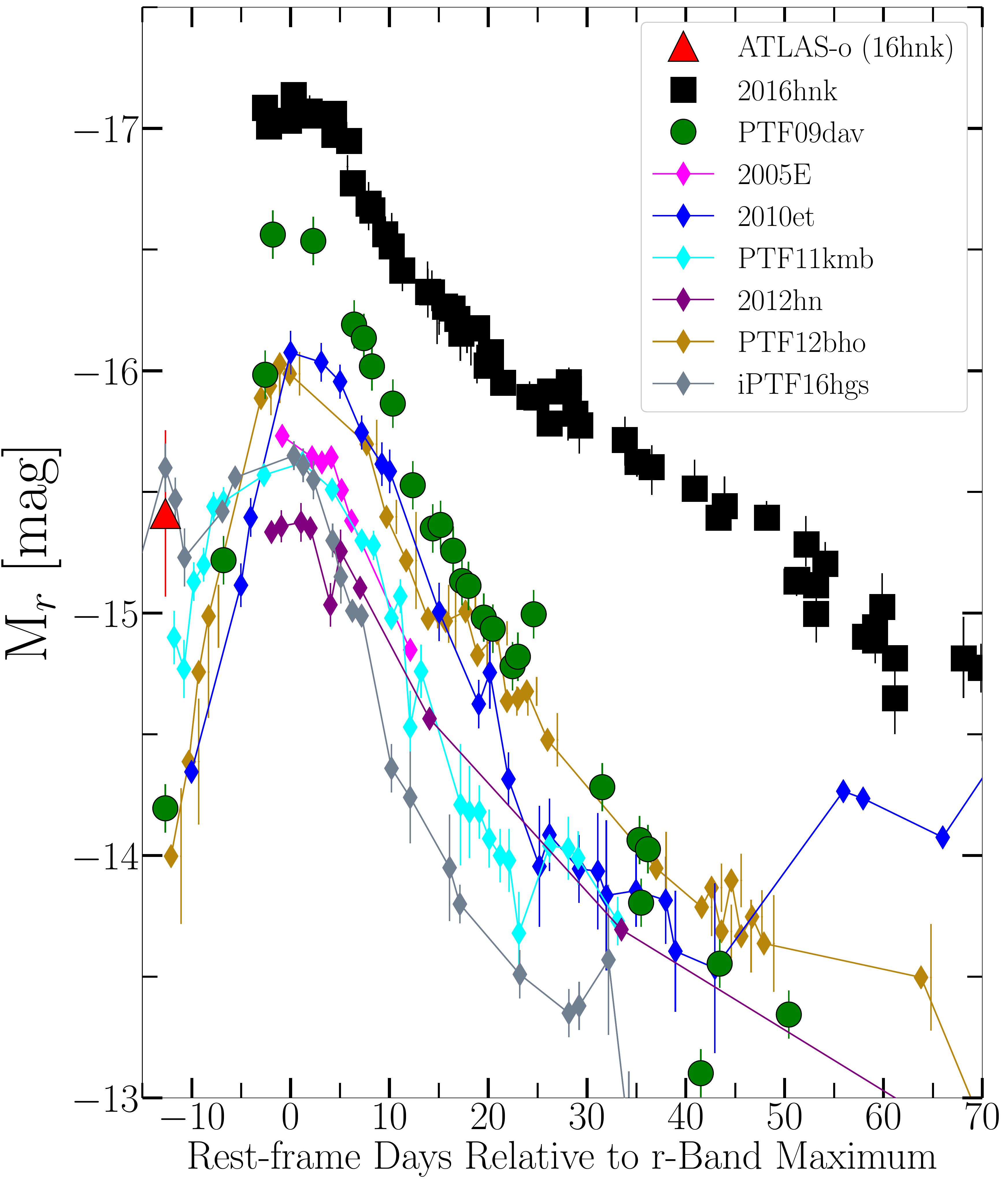}}\\[1ex]
\subfigure[]{\includegraphics[width=.8\textwidth]{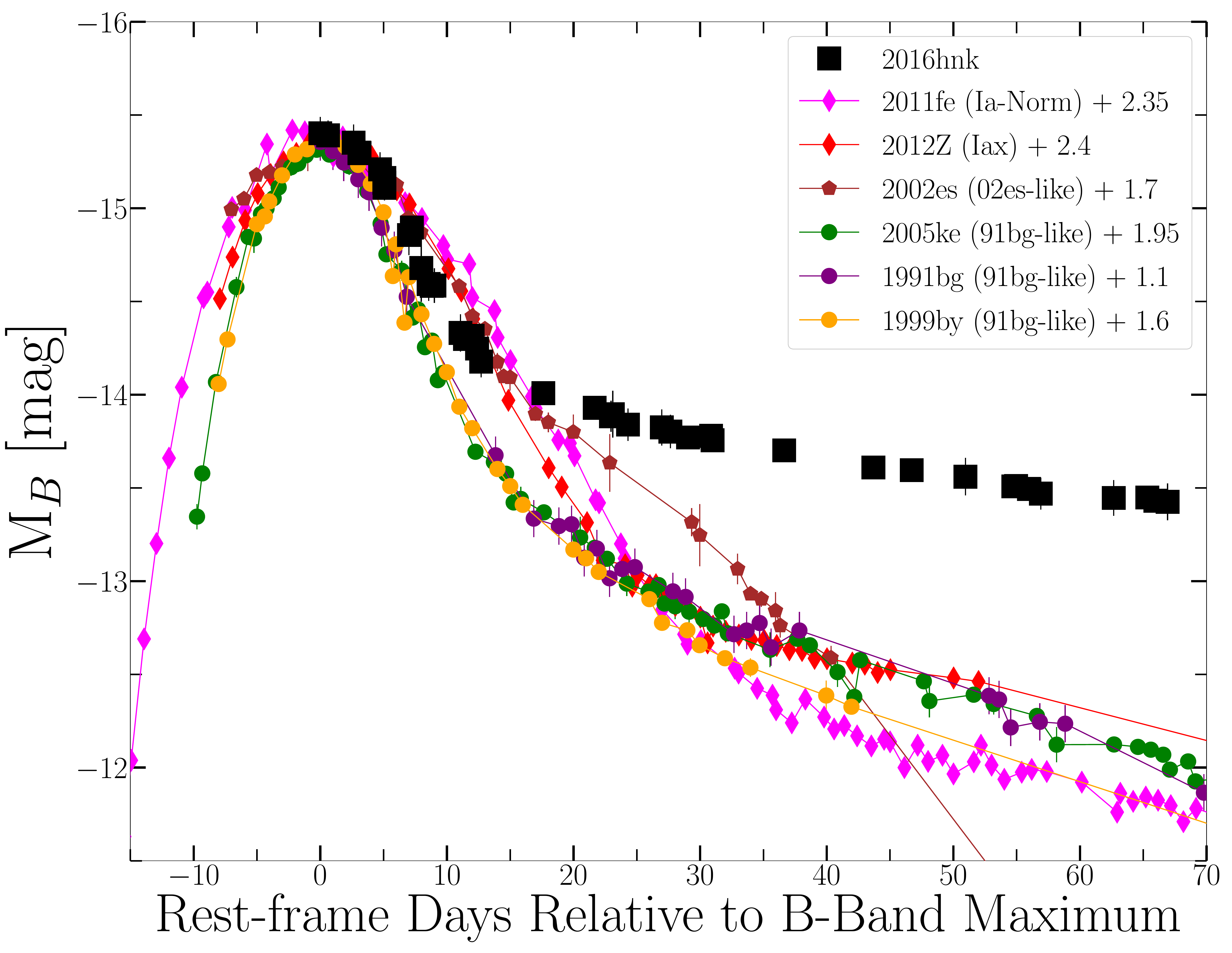}}
\caption{(a) \textit{g}-band comparison of SN~2016hnk and classified Ca-rich transients. (b) \textit{r}-band comparison of SN~2016hnk and known Ca-rich transients. (c) B-band comparison of SN~2016hnk and various sub-classes of SNe Ia, all aligned to the M$_{\textrm{B, max}}$ of SN~2016hnk. The increased flux in SN~2016hnk at t>20d is unlike anything observed in Ca-rich objects or SNe Ia.  \label{fig:LC_compare}}
\end{figure*}

SN~2016hnk has a lower peak $M_B$ than normal SNe~Ia as well as sub-luminous SN~Ia varieties such as SNe~Iax, 91bg-like and 02es-like objects. SN~2016hnk does have a similar $\Delta \rm{m}_{15}(B)$ to SNe~Iax and SN~2002es, but it is significantly slower fading than other Ca-rich transients and 91bg-like objects. $\Delta \rm{m}_{15}(B)$ vs. $M_B$ comparison of these objects and SN~2016hnk is shown in Figure \ref{fig:MB_dm15}.

We present light curve comparisons of SN~2016hnk to peculiar thermonuclear SNe in Figure \ref{fig:LC_compare}. As shown in Figure \ref{fig:LC_compare}(a) \& (b), SN~2016hnk is more luminous than Ca-rich transients SN~2005E \citep{perets05}, PTF 09dav \citep{sullivan11}, SN~2010et \citep{kasliwal12}, PTF11kmb \citep{galyam11}, SN~2012hn \citep{valenti14}, PTF12bho \citep{kasliwal12} and iPTF16hgs \citep{de18}, in addition to having a slower decline rate. The rise time of these Ca-rich objects is faster than that observed in SN~2016hnk, the most similar objects being SN~2007ke and PTF11kmb, each having a rise time of 15 days \citep{kasliwal12,lunnan17}. 

Additionally, we present the absolute magnitude \textit{B}-band light curves of the following SNe~Ia with respect to SN~2016hnk: SN~1991bg \citep{filippenko92AJ,leibundgut93}, SN~1991by \citep{garnavich04}, SN~2002es \citep{ganeshalingam12}, SN~2005ke \citep{galyam05}, SN~2011fe \citep{nugent11, li11} and SN~2012Z \citep{mccully14, stritzinger15}. In addition to a difference in peak magnitude, the B-band decline of SN~2016hnk after 20 days is unlike that of normal/sub-luminous SNe~Ia or SNe~Iax. G19 propose that this late-time light curve excess could be the result of either a light echo or intervening ISM dust at $\approx 1$ parsec from the explosion site. However, G19 show that there is no evidence of a light echo in SN~2016hnk spectra and the lack of significant changes to the \textit{r-i} color evolution at these times also disfavors this scenario.

\begin{figure*}
\centering
\subfigure[]{\includegraphics[width=.6\textwidth]{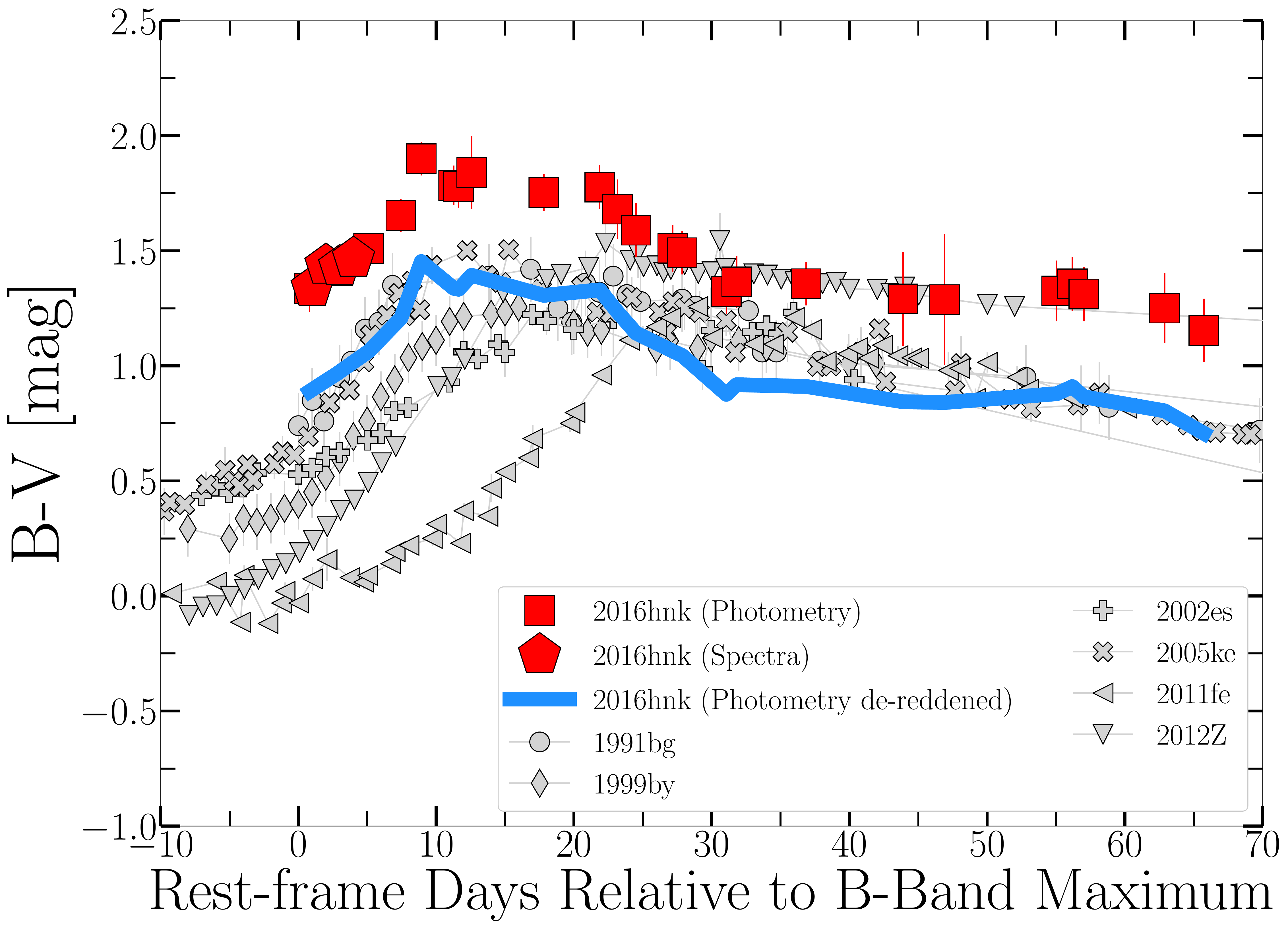}}\\[1ex]
\subfigure[]{\includegraphics[width=.45\textwidth]{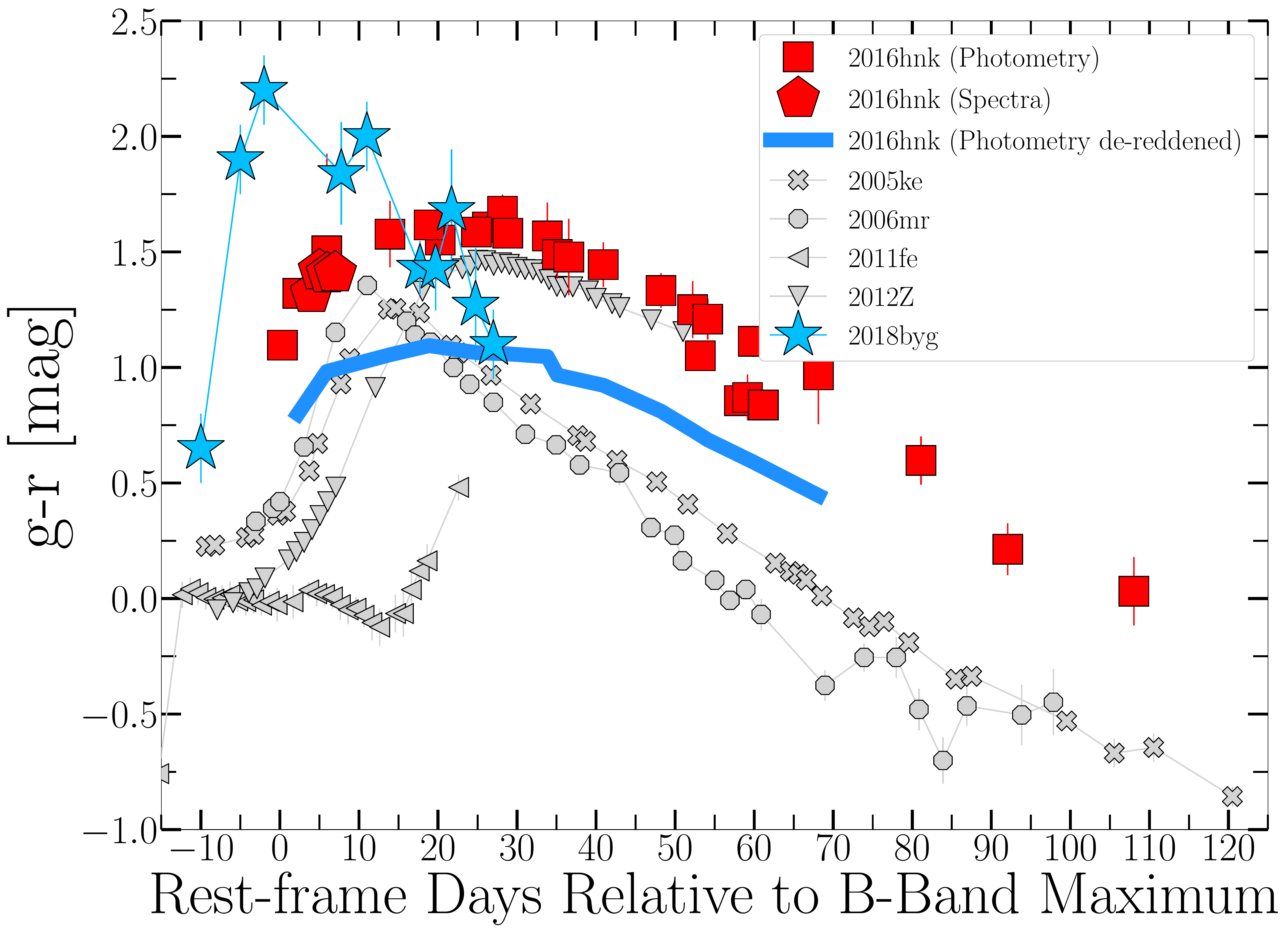}}
\subfigure[]{\includegraphics[width=0.45\textwidth]{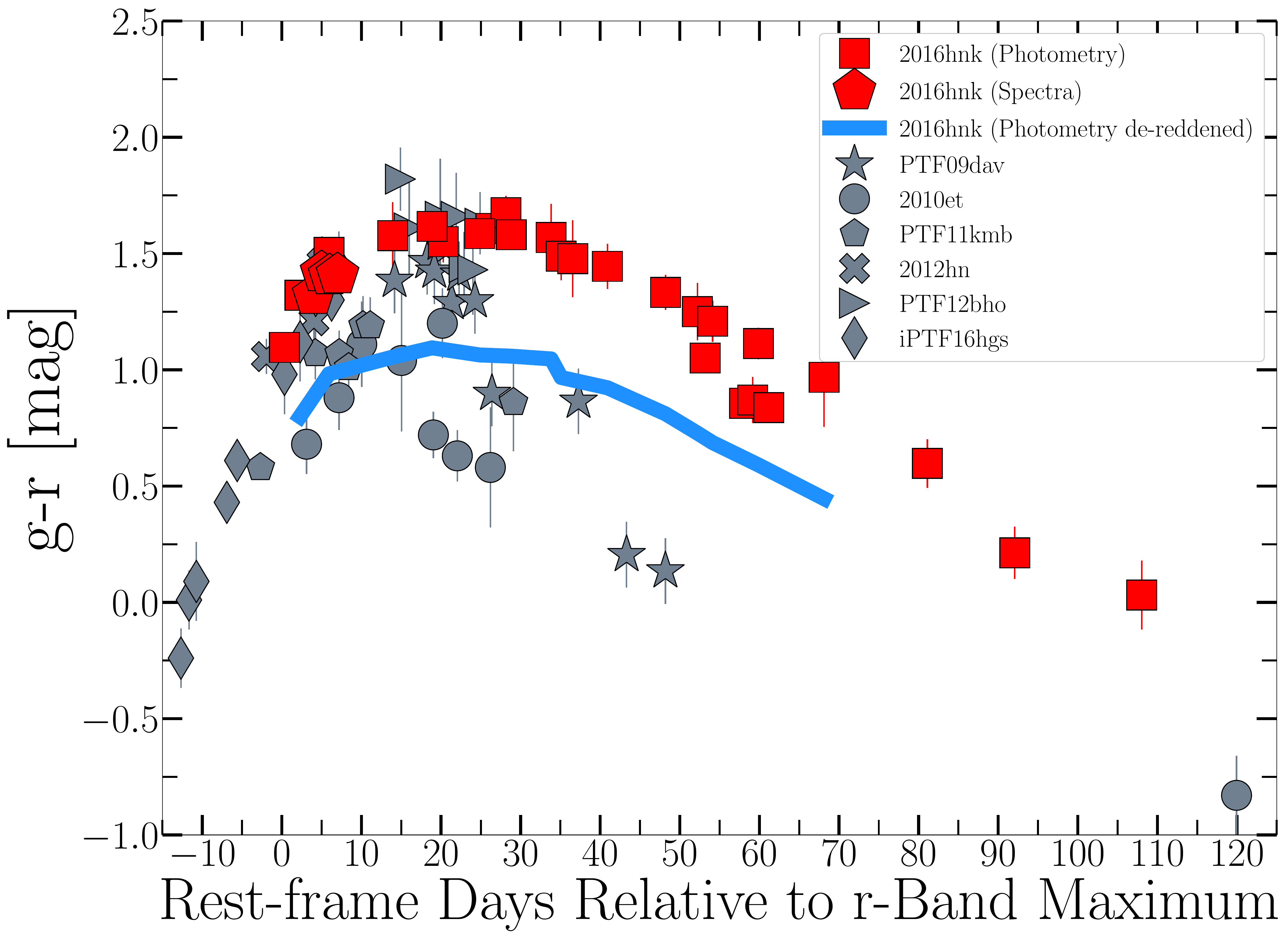}}\\[1ex]
\subfigure[]{\includegraphics[width=.45\textwidth]{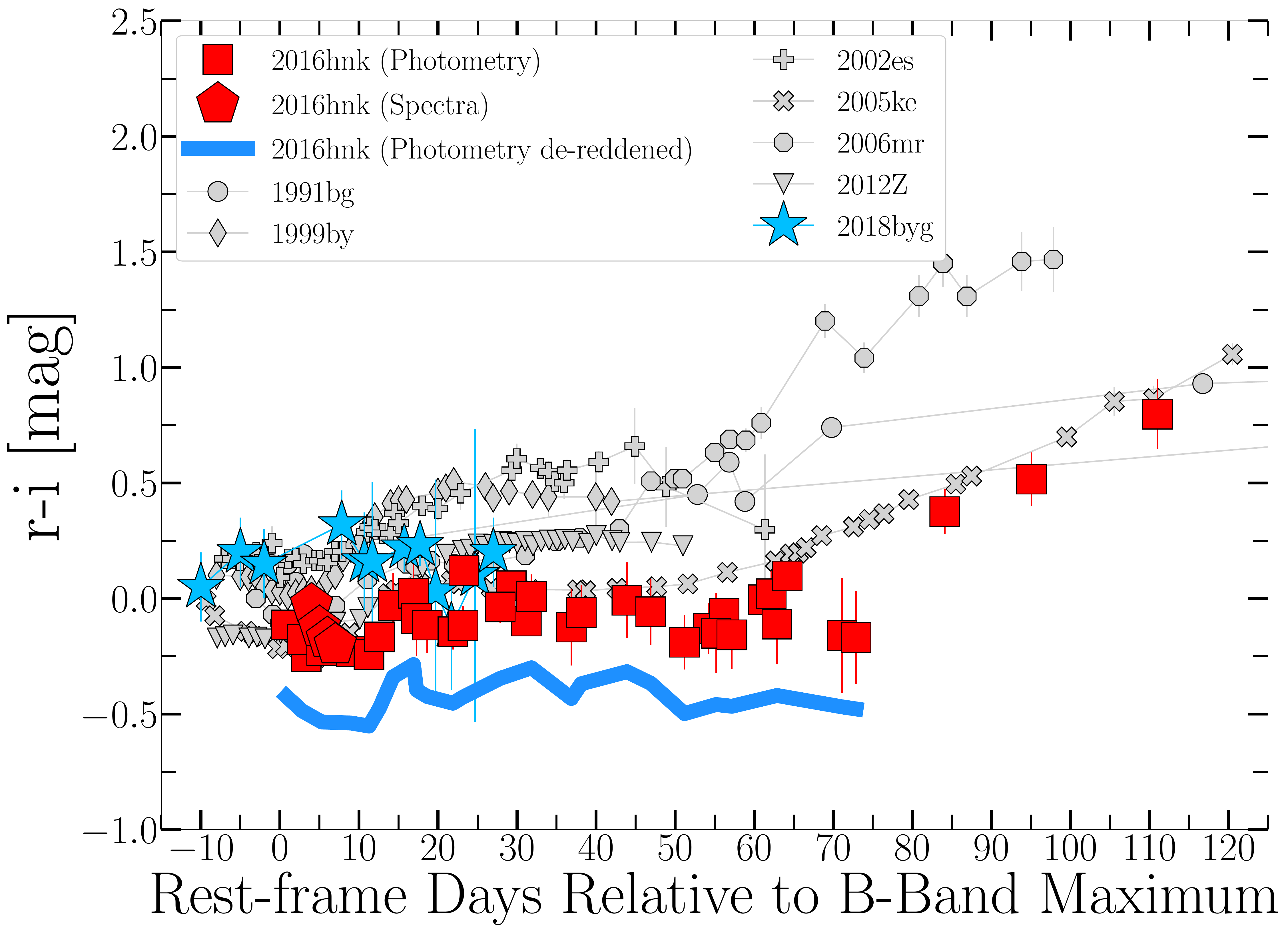}}
\subfigure[]{\includegraphics[width=0.45\textwidth]{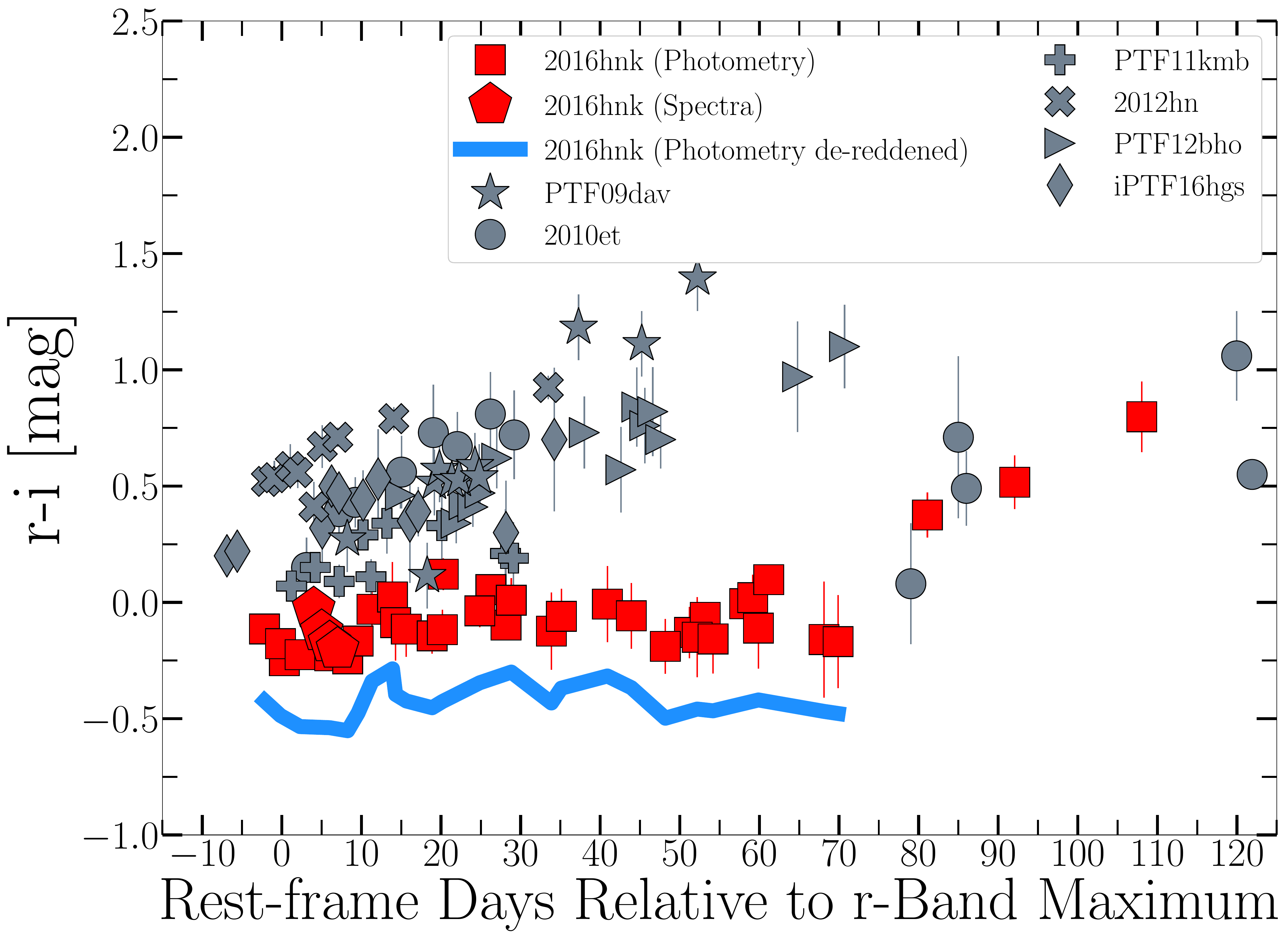}}\\[1ex]
\caption{(a) \textit{B-V} color comparison of SN~2016hnk and various types of SNe Ia. SN~2016hnk colors from photometry presented as red squares and colors from spectra as red polygons. The blue line represents the photometric colors that have been de-reddened to match SNe~Ia color curves. With this artificial de-reddening, SN~2016hnk does not match any SNe Ia in \textit{r-i}. (b) \textit{g-r} color comparison of SN~2016hnk and assorted SNe Ia. SN~2018byg shown as light blue stars. (c) \textit{g-r} color comparison of SN~2016hnk and Ca-rich transients. (d) \textit{r-i} color comparison of SN~2016hnk and different SNe Ia. (e) \textit{r-i} color comparison of SN~2016hnk and Ca-rich transients. \label{fig:colors}}
\end{figure*}

In Figure \ref{fig:colors}, we present \textit{B-V}, \textit{g-r} and \textit{r-i} color comparison plots for SN~2016hnk, Ca-rich transients and SNe~Ia sub-classes. In Figure \ref{fig:colors}(a), SN~2016hnk's \textit{B-V} colors are generally redder than all other varieties of normal and sub-luminous SNe~Ia as well as SNe~Iax. In relation to SNe~Ia, SN~2016hnk's \textit{g-r} and \textit{r-i} color evolution is also significantly different: SN~2016hnk is $\approx 0.3-0.6$ mags redder in \textit{g-r} and $\approx0.25-0.5$ mags bluer in \textit{r-i} than even the most similar sub-luminous SNe~Ia. Additionally, we present \textit{g-r} and \textit{r-i} colors of SN~2018byg as light blue stars in Figure \ref{fig:colors}(b) \& (d). The color evolution of SN~2016hnk is consistent with SN~2018byg to within 0.5 mag in \textit{r-i} and \textit{g-r}. However, SN~2018byg is slightly redder at most epochs. Comparing to Ca-rich objects, SN~2016hnk is consistent to within 0.2 mags in \textit{g-r} colors, but is $\approx0.5$ mag bluer than the typical Ca-rich object in \textit{r-i}. As shown by the pink squares in Figure \ref{fig:colors}, if there were dust reddening from the host-galaxy, SN~2016hnk would be even bluer in \textit{r-i}, making it an exceptionally odd object. 

\subsection{Pseudo-Bolometric Light Curve}\label{subsec:LC_bolometric}
\begin{figure*}
\centering
\subfigure[]{\includegraphics[width=0.48\textwidth]{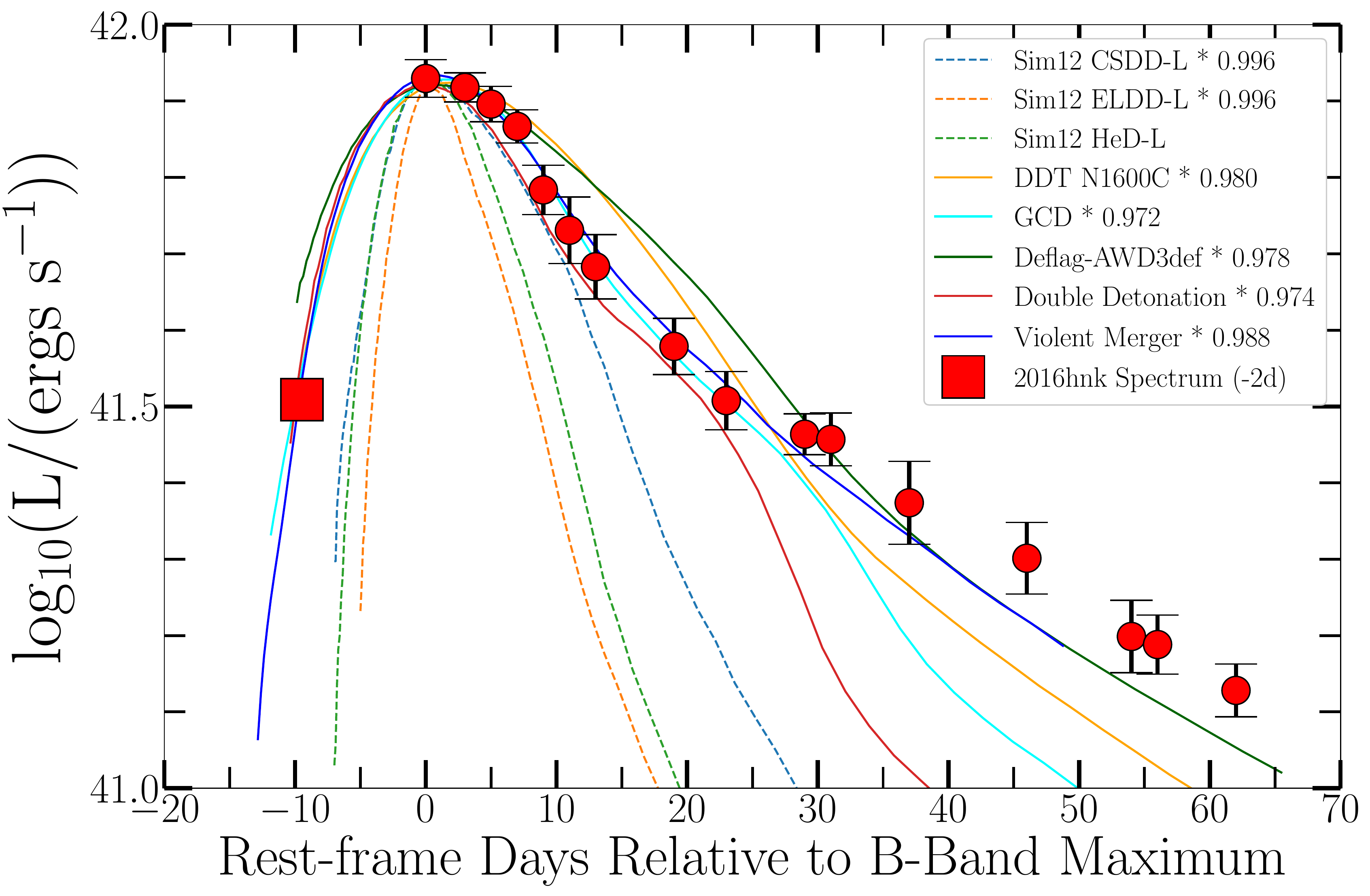}}
\subfigure[]{\includegraphics[width=0.48\textwidth]{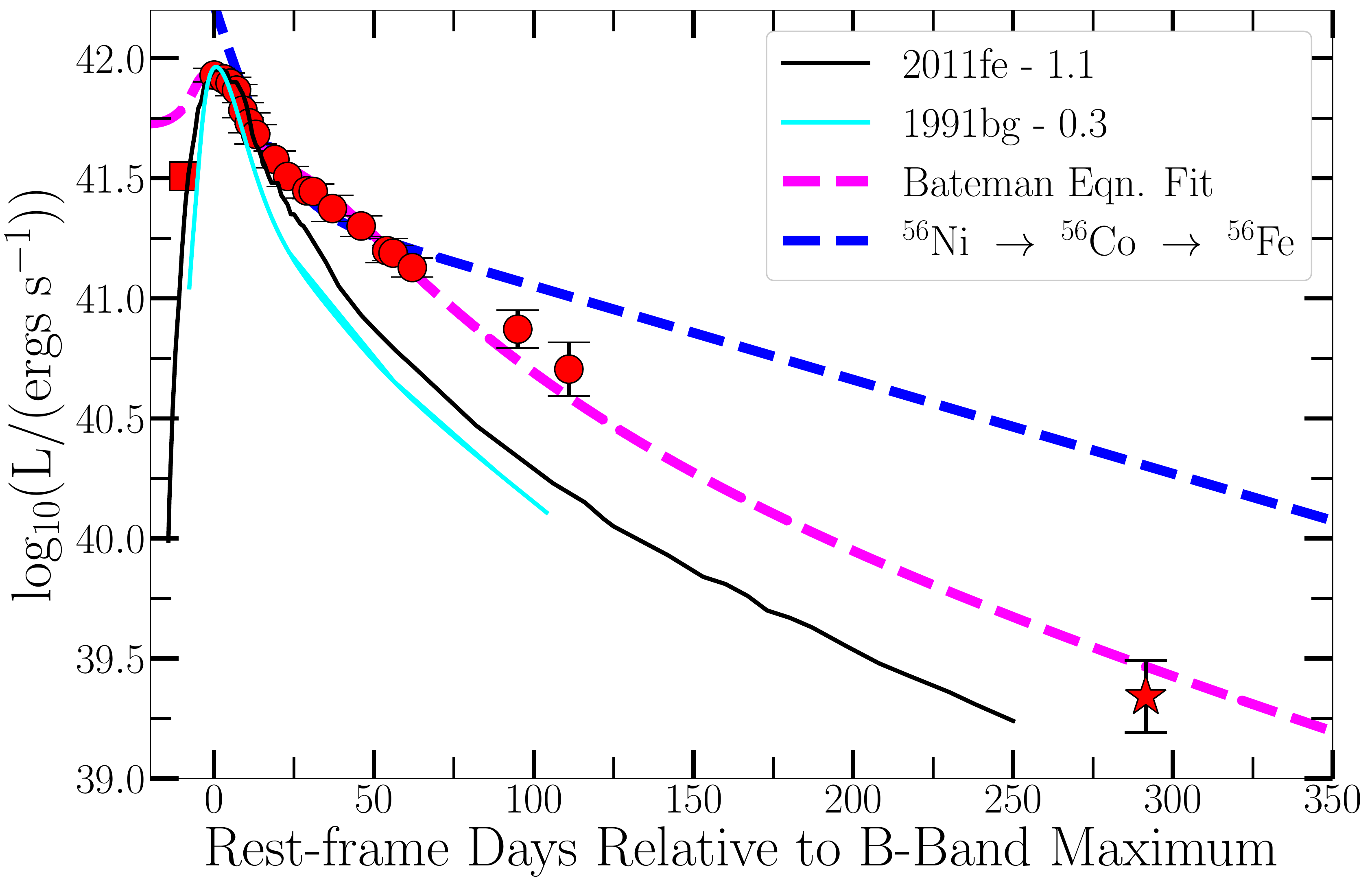}}
\caption{(a) Early-time pseudo-bolometric light curve of SN~2016hnk. The detection luminosity has been estimated using the ATLAS o-band observation and pre-maximum spectra of SNe 2016hnk and 2018byg. Spectral phases relative to \textit{r}-band maximum and plotted with respect to \textit{B}-band. We also show explosion models from the Heidelberg Group for reference. (b) Full pseudo-bolometric light curve of SN~2016hnk with the $\sim$300 day luminosity estimate calculated using the Keck \textit{I}-band detection and nebular spectra from a similar phase. Psuedo-bolometric light curves of SN~1991bg and 2011fe are shown in cyan and black, respectively \citep{stritzinger06, zhang16}. The Bateman equation fit is shown in magenta, while the standard ${}^{56}\textrm{Co}$ decay with complete $\gamma$-ray trapping shown in blue. \label{fig:bol_LC}}
\end{figure*}

We construct a pseudo-bolometric light curve for SN~2016hnk using a combination of multi-color optical photometry from Swope, PS1 and LCOGT observations. Luminosities are calculated by trapezoidal integration of SN flux in \textit{BVgriz} filters (3000-9000 \AA). In regions without complete color information, we extrapolated between light curve data points using a low-order polynomial spline. We present the early-time, pseudo-bolometric light curve in Figure \ref{fig:bol_LC}(a). For reference, we display models from the Heidelberg Supernova Model Archive (HESMA)\footnote{\url{https://hesma.h-its.org/doku.php?id=start}} that include various binary configurations and explosion mechanisms. We also include an estimated pre-maximum bolometric luminosity at -9.5d by integrating the pre-maximum brightness spectra of SN~2016hnk and a similar object SN~2018byg. These spectra are scaled to the first ATLAs-o detection and each phase is relative to \textit{r}-band peak brightness. 

Using the constructed light curve, we find a peak luminosity of $(8.5 \pm \: 0.5) \times 10^{41} \: \mathrm{erg\:s^{-1}}$. This value is smaller than that calculated in G19 who apply a significant host-galaxy reddening correction. We, however, do not apply such a correction because we find no evidence for host extinction (see Section \ref{subsec:photometry}). We can then use $L_{\textrm{bol}}$ at peak to estimate the total mass of synthesized ${}^{56}\textrm{Ni}$, $M_{\textrm{Ni}}$, assuming that the radiated luminosity is generated via thermalization of $\gamma$-rays from the $\beta$-decay of ${}^{56}\textrm{Ni}$ $\rightarrow$ ${}^{56}\textrm{Co}$ \citep{arnett82}. We use the following relation to calculate ${}^{56}\textrm{Ni}$ mass:
\begin{equation}\label{eq:m_ni56}
M_{\textrm{Ni}} = \frac{L_{\textrm{bol}}}{\gamma \dot{S}(\tau_{\textrm{r}})}
\end{equation}
where $\tau_{\textrm{r}}$ is the rise time and $\gamma$ is the ratio of bolometric to radioactivity luminosities. As in \cite{nugent95}, we adopt $\gamma = 1.2 \pm 0.2$. $\dot{S}$ is the radioactivity luminosity per solar mass of ${}^{56}\textrm{Ni}$ decay:

\begin{equation}\label{eq:sdot}
\dot{S} = \Big(6.31e^{-\tau_{\textrm{r}}/8.8\textrm{d}} + 1.43e^{-\tau_{\textrm{r}}/111\textrm{d}}\Big) \times 10^{43} \ \textrm{erg} \ \textrm{s}^{-1} \ \textrm{M}_{\odot}^{-1} 
\end{equation}

Using $\tau_{\textrm{r}} = 15 \pm 2$ days (Section \ref{subsec:LC_properties}), we calculate $M_{\textrm{Ni}} = 0.03 \pm 0.01 M_{\odot}$. This value is consistent to 1$\sigma$ with that reported for PTF09dav ($0.019\pm0.003M_{\odot}$; \citealt{sullivan11}) and consistent to $\sim$2$\sigma$ with the lowest luminosity 91bg-like object, SN~2007ax ($0.038\pm0.008M_{\odot}$). The total ${}^{56}\textrm{Ni}$ mass mass in SN~2016hnk is also lower than that found for SN~2018byg ($\approx 0.11\Msun$; \citealt{de19}), which is consistent with the observed difference in luminosity between the two objects at peak. The total ${}^{56}\textrm{Ni}$ mass in SN~2016hnk is still significantly less than the majority of sub-luminous SNe~Ia. 

We then use the relation from \cite{foley09} to estimate the ejecta mass of SN~2016hnk:

\begin{equation}\label{eq:ejecta_mass}
M_{\rm{ej}} = 0.16 \Big(\frac{\tau_{\textrm{r}}}{10\rm{d}}\Big)^2 \Big(\frac{0.1 \rm{cm}^2\rm{g}^{-1}}{\kappa}\Big)\Big(\frac{v}{2\times10^8 \rm{cm \ s^{-1}}}\Big)
\end{equation}
where $\kappa$ is the opacity and $v$ is the ejecta velocity. Using a standard SN~Ia opacity of $0.2 \ \rm{cm}^2\rm{g}^{-1}$ and a calculated ejecta velocity of $-10,300 \ \kms$ from Si II absorption, we find $M_{\rm{ej}} = 0.9 \pm 0.3 \ \Msun$. Given the strength of Ca II in the SN~2016hnk spectra, the true opacity could be larger than a normal SN~Ia, which in turn would yield a smaller ejecta mass than our current estimate.


We also examine the decline of pseudo-bolometric light curve at late-times to check the consistency with standard ${}^{56}\textrm{Co}$ decay. As shown in Figure \ref{fig:bol_LC}(b), the lack of multi-filter, photometric data after $\sim$100 days makes it difficult to calculate precise late-time bolometric luminosities for SN~2016hnk. However, we can place an estimate on the bolometric luminosity by scaling nebular spectra of SN~2016hnk and similar objects to the Keck \textit{I}-band flux at +291d. If we first assume that all the flux at late-times is derived from [Ca II] emission, we can scale the Keck LRIS spectrum at +264d to the \textit{I}-band flux and integrate over the wavelength range of the Keck \textit{I}-band transmission function. With this method, we calculate a luminosity of $2.2 \pm 0.90 \times10^{39} $ ergs/s (red star in Figure \ref{fig:bol_LC}(b)). The phase of the Keck spectrum and the \textit{I}-band detection are not identical, but this does not add extra uncertainty on the luminosity if we assume that the spectral evolution between epochs is small. Additional uncertainties on this measurement originates the lack of photometric detections in other Keck filters; this may indicate additional flux outside of the \textit{I}-band transmission curve. However, we consider this unlikely given that the only detectable feature in the Keck spectrum is [Ca II]. We place additional constraints on SN~2016hnk's bolometric decline by estimating its luminosity using SNe~Iax 2002cx and 2008A, which have a similar Ca-dominated spectral composition at nebular times. Scaling the spectra of these objects at a similar phase to the Keck \textit{I}-band flux, we calculate additional bolometric luminosities for SN~2016hnk. Luminosities from 2002cx and 2008A SEDs are within the errorbars present in \ref{fig:bol_LC}(b).

We then use our estimates for the late-time bolometric luminosity of SN~2016hnk to explore the standard ${}^{56}\textrm{Ni}$-powered decline that is typically observed in SNe~Ia. As shown in Figure \ref{fig:bol_LC}(b), the bolometric decline of SN~2016hnk is slower than that of SN~2011fe (black; \citealt{zhang16}) and SN~1991bg (cyan; \citealt{stritzinger06}). The SN~2016hnk decline rate is also inconsistent with a simple ${}^{56}\textrm{Co}$ $\rightarrow$ ${}^{56}\textrm{Fe}$ decay (i.e. complete $\gamma$-ray trapping) as presented by the dashed blue line in \ref{fig:bol_LC}(b). 

We model the radioactive decay powered light curve in a similar method to late-time studies of SNe~Ia \citep{seitenzahl13a, Graur16, shappee17, dimitriadis17, jacobson-galan18}. We employ the Bateman Equation to fit the bolometric decline produced by radioactive isotopes: 

\vspace*{1mm}
\begin{equation}\label{eq:bateman}
\begin{split}
&L_A(t) = 2.221 \frac{\lambda_A}{A} \frac{M(A)}{M_{\odot}} \frac{q^x_A + q^l_Af^{l}_A(t) +  q^{\gamma}_Af^{\gamma}_A(t)}{\mathrm{keV}}\\
& \ \ \ \ \ \ \ \ \ \ \ \ \ \ \ \ \ \ \ \ \ \ \ \ \ \ \ \ \ \ \textrm{exp}(-\lambda_{A}t) \ \times \ \ 10^{43} \ \textrm{erg} \ \textrm{s}^{-1}
\end{split}
\end{equation}
where $t$ is time since explosion, $\lambda_A$ is the decay constant, $A$ is the atomic number, and $q^l$, $q^\gamma$, and $q^x$ are the average energies of charged leptons, $\gamma$-rays, and X-rays, respectively, per decay. All decay energies and constants as presented in Table 2 of \cite{seitenzahl14}. We assume complete trapping of energy deposited by charged leptons (i.e., $f^{l}_{A} = 0$) and do not include X-ray energies in our fits. We use the following expression for $\gamma$-ray trapping:

\begin{equation}\label{eq:trapping}
f^{\gamma}_{A} = 1- \textrm{exp}\Big[-\Big(\frac{t^{\gamma}_{A}}{t}\Big)^2 \Big]
\end{equation}
where $t^{\gamma}_{A}$ is the $\gamma$-ray trapping timescale for a given radioactive isotope produced in the explosion \citep{woosley89, seitenzahl14}. 

In our model, we fit for the mass and $\gamma$-ray trapping timescale of ${}^{56}\textrm{Co}$ as well as the mass of ${}^{57}\textrm{Co}$. However, the luminosity contribution is dominated by synthesized ${}^{56}\textrm{Co}$ until $\approx 300$ days after explosion thus making the ${}^{57}\textrm{Co}$ contribution negligible on this time scale. Modeling the data with Equation \ref{eq:bateman}, we find ${}^{56}\textrm{Co}$ and ${}^{57}\textrm{Co}$ masses of $0.029 \pm 0.001$ and $\approx 0 \ \Msun$, respectively. This result is consistent with our total ${}^{56}\textrm{Ni}$ mass estimate of $M_{\textrm{Ni}} = 0.03 \pm 0.006 M_{\odot}$. We also calculate a trapping timescale of $t^{\gamma}_{56} = 60.06^{+4.12}_{-3.71}$ days. We display our model fit as the dashed magenta line in Figure \ref{fig:bol_LC}(b).


\section{Spectral Analysis}\label{sec:spectra_analysis}

\subsection{Line Identification}\label{subsec:spectra_compare}

\begin{figure}
\centering
\vspace*{-2mm}
\subfigure[]{\includegraphics[width=.47\textwidth]{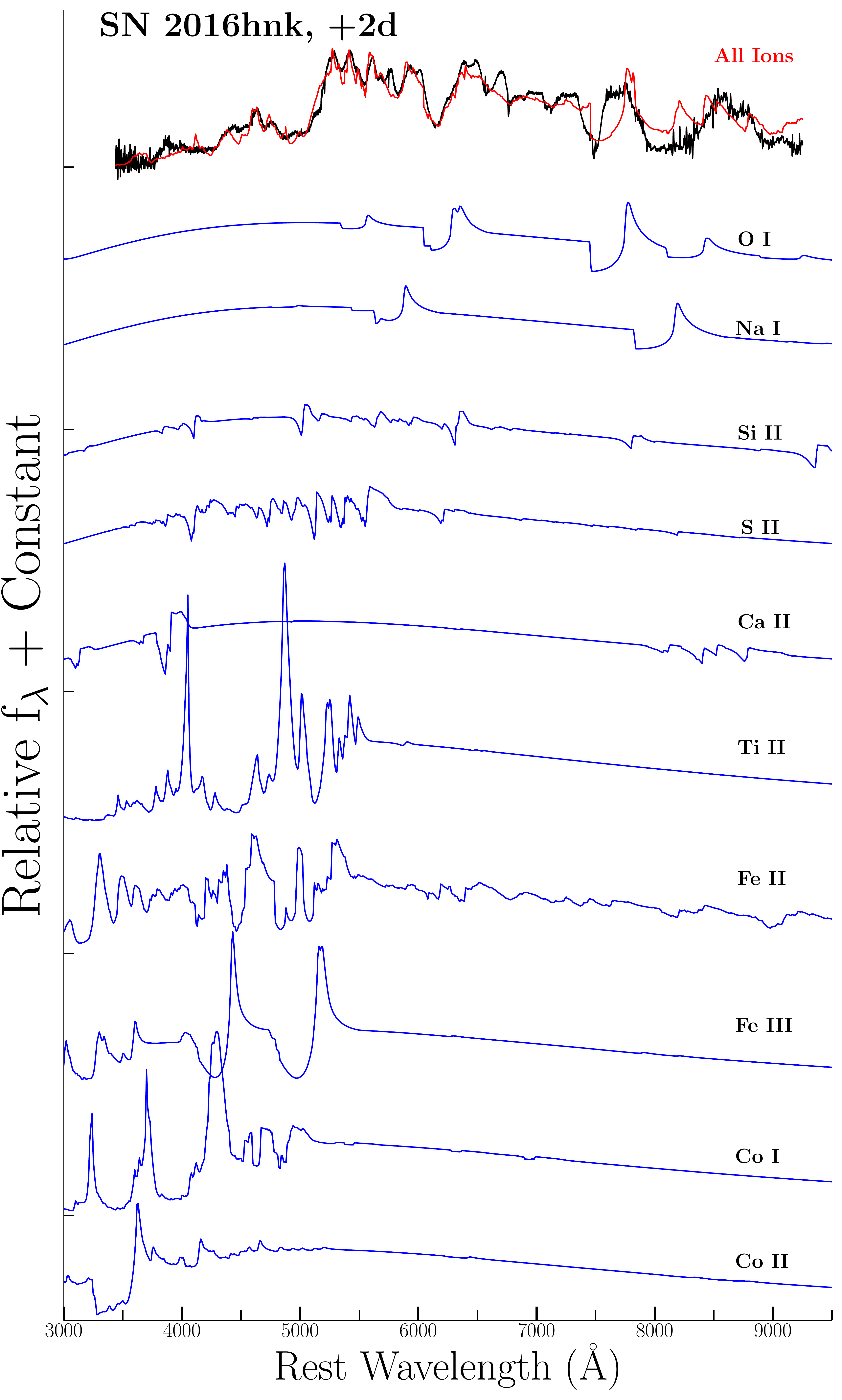}}
\vspace*{-2mm}
\subfigure[]{\includegraphics[width=.47\textwidth]{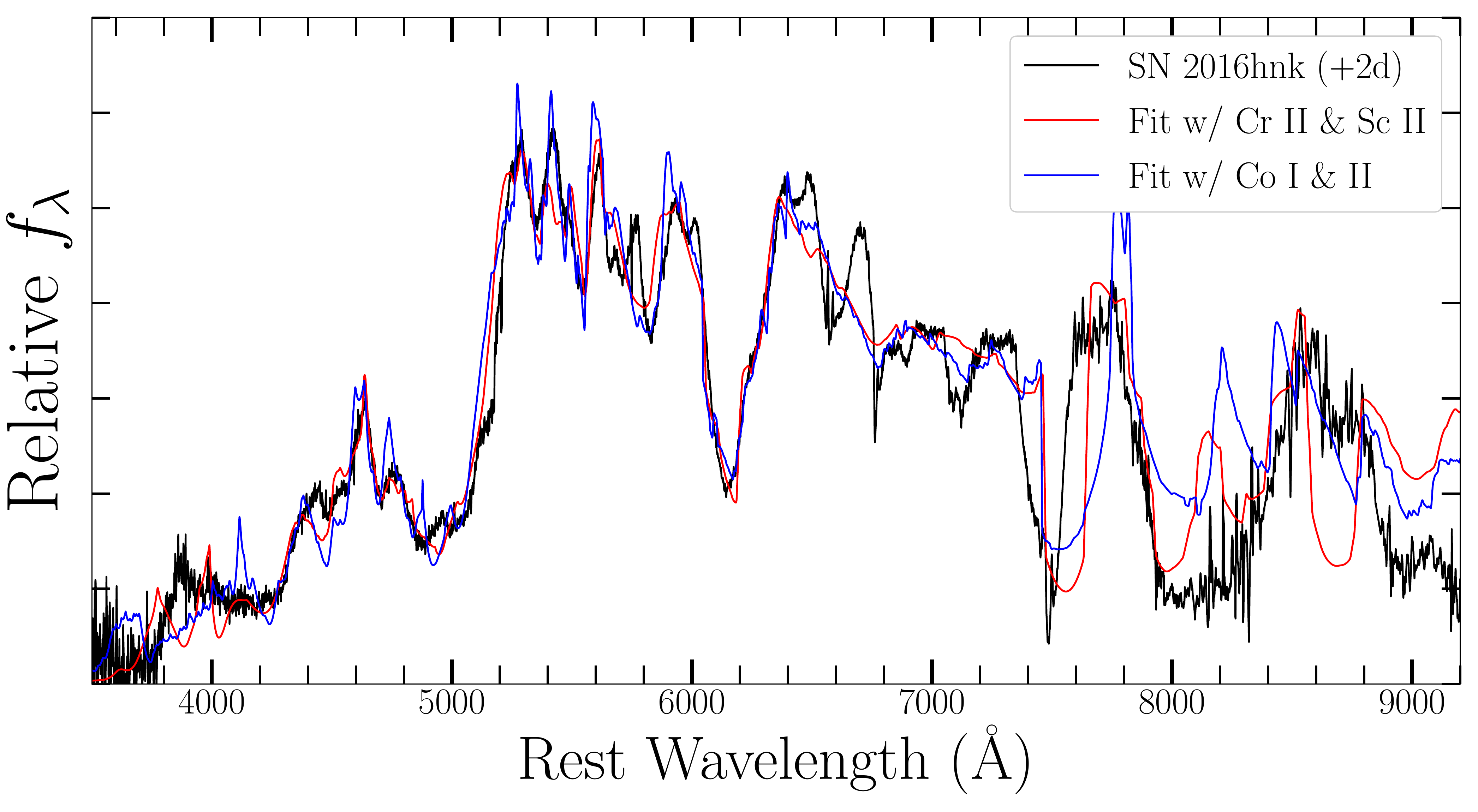}}
\caption{(a) Decomposition of active ions in \texttt{SYNAPPS} fit. Phase relative to \textit{B}-band maximum. Total fit is shown in red, while blue lines mark each individual ion's contribution. (b) Similar \texttt{SYNAPPS} fits to SN~2016hnk. In red, the fit includes all active ions shown in (a) except with Sc II and Cr II instead Co I and Co II.  \label{fig:synapps}}
\end{figure}

We model the early-time spectra of SN~2016hnk with spectral synthesis code \texttt{SYNAPPS} \citep{thomas11} in order to understand the elements produced in the explosion. \texttt{SYNAPPS} is utilized primarily for line identification and relies on generalized assumptions about the SN such as spherical symmetry, local thermal equilibrium, and homologous expansion of ejecta. As shown in Figure \ref{fig:synapps}(a), we identify the following species most commonly found in thermonuclear SNe: O I,  Si II, S II, Ca II, Ti II, Fe II/III and Co I/II. We do not detect C I nor Mg I in our fits and the detection of Na I is probable but not definite. We also find no detectable He I in any \texttt{SYNAPPS} modeled spectra. This finding is unlike other Ca-rich objects, which typically show clear signatures of photospheric helium near peak. 

We ran various \texttt{SYNAPPS} fits in an attempt to identify more exotic elements such as Cr II, Sc II and Sr II, all of which are claimed to be robustly detected in \texttt{SYNAPPS} models of PTF09dav by \cite{sullivan11}. We present a model comparison in Figure \ref{fig:synapps}(b) of a fit including typical thermonuclear species (e.g., Fe II/III, Ti II, Co I/II; (blue line)) and a fit with both typical species (e.g., Fe II/III and Ti II; (red line)) in additional to exotic species (e.g., Cr II and Sc II). The two line profiles at $\lambda\lambda5400, 5550 \AA$ are more accurately fit with blends of S II and Fe-group elements than Sc II. We also find that adding Cr II does not improve our \texttt{SYNAPPS} fits. Despite the visual similarities between near-peak spectra of SN~2016hnk and PTF09dav, we cannot place the same level of confidence on the detection of Sc II, Cr II and Sr II in SN~2016hnk as \cite{sullivan11} did for PTF09dav.

\subsection{Spectral Comparisons}\label{subsec:spectra_compare}

\begin{figure}[h]
\centering
\includegraphics[width=0.45\textwidth]{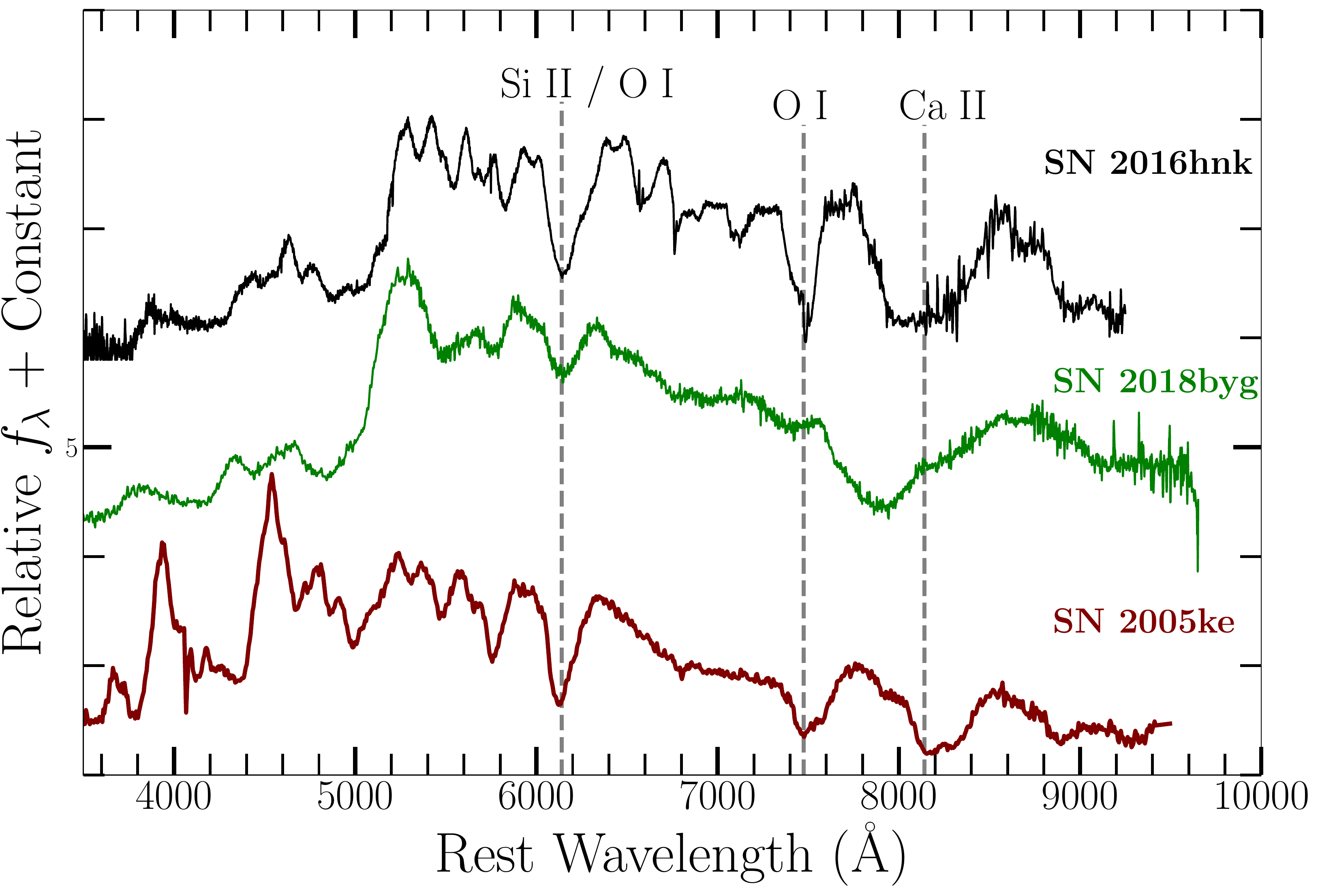}
\caption{Spectral comparison of SNe~2016hnk (black), 2018byg (green) and 2005ke (maroon) at similar phases near peak magnitude. SNe~2016hnk and 2018byg have significant bluewards line-blanketing amongst Fe-group elements and higher velocity Ca II absorption than 91bg-like SN~2005ke. \label{fig:16hnk_91bg_18byg}}
\end{figure}

SN~2016hnk is extremely similar to SN~2018byg at each phase of spectroscopic observations presented in \ref{fig:2018byg_compare}(b). At similar phases near maximum light, both objects contain broad, high-velocity Ca II absorption features with velocities of $-17151.41 \pm 540.10 \ \kms$ in SN~2016hnk and $-22221.70 \pm 706.0 \ \kms$ in SN~2018byg. Ca II and Si II velocities were calculated by fitting the minimum of each absorption feature using a low-order polynomial. A more complete velocity evolution for both objects is presented in Figure \ref{fig:models_spectra_vels}. Furthermore, the spectra of SNe~2016hnk and 2018byg both show suppressed emission from line blanketing of Fe-group elements. As indicated by the colors shown in Figure \ref{fig:colors}, SN~2016hnk's ``redder'' \textit{B-V} colors with respect to its ``bluer'' \textit{r-i} colors is the result of this blue-wards line blanketing. This characteristic is unique to these two SNe and sets them apart from other sub-luminous thermonuclear objects such as 91bg-like objects (Figure \ref{fig:16hnk_91bg_18byg}).

Spectroscopic differences between the two objects include higher overall elemental velocities in SN~2018byg and deeper Si II / O I absorption in SN~2016hnk. There are no published nebular spectra of SN 2018byg to compare with our late-time SN~2016hnk spectral observations. 

\begin{figure}
\centering
\vspace*{-2mm}
\subfigure[]{\includegraphics[width=.47\textwidth]{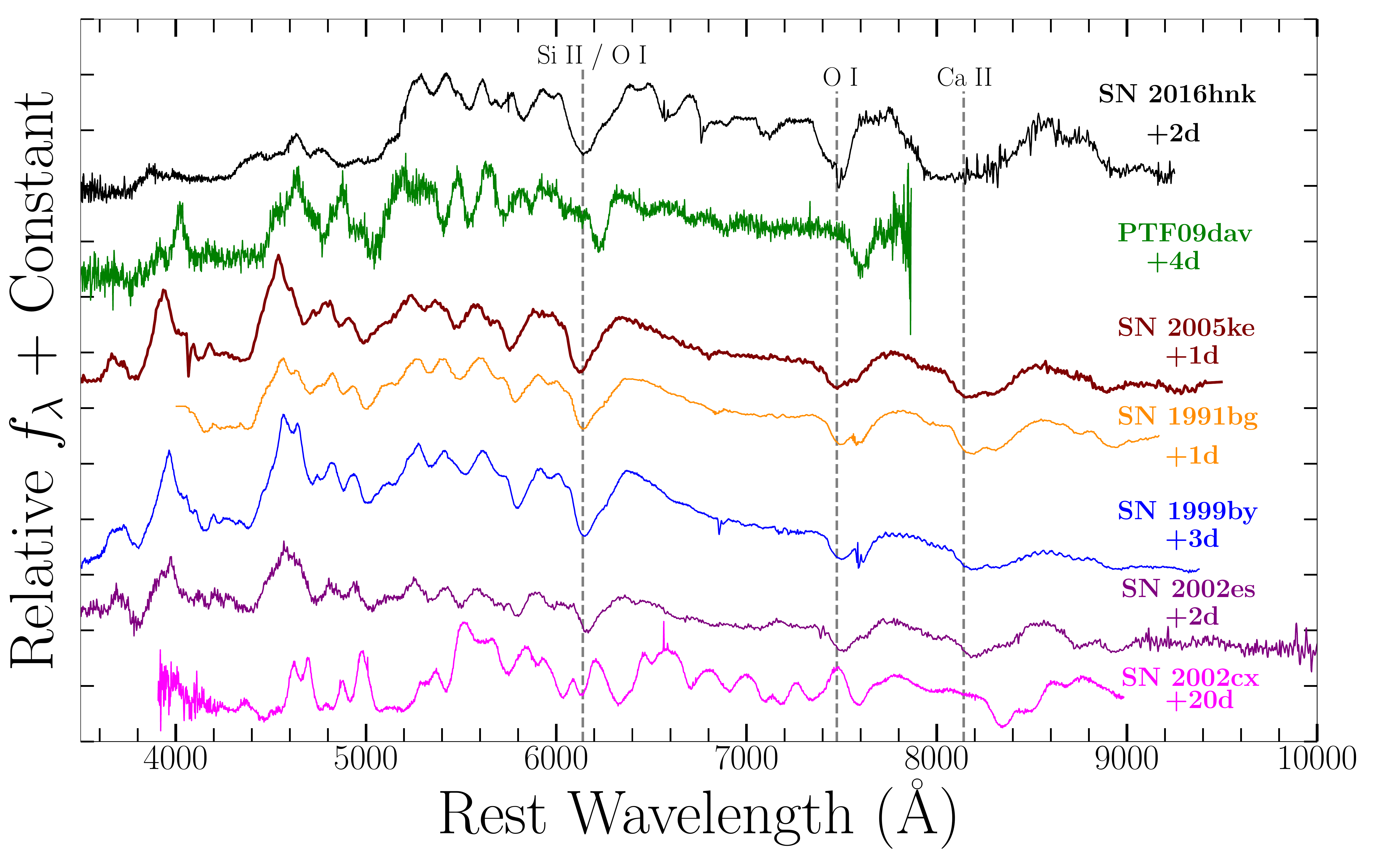}}
\vspace*{-2mm}
\subfigure[]{\includegraphics[width=.47\textwidth]{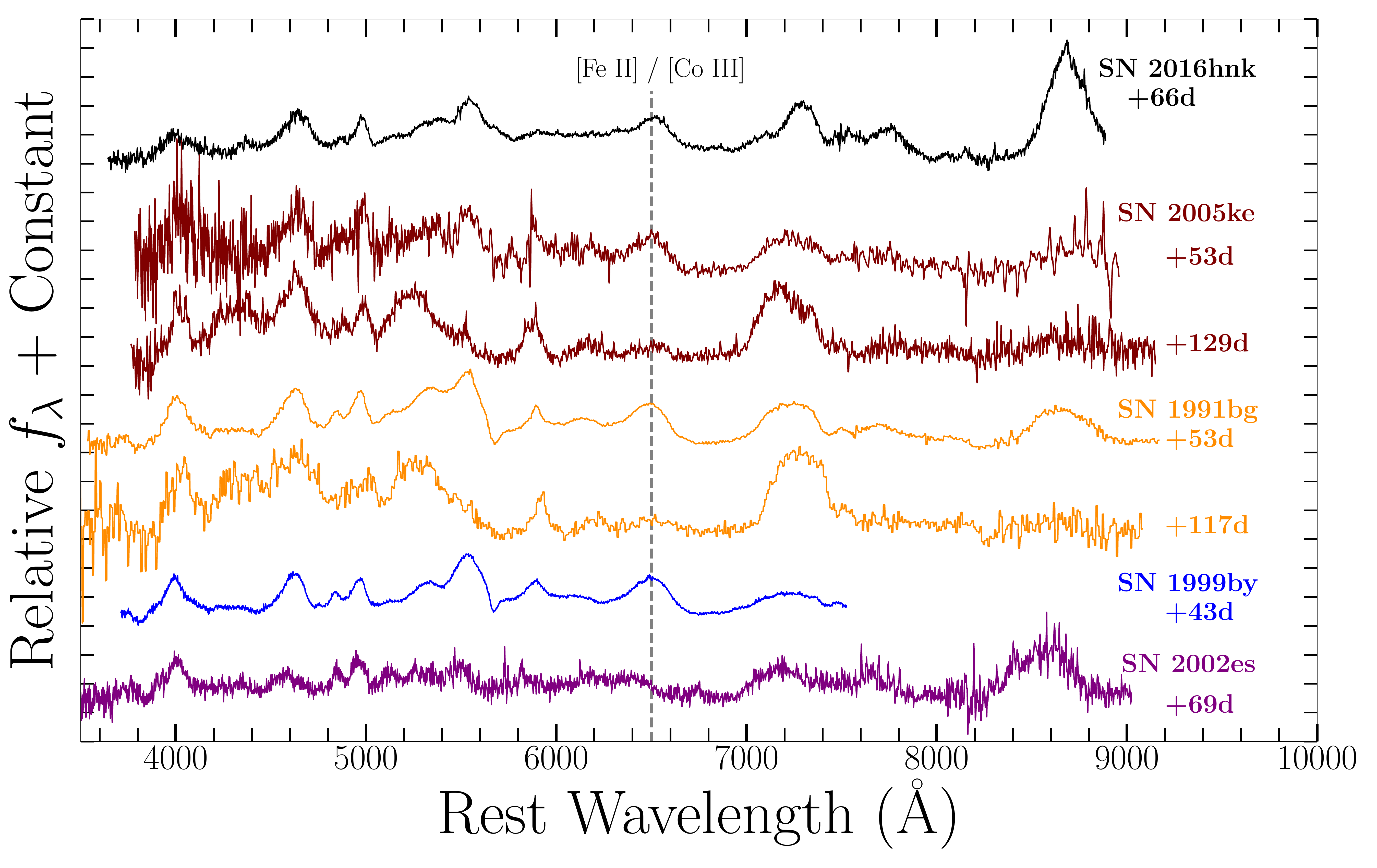}}
\subfigure[]{\includegraphics[width=.47\textwidth]{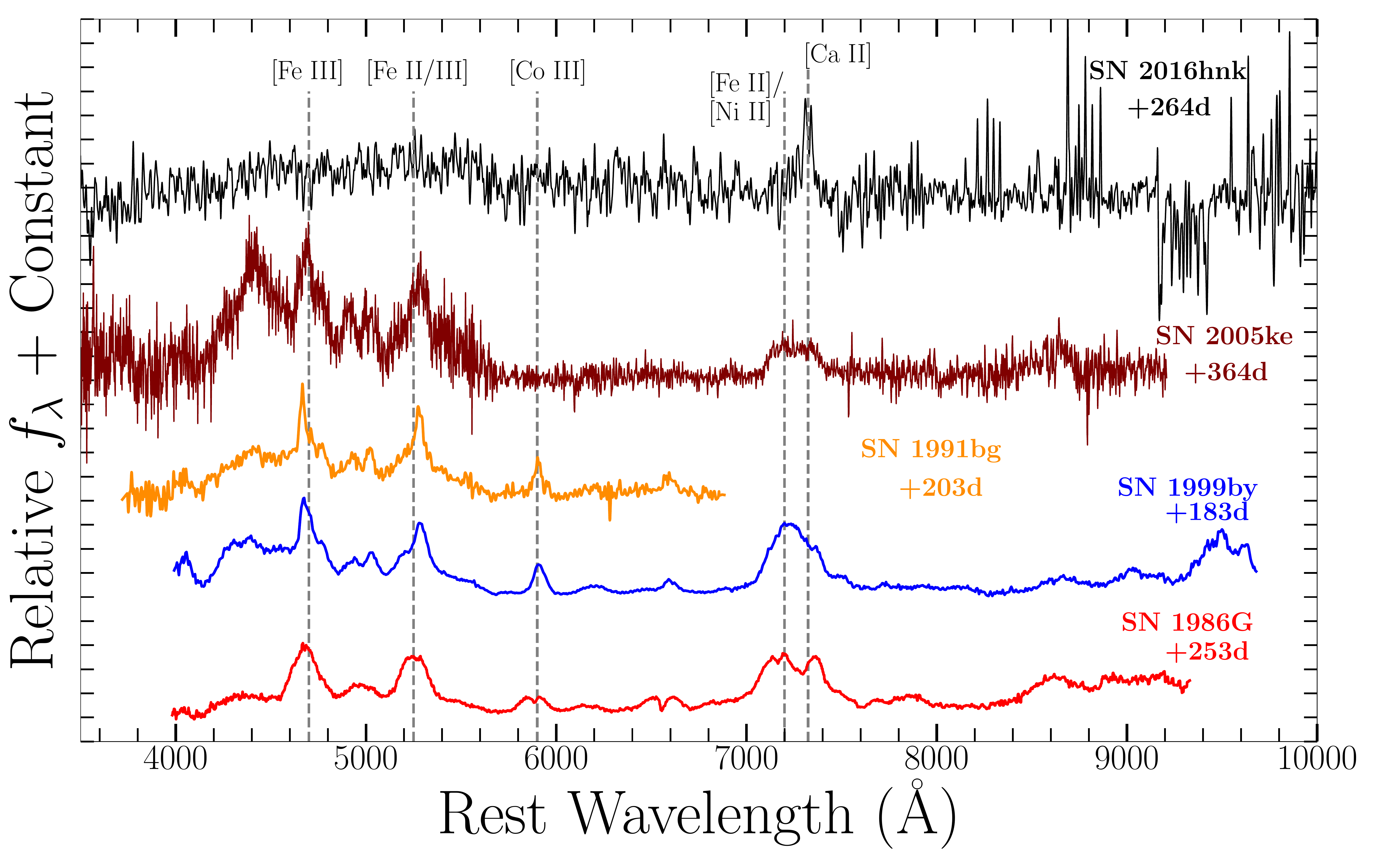}}
\caption{(a) Early-time spectral comparison of SN~2016hnk, PTF09dav, and assorted sub-luminous SNe Ia. All phases with respect to \textit{B}-band maximum. (b) Pre-nebular spectral comparison of SN~2016hnk and sub-luminous SNe Ia. (c) Comparison of nebular spectra of SN~2016hnk and 91bg-like objects at approximately the same phase. SN~2016hnk spectrum (black) has been smoothed. \label{fig:Ia_spectra_compare}}
\end{figure}

Initial spectral classifications of SN~2016hnk noted similarities between both sub-luminous, 91bg-like objects as well as Ca-rich transient PTF09dav. We show a time-series spectral comparison in Figure \ref{fig:Ia_spectra_compare}. As shown in Figure \ref{fig:Ia_spectra_compare}(a), SN~2016hnk has matching Si II, O I and Ca II absorption to 91bg- and 02es-like objects. Similar features are also observed in PTF09dav, but with slower expansion velocities than those in the SN~2016hnk spectra. Fe II/III emission and Ti II absorption profiles are shared between both objects. SN~2016hnk, however, has significantly suppressed emission in blue wavelengths compared to PTF09dav.

The primary differences between SN~2016hnk and other sub-luminous SNe~Ia are the relative velocities and the prominence of Fe-group elements. While the Si II velocity of SN~2016hnk does match 91bg-like objects, the Ca II in SN~2016hnk is moving $\sim 7000 \ \kms$ faster than the velocities observed in  ``91bg-like'' SNe~1991bg ($10346.30 \pm 378.08 \ \kms$), 1999by ($9285.32 \pm 447.25 \ \kms$) and 2005ke ($11110.36 \pm 684.15 \ \kms$) near peak. This variation in Ca II velocity might suggest a different explosion mechanism for SN~2016hnk than that of 91bg-like objects. Furthermore, at early-times, the observed line blanketing of Fe-group elements in the SN~2016hnk spectra is not observed in other comparison objects presented in Figure \ref{fig:Ia_spectra_compare}(a).

Despite the similar prevalence of [Ca II] lines in the nebular phase, the dominance of Fe-group elements in 91bg-like objects is not observed in the Keck spectrum of SN~2016hnk at +264d. As shown in Figure \ref{fig:Ia_spectra_compare}(c), the nebular spectra of all 91bg-like objects have prominent [Fe II], [Fe III] and [Co II] emission features in blue wavelengths. These features are not present in the SN~2016hnk spectrum at a similar phase, which suggests that significantly less ${}^{56}\textrm{Ni}$ was synthesized during explosion. This observed contrast is supported by SN~2016hnk's low peak luminosity as compared to other sub-luminous SN~Ia varieties. Furthermore, the [Ca II] profile in SN~2016hnk is much narrower than that shown in the 91bg-like objects, all of which possess a blend of [Ca II] and [Ni II] emission at $\sim$7300 \AA. 

\begin{figure*}
\centering
\includegraphics[width=\textwidth]{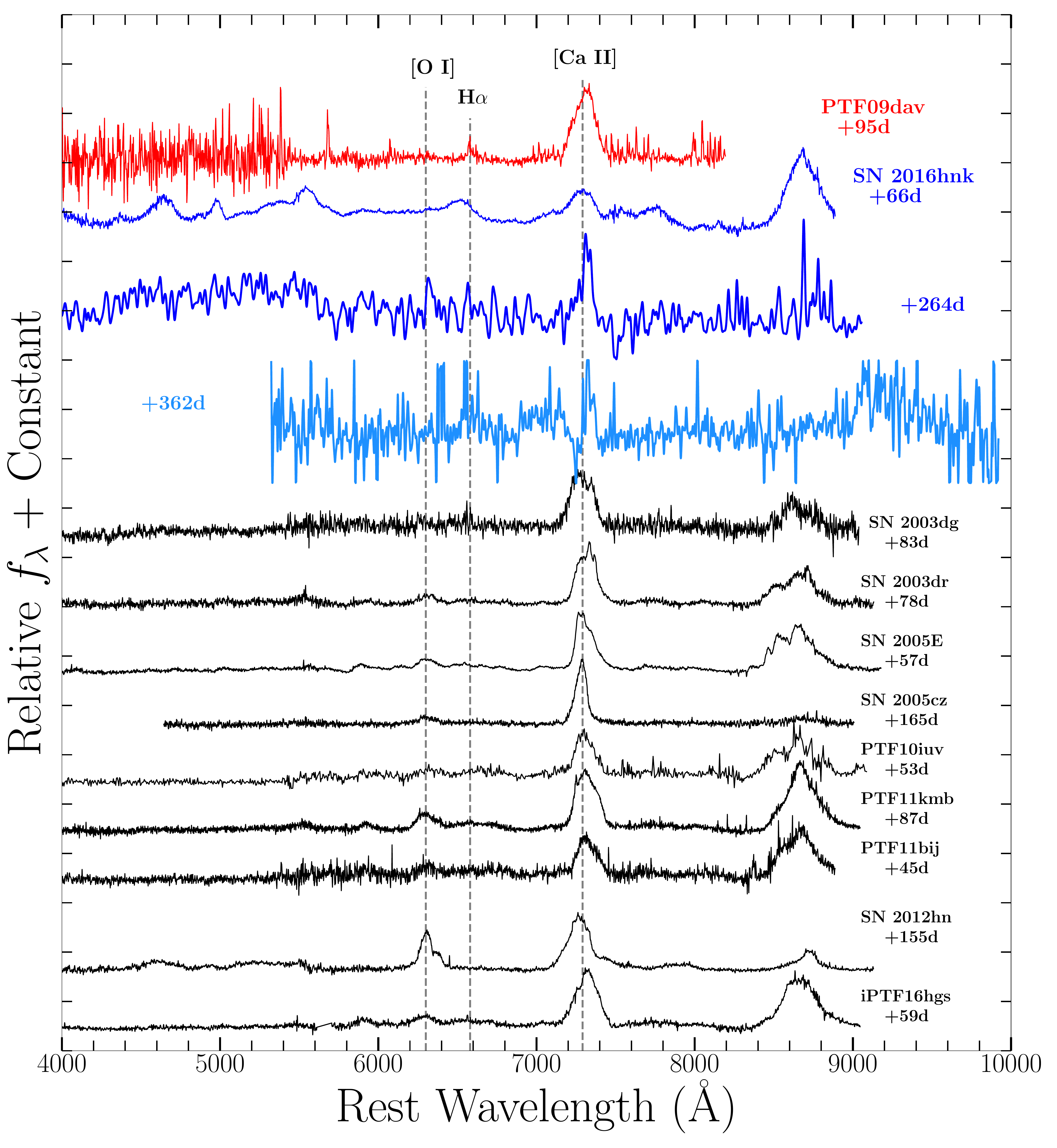}
\caption{Nebular spectra of all classified ``Ca-rich'' transients. The nebular spectra of PTF09dav and SN~2019hnk are shown in red and blue, respectively. Note that the +66d spectrum of SN~2016hnk is still photospheric and thus cannot be directly compared with Ca-rich objects at a similar phase. The +264d spectrum of SN~2016hnk has been smoothed with a Gaussian-filter. The light blue +362d spectrum has also been smoothed and was presented originally in G19. Prominent [O I] and [Ca II] lines marked by vertical dashed grey lines. \label{fig:ca-rich_nebular}}
\end{figure*}

\begin{figure}[h]
\centering
\subfigure[]{\includegraphics[width=0.45\textwidth]{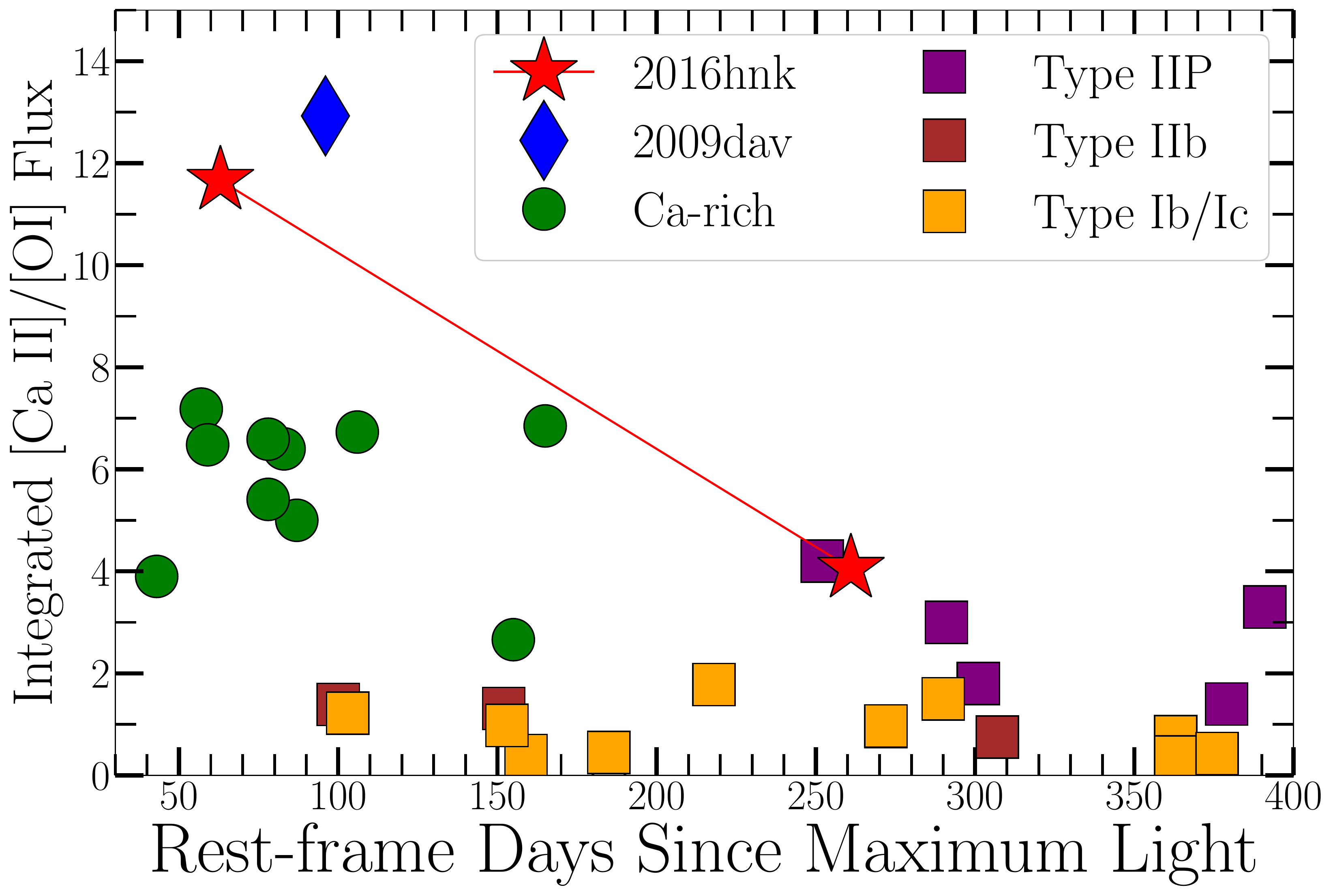}}
\subfigure[]{\includegraphics[width=0.44\textwidth]{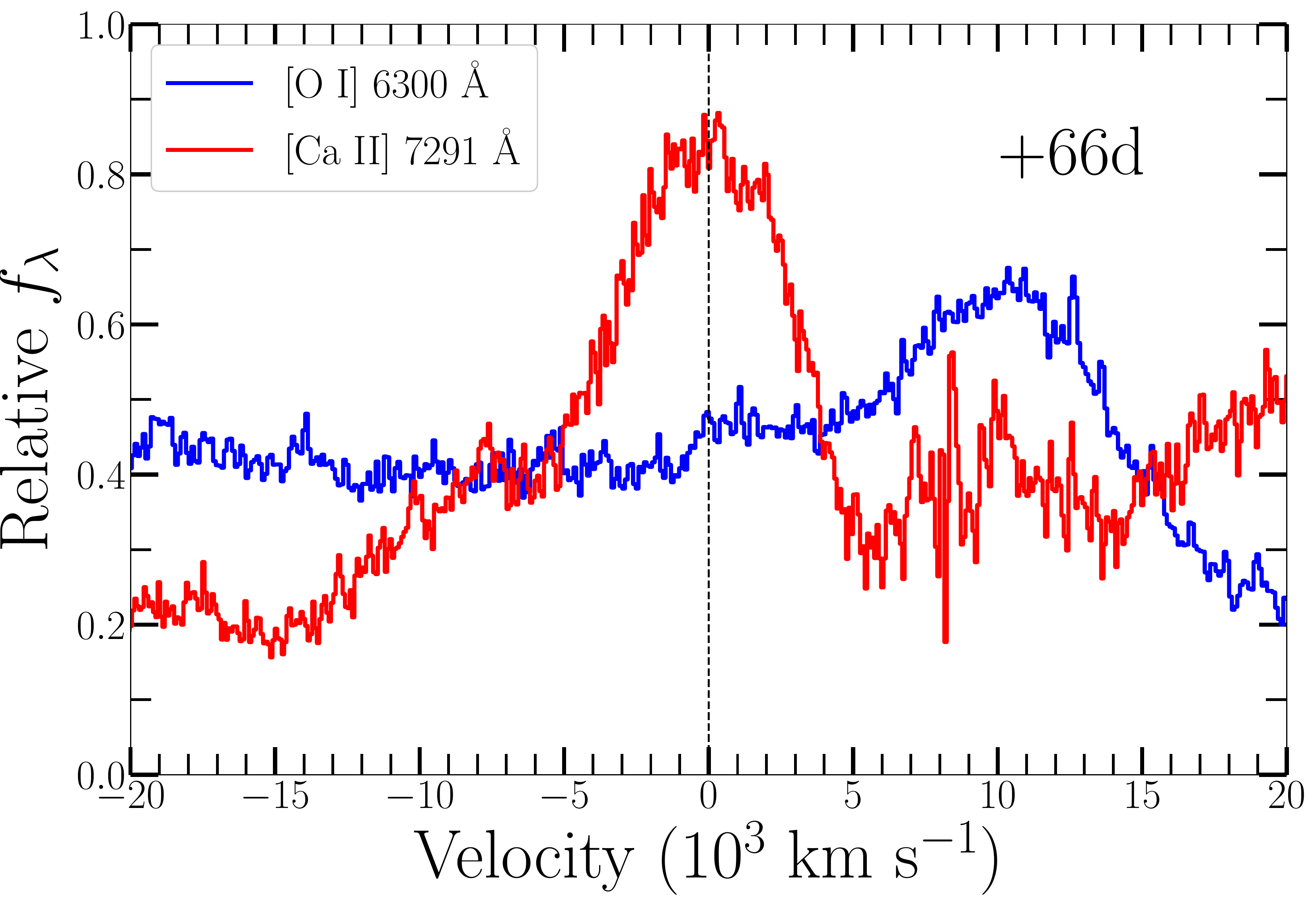}}
\subfigure[]{\includegraphics[width=0.44\textwidth]{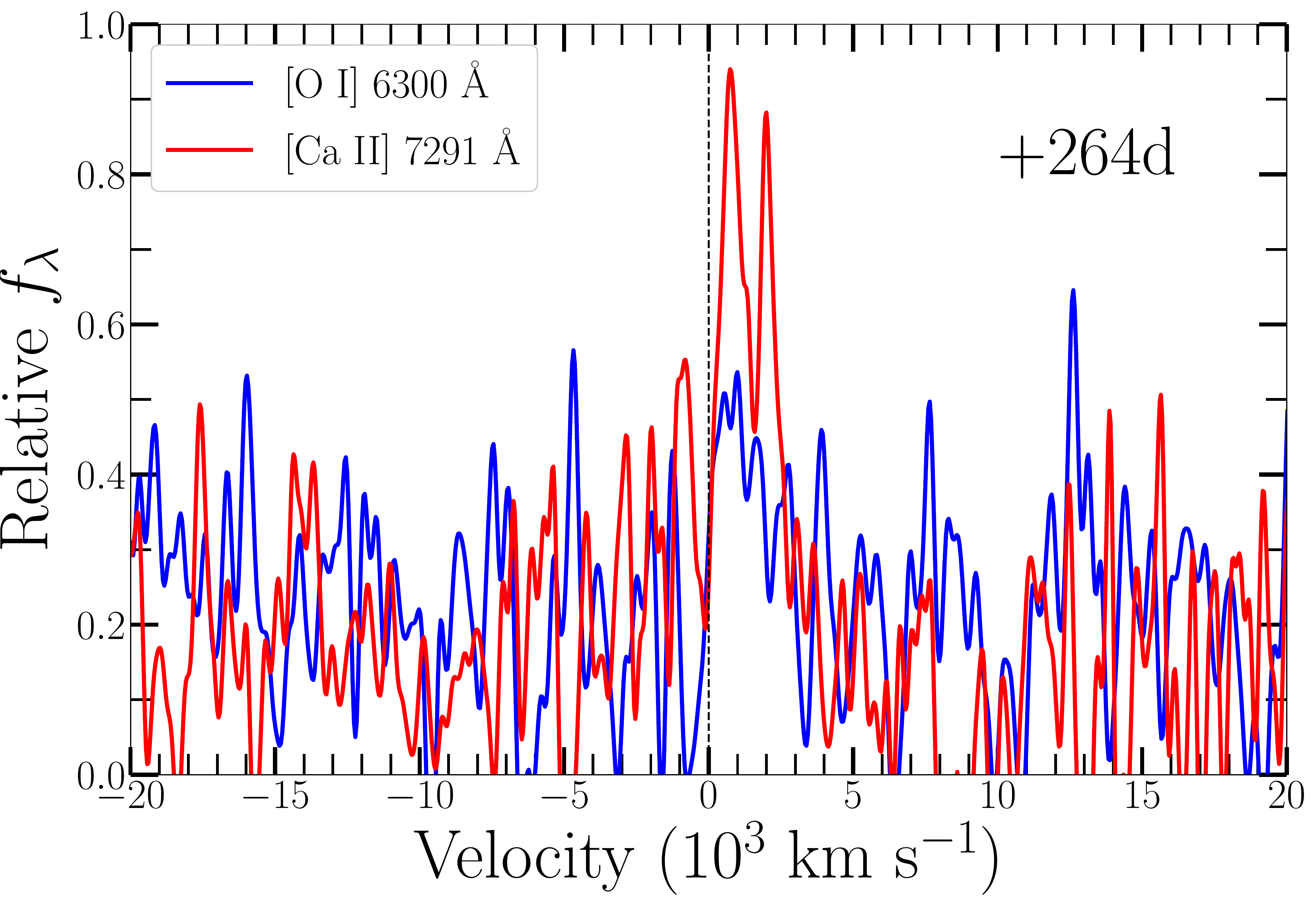}}
\caption{(a) Ratio of integrated [Ca II] and [O I] flux with respect to phase for SN~2016hnk, PTF09dav, Ca-rich transients and assorted types of core-collapse SNe. [Ca II]/[O I] values for all Type II/Ibc objects from \cite{milisavljevic17}. (b)/(c) Velocity profiles of [O I] $\lambda\lambda$ 6300, 6364 (blue) and [Ca II] $\lambda\lambda$ 7291,7324 (red) in SN~2016hnk at +66d and +264d, respectively. \label{fig:caii_oi}}
\end{figure}

Nebular spectra of SN~2016hnk does show some similarities to observations of other Ca-rich objects at late-times. In Figure \ref{fig:ca-rich_nebular}, we present SN~2016hnk spectra at +66d and +264d with relation to most nebular spectra of Ca-rich transients. There is a noticeable parallel between the [Ca II] emission features in all presented objects, in addition to the minor or non-existent presence of Fe-group elements. [Ca II] line profiles are similar in all objects and do not show significant [Ni II] line blending. Like PTF09dav, SN~2016hnk does not have detectable [O I] and is thus significantly more O-poor than other objects such as 2003dr, 2005E, PTF11kmb and 2012hn. We present direct comparison of [O I] and [Ca II] line profiles in Figure \ref{fig:caii_oi}(a). Furthermore, SN~2016hnk does not show narrow H$\alpha$ emission at late-times as is detected in PTF09dav. 

We compare the [Ca II]/[O I] ratio of SN~2016hnk to all Ca-rich objects and Type Ibc/IIP/IIb SNe in \ref{fig:caii_oi}(b). Given the lack of observed [O I] in both objects, PTF09dav and SN~2016hnk have the highest [Ca II]/[O I] ratio with respect to other SNe. This indicates a relatively oxygen-poor explosion for both PTF09dav and SN~2016hnk as compared to most other Ca-rich objects. 

These spectroscopic comparisons have demonstrated that SN~2016hnk is most similar to SN~2018byg given its observed bluewards line-blanketing and its strong, high-velocity Ca II features. SN~2016hnk has some similar line profiles to 91bg-like SNe (e.g., Si II \& S II), but has significantly faster Ca II velocities and substantial line-blanketing than this sub-class of SNe Ia. SN~2016hnk also has weaker Fe-line transitions than these objects at nebular times. Like other Ca-rich transients, the nebular spectra of SN~2016hnk is dominated by [Ca II] emission. While the ratio of [Ca II]/[O I] line fluxes for Ca-rich transients is typically greater than 2, this ratio for SN~2016hnk and PTF09dav is almost twice that of typical Ca-rich objects. This suggests either stronger calcium emission in SN~2016hnk or a limited oxygen presence in this SN than Ca-rich SNe. 

\subsection{Comparison to SNe~Iax Nebular Spectra}\label{subsec:Iax_caii_nebular}
\begin{figure}[h]
\centering
\includegraphics[width=0.45\textwidth]{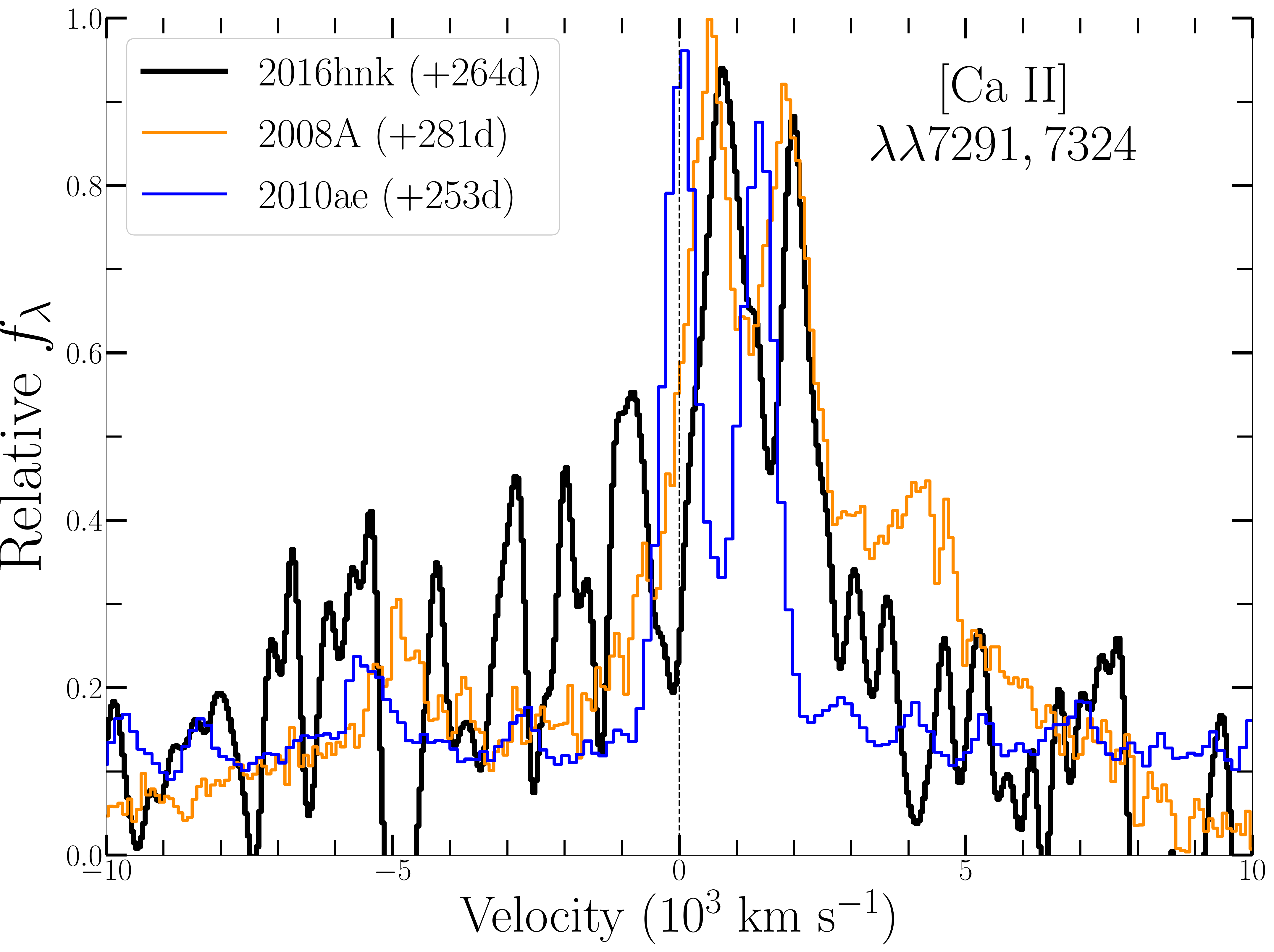}
\caption{[Ca II] $\lambda\lambda$ 7291,7324 velocity profiles of SN~2016hnk (black) and SNe~Iax, 2008A (orange) and 2010ae (blue) at similar phases. Like SN~2016hnk, SNe~Iax are also Ca-rich in nebular phases and their [Ca II] emission may be a signature of wind from a bound remnant \citep{foley16}. \label{fig:16hnk_Iax_velocity}}
\end{figure}

As shown in Figure \ref{fig:16hnk_Iax_velocity}, SN~2016hnk's [Ca II] emission profile is quite similar to that found in nebular phase spectra of SNe~Iax. Also "Ca-rich" at late-times, SNe~Iax exhibit strong [Ca II] and variable [Fe II]/[Ni II] emission features, both of which being tracers of initial WD mass and the possible wind from a bound SN remnant \citep{foley16}. Unlike most SNe~Iax, SN~2016hnk does not have visible [Ni II] emission, which suggests that there is very little stable Nickel isotopes, such as ${}^{58}\textrm{Ni}$, present in the ejecta. This is indicative of a sub-Chandrasekhar explosion in which the WD density is too low for sufficient electron capture. This conclusion is supported by our ejecta mass estimate of $\sim$0.9 $\Msun$ (Section \ref{subsec:LC_bolometric}). Furthermore, as shown in \cite{foley16}, narrow forbidden lines in nebular spectra may result from wind from a bound remnant. Such a model could be a viable explanation for SN~2016hnk at late-times given its [Ca II] profile shape is consistent with SNe~Iax such as 2008A and 2010ae.

\subsection{Stripped Mass Calculation}\label{subsec:strip_mass}

The single-degenerate (SD) progenitor channel involves a WD in a binary system with a non-degenerate companion star. In this scenario, it is predicted that ablated H- or He-rich material from the non-degenerate companion will be swept up by the SN ejecta and should be detectable in late-time spectra as narrow emission lines with velocities of $\approx$1000 $\kms$ \citep{marietta00, mattila05, liu12, liu13b, pan12, lundqvist13}. We test this scenario by calculating upper limits on the mass of stripped hydrogen and helium potentially present in the nebular spectrum of SN~2016hnk. 

We preface this analysis by stating that all stripped mass models used have been designed specifically for SNe~Ia and not Ca-rich or low luminosity thermonuclear objects. Thus these models are most notably different from SN~2016hnk in their energetics and total Ni mass produced, in addition to possible asymmetries. Consequently, the lower overall energies and SN densities in Ca-rich objects may affect the mixing of H/He CSM, but the differences in Ni mass between each SN type will scale with the luminosity. Nonetheless, these models and radiative transfer simulations are applicable to this analysis as a means of testing a SD, companion interaction scenario for SN~2016hnk.

We visually examine the +264d nebular spectrum of SN~2016hnk and find no obvious detections of narrow H$\alpha$ or He I emission. We do, however, observe a narrow feature near $\lambda6563$ in the nebular spectrum, which has a FWHM velocity of $203.65 \pm 55.8 \ \kms$. This velocity is low for SN-related material e.g., the H$\alpha$ emission in PTF09dav has a velocity of $1315.17 \pm 241.05 \ \kms$. Furthermore, this feature does not have significant S/N compared to the overall spectral noise. Therefore we perform the analysis of potential ablated material by assuming a non-detection of nebular H$\alpha$ emission.

\begin{figure}[h]
\centering
\subfigure[]{\includegraphics[width=0.45\textwidth]{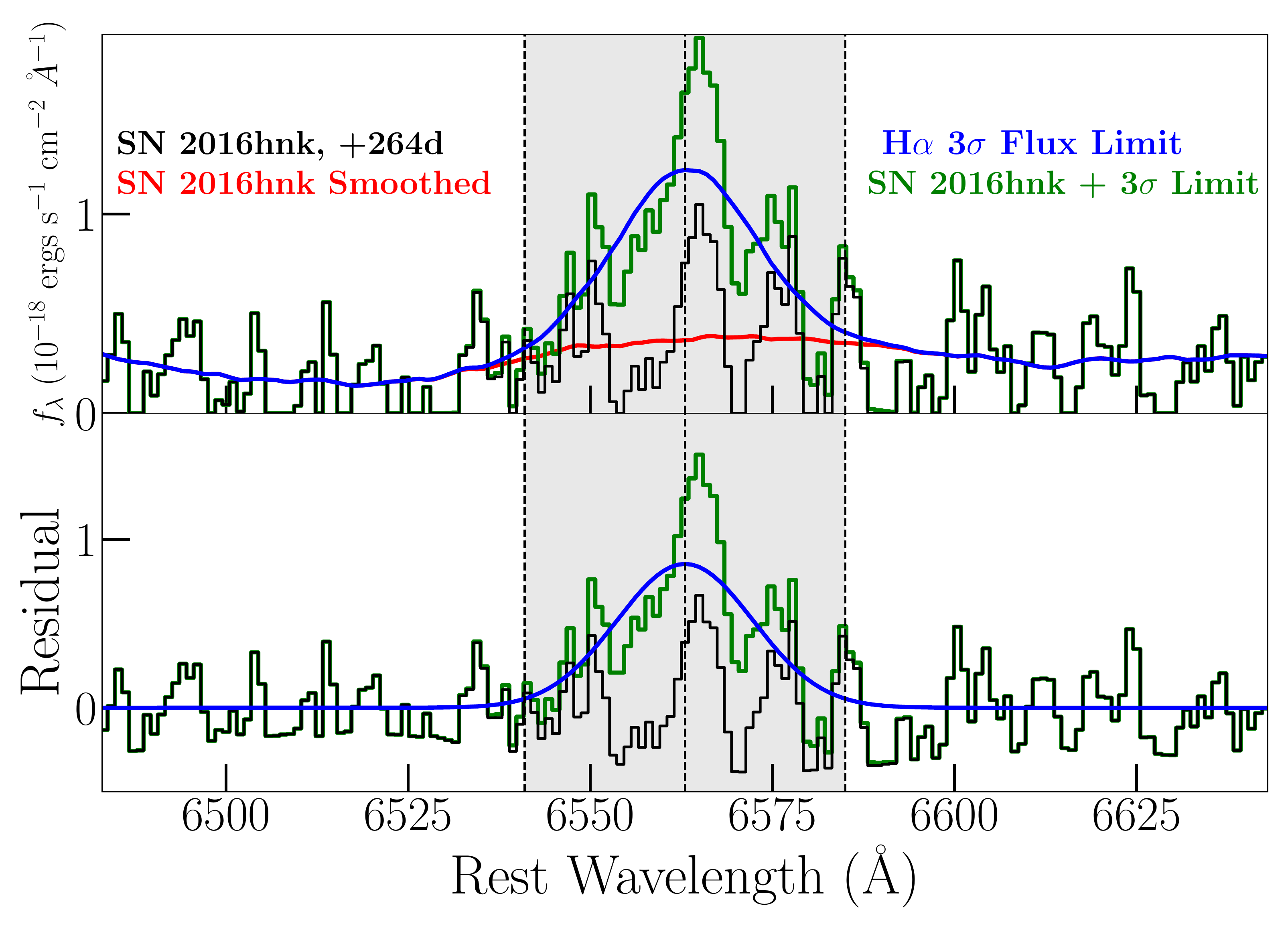}}
\subfigure[]{\includegraphics[width=0.44\textwidth]{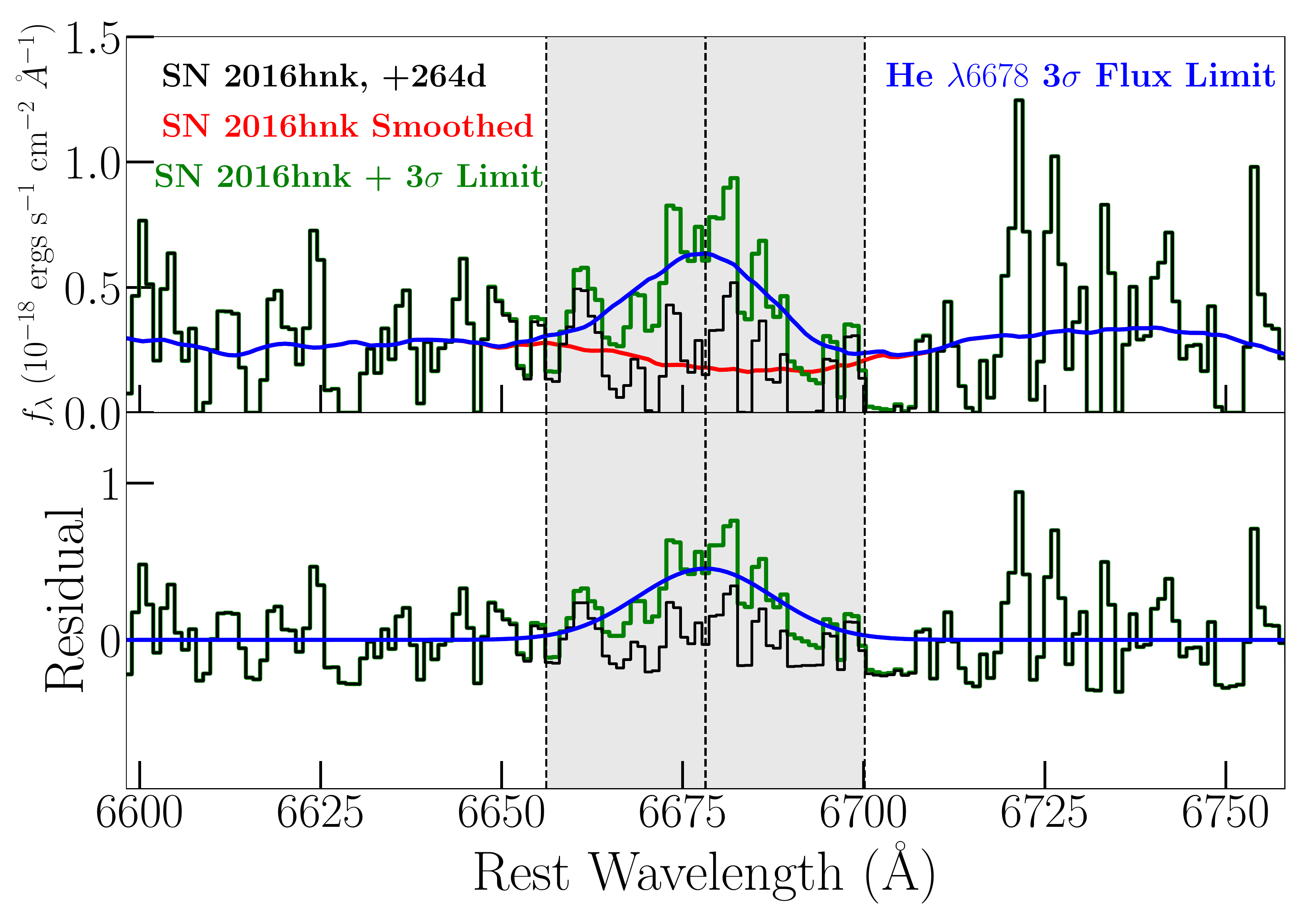}}
\caption{(a) \textit{Upper Panel}: In black, flux calibrated late-time data of SN 2016hnk at +264d with no apparent H$\alpha$ emission. Shown in
red is the continuum that has been smoothed with a Savitzky-Golay filter. The $3\sigma$ RMS flux limit for marginal detection of H$\alpha$ is shown in blue. The data plus $3\sigma$ flux limit is
shown in green. The grey shaded region represents a wavelength range
of 22\AA \ ($\sim$ 1000$\kms$). Lower Panel: In green, residuals of data plus
the $3\sigma$ limit, minus smoothed data. In black, residuals of data minus
smoothed continuum. H$\alpha$ $3\sigma$ flux limit shown in blue. (b) Same
method as for H$\alpha$, but with marginal detection of the $6678\AA$ He I
emission line. \label{fig:stripped_mass}}
\end{figure}

To calculate the stripped hydrogen luminosity limit, we simulate a marginal detection by modeling the H$\alpha$ emission as a Gaussian profile (FWHM = $1000\kms$) with a peak flux of three times the spectrum's root-mean-square (RMS) flux \citep{sand18,dimitradis19}. We present the simulated limit for marginal detection of H$\alpha$ emission in Figure \ref{fig:stripped_mass}(a).

We use the 3-$\sigma$ flux limit for marginal detection to calculate the luminosity limit of stripped hydrogen in the nebular spectrum. We then convert the H$\alpha$ luminosity to stripped mass using the following relation from \cite{botyanszki18}:

\begin{equation}\label{eq:ha_lum}
  \log_{10}(L_{\rm{H\alpha}}) = -0.2 M_{1}^2 + 0.17 M_{1} + 40,
\end{equation}
where $L_{\rm{H\alpha}}$ is the H$\alpha$ luminosity in cgs units at
200~days after peak, $M_{1} = \log_{10}(M_{\rm H}/M_{\odot})$, and
$M_{\rm H}$ is the stripped hydrogen mass. We re-scale our estimated H$\alpha$ luminosity limit because this relation is derived at exactly 200 days after peak. We calculate the decline in luminosity of SN~2016hnk to be a factor of 4 between 200 and 300 days. We find an H$\alpha$ luminosity limit luminosity limit of $L_{\textrm{H$\alpha$}} < 1.3\times 10^{38}$ erg s$^{-1}$, which corresponds to $4.6 \times 10^{-3} \Msun$ of undetected, stripped H-rich material in SN~2016hnk. This mass is an order of magnitude lower than model predictions for stripped H-rich material that is swept-up by a SN Ia in a Roche-lobe filling progenitor system \citep[$1.4 \times 10^{-2}$ to $0.25 \ \Msun$;][]{pan12,
liu12, boehner17}. 

To calculate the limit of stripped helium in SN~2016hnk, we mimic the procedure outlined in \cite{jacobson-galan19} for He I $\lambda6678$. Our marginal detection calculation follows the same procedure as H$\alpha$ and we display the 3$-\sigma$ limit for He I emission in Figure \ref{fig:stripped_mass}(b). As shown in Figure 4 of \cite{botyanszki18}, the MS38 model produces an H$\alpha$ emission line that is $\sim$5 times more luminous than He I $\lambda6678$. Because of this, we modify Equation \ref{eq:ha_lum} to have the following form: 

\begin{equation}\label{eq:he_lum}
  \log_{10}(L_{\rm{He}}) = -0.2 M_{1}^2 + 0.17 M_{1} + 39.3,
\end{equation}
where $M_{1} = \log_{10}(M_{\rm He}/M_{\odot})$ and $M_{\rm He}$ is the stripped helium mass. We calculate a stripped helium luminosity limit of $L_{\textrm{He}} < 6.8\times 10^{37}$ erg s$^{-1}$ and a corresponding maximum stripped mass of $1.2 \times 10^{-2} \Msun$ found using Equation \ref{eq:he_lum}. In SD, He-star companion interaction models, the predicted ranges of stripped He-rich masses are $2.5 \times
10^{-3} - 1.3 \times 10^{-2} \ \Msun$ \citep{pan12} and $9.5 \times 10^{-3} - 2.8 \times 10^{-2} \ \Msun$ \citep{liu13}; both of which are consistent with the helium mass limit of SN~2016hnk.

\section{Explosion Models} \label{sec:models}


The observed spectroscopic and photometric properties of SN~2016hnk suggest a different explosion scenario than typical sub-luminous SNe Ia. The prominence of high velocity ($v\sim$18,000$ \kms$) intermediate mass elements such as Ca II in SN~2016hnk indicates a potential helium detonation \citep{fink10, kromer10}. Such a scenario can produce a large amount of Fe-group elements in the outer regions of the SN ejecta that causes significant line blanketing near maximum light, in addition to a fast rising light curve \citep{shen18, polin19}. These expected signatures of helium shell detonation on a C/O WD are consistent with the observed spectral features and light curve evolution of SN~2016hnk. 


We model the light curve and spectra of SN~2016hnk with a grid of different helium shell detonation scenarios derived from \cite{polin19}. Our simulations track the double detonation of a 50\% carbon + 50\% oxygen, $\sim0.8 \Msun$ WD with $\sim0.03 \Msun$ of helium on the surface. Each progenitor profile is constructed using a semi-analytic method which ensures the WD and helium shell of our chosen masses begin in hydrostatic equilibrium \citep{zingale2013}. We then use the Eulerian hydrodynamics code \texttt{Castro} \citep{almgren10} to follow each simulations hydrodynamical evolution and nucleosynthetic reactions from initial helium ignition through homologous expansion. Once the ejecta have reached homology we generate synthetic light curves and spectra from each model using the radiative transfer code \texttt{Sedona} \citep{kasen06}.  

\begin{figure*}
\centering
\subfigure[]{\includegraphics[width=.9\textwidth]{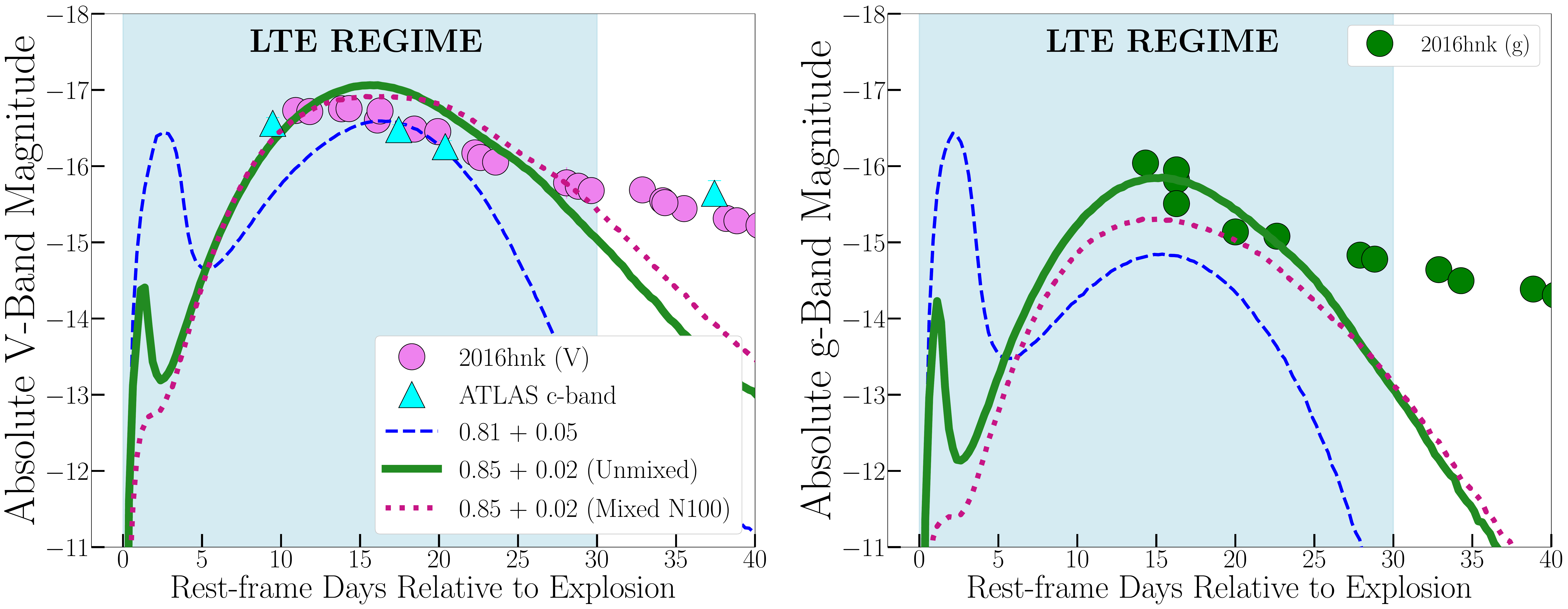}}\\[1ex]
\subfigure[]{\includegraphics[width=.9\textwidth]{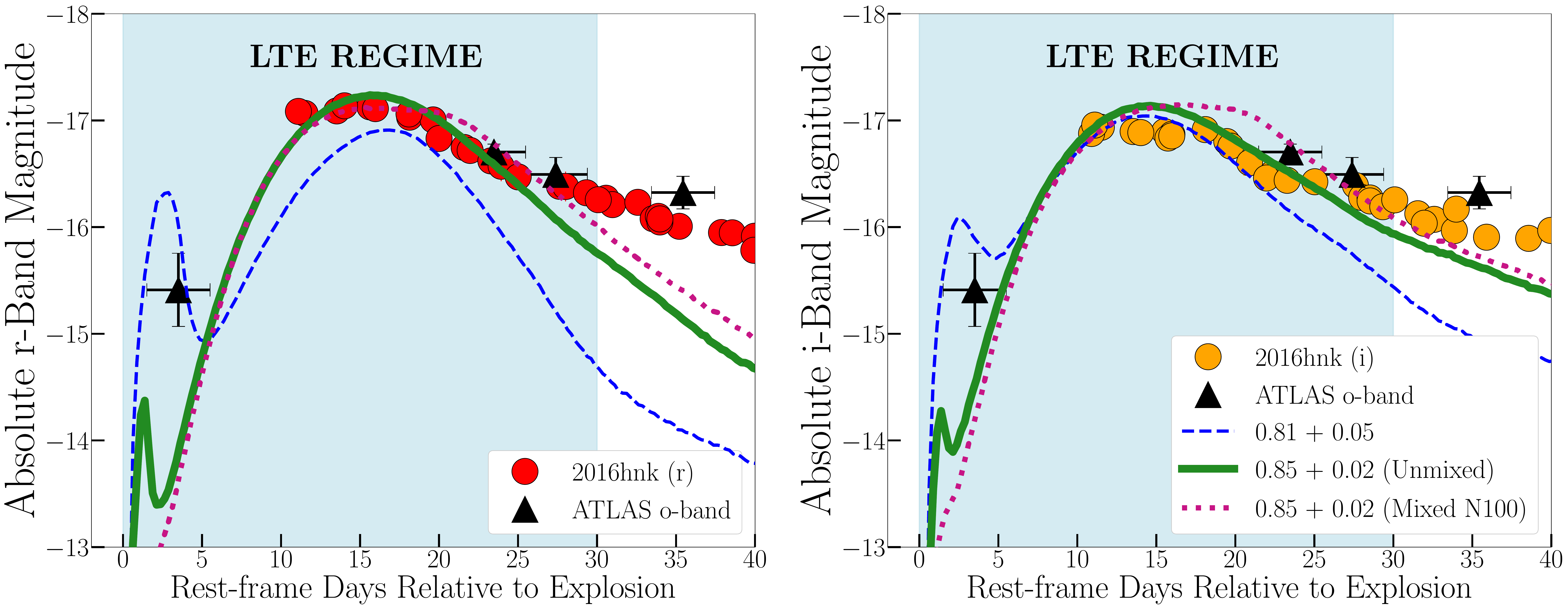}}\\[1ex]
\subfigure[]{\includegraphics[width=.9\textwidth]{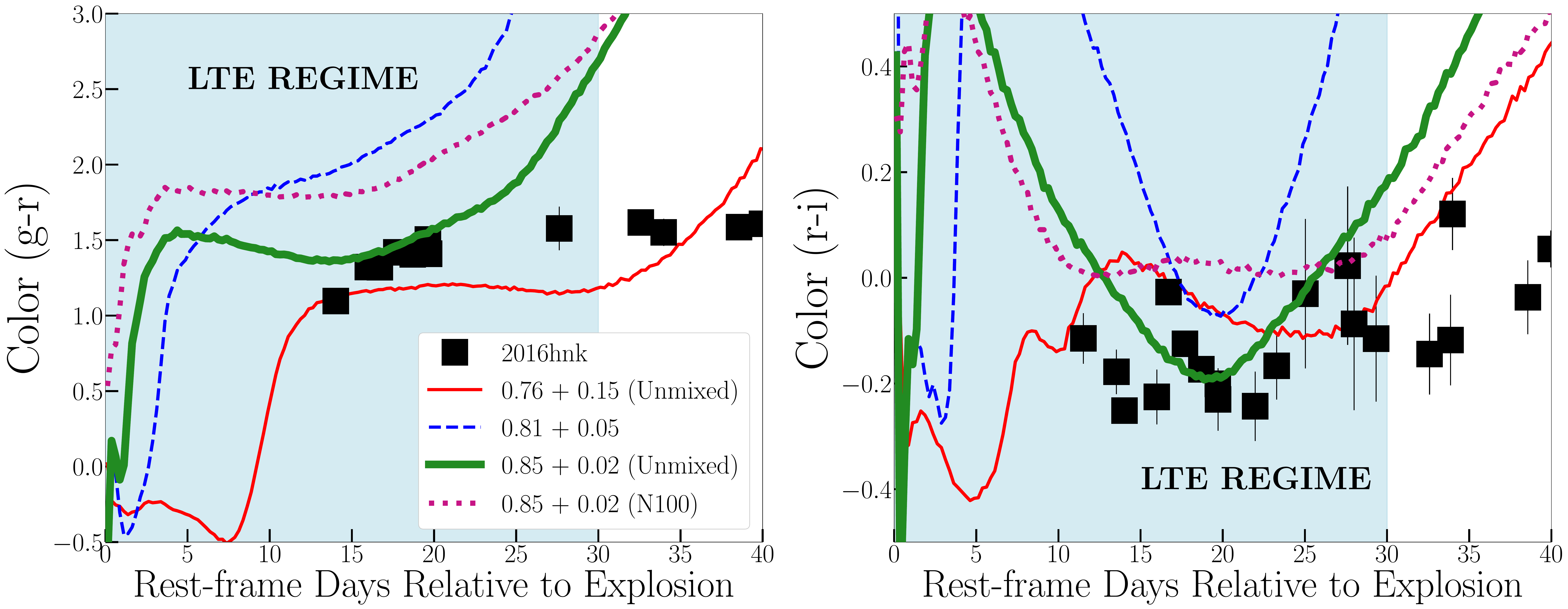}}
\caption{(a)/(b) Light curve comparison to double detonation helium shell models presented in \cite{polin19}. LTE regime (i.e. where the models are most reliable) shown by shaded light blue region. Best fitting 0.85+0.02 model has no mixing of ejecta and is shown in forest green. Same model but with mixing (N100) shown in violet-red. Left to right: \textit{V, g, r, i} band photometry. ATLAS photometry included for early-time light curve information. We put errorbars on the phases of the ATLAS data to illustrate that the date of explosion is uncertain to within $\sim$2 days for all photometry.} (c) \textit{g-r} and \textit{r-i} color comparison to helium shell models. \label{fig:DD_LC}
\end{figure*}

In Figure \ref{fig:DD_LC}(a)/(b), we compare the helium shell model light curves to our \textit{V, g, r, i} band photometry. We find that helium shell detonation with a total mass of $0.87 \ \Msun$ (WD + helium shell) can reproduce the peak absolute magnitudes in all bands as well as the early-time light curve decline. The first, $^{48}$Cr-powered light curve peak is eliminated when the mixing of outer ejecta is included in our models. Such a model (N100) is presented as a dotted line in Figure \ref{fig:DD_LC}(a) and include mixing of 0.1~$\Msun$ of outer ejecta. The mixing of outer ejecta produces a slightly better fit to some photometric observations, but additional constraints cannot be placed on the very early-time light curve evolution due to the limited pre-maximum data for SN~2016hnk. However, we do include ATLAS $o$- and $c$-band data for comparison with $r$/$i$- and $V$-bands, respectively. The pre-maximum $o$-band data point in Figure \ref{fig:DD_LC} appears to be most consistent with the best fitting, non-mixed 0.85+0.02 shell model (green line), but it is also within the phase uncertainties of the 0.85+0.02 mixed N100 model (dotted violet-red line).

The divergence between the models and SN~2016hnk at $\sim$25 days post explosion may be attributed a variety of factors within the explosion. Firstly, for a similar model, the LTE assumptions within the \texttt{Sedona} code may be less representative of the conditions within the ejecta of SN~2016hnk than SN~2018byg, which is modeled to have a higher total mass than SN-2016hnk (0.91 $\Msun$ vs 0.82 $\Msun$), and may take longer to become optically thin. A LTE condition is not applicable once the ejecta beings to become optically thin, which typically occurs $\sim$30 days after explosion for sub-Chandrasekhar mass ejecta (light blue shaded region in Figure \ref{fig:DD_LC}). A more detailed treatment of Non-LTE conditions could explain the slower light curve decline in SN~2016hnk. Furthermore, our 1D helium shell model does not account for additional physics such as asymmetries or external emission components that could have influenced the increased light curve flux relative to the models. Nonetheless, while the model is an approximation of the SN explosion physics, it provides a reasonable match to the observables with only two free parameters (the mass of the Helium shell and the mass of the underlying WD).

In Figure \ref{fig:DD_LC}(c) we compare \textit{g-r} and \textit{r-i} colors of SN~2016hnk to various helium shell models from \citep{polin19}. Overall both mixed and unmixed models have bluer colors than those observed in SN~2016hnk. However, in the LTE regime the models are consistent to within 0.3 mag of SN~2016hnk's \textit{g-r} and \textit{r-i} color evolution. We cannot constrain the helium shell scenario further given the lack of pre-maximum color information. 

\begin{figure*}[t!]
\centering
\subfigure[]{\includegraphics[width=.45\textwidth]{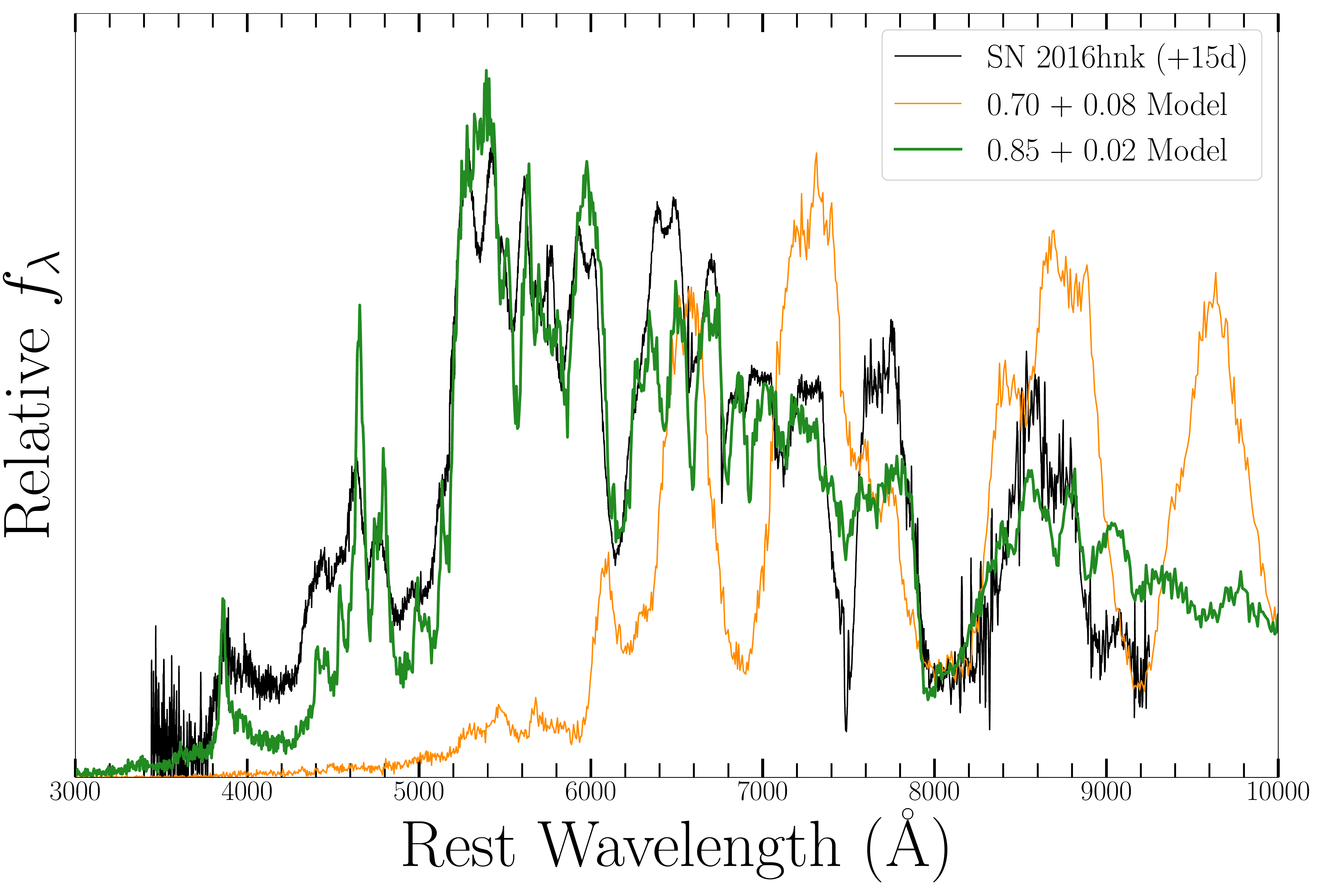}}
\subfigure[]{\includegraphics[width=.45\textwidth]{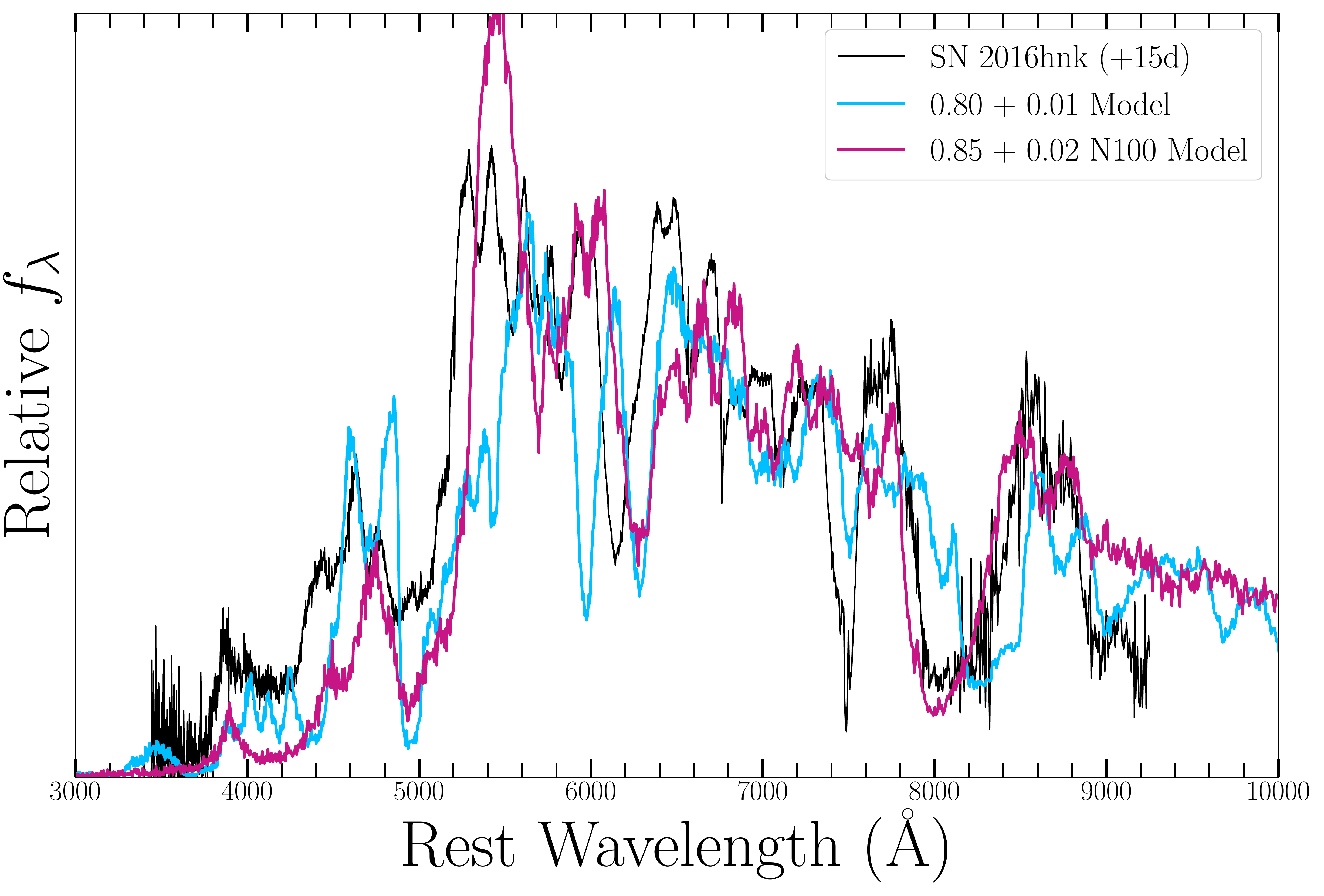}}\\[1ex]
{\includegraphics[width=.9\textwidth]{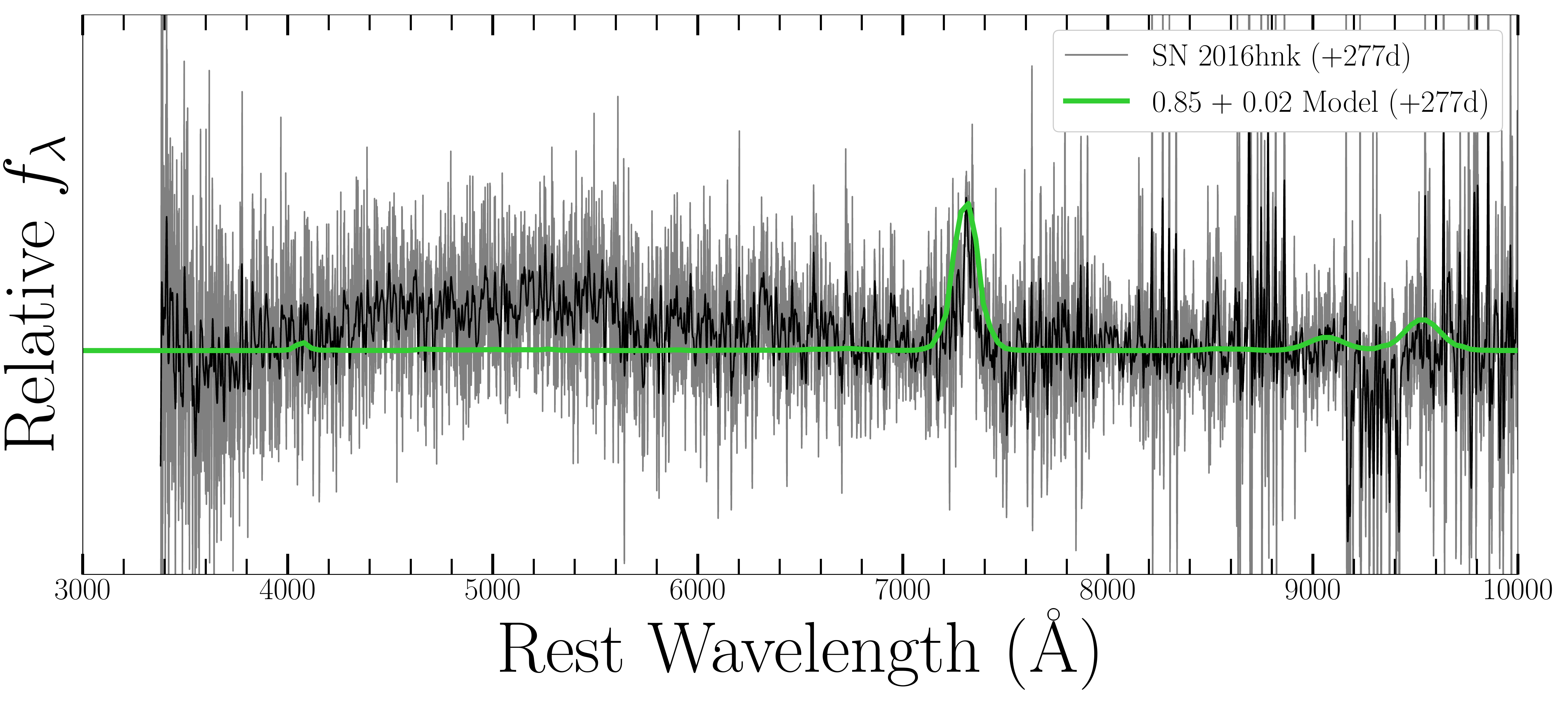}}
\caption{(a)/(b) Early-time model comparison of SN~2016hnk (black) and double detonations of $0.70-0.85\Msun$ WDs with $0.01-0.08\Msun$ helium shells. Best fitting, non-mixed 0.85+0.02 model shown in forest green. Only the 0.80+0.02 N100 Model spectrum (shown in violet-red) contains mixing of ejecta. Models presented are from \cite{polin19}. Phases presented are with respect to explosion. (c) Nebular comparison of SN~2016hnk and best fitting 0.85+0.02 helium shell nebular model from \cite{polin2019b}. Phases are also with respect to estimated explosion date.} \label{fig:DD_spectra}
\end{figure*}

\begin{figure*}
\centering
\subfigure[]{\includegraphics[width=.49\textwidth]{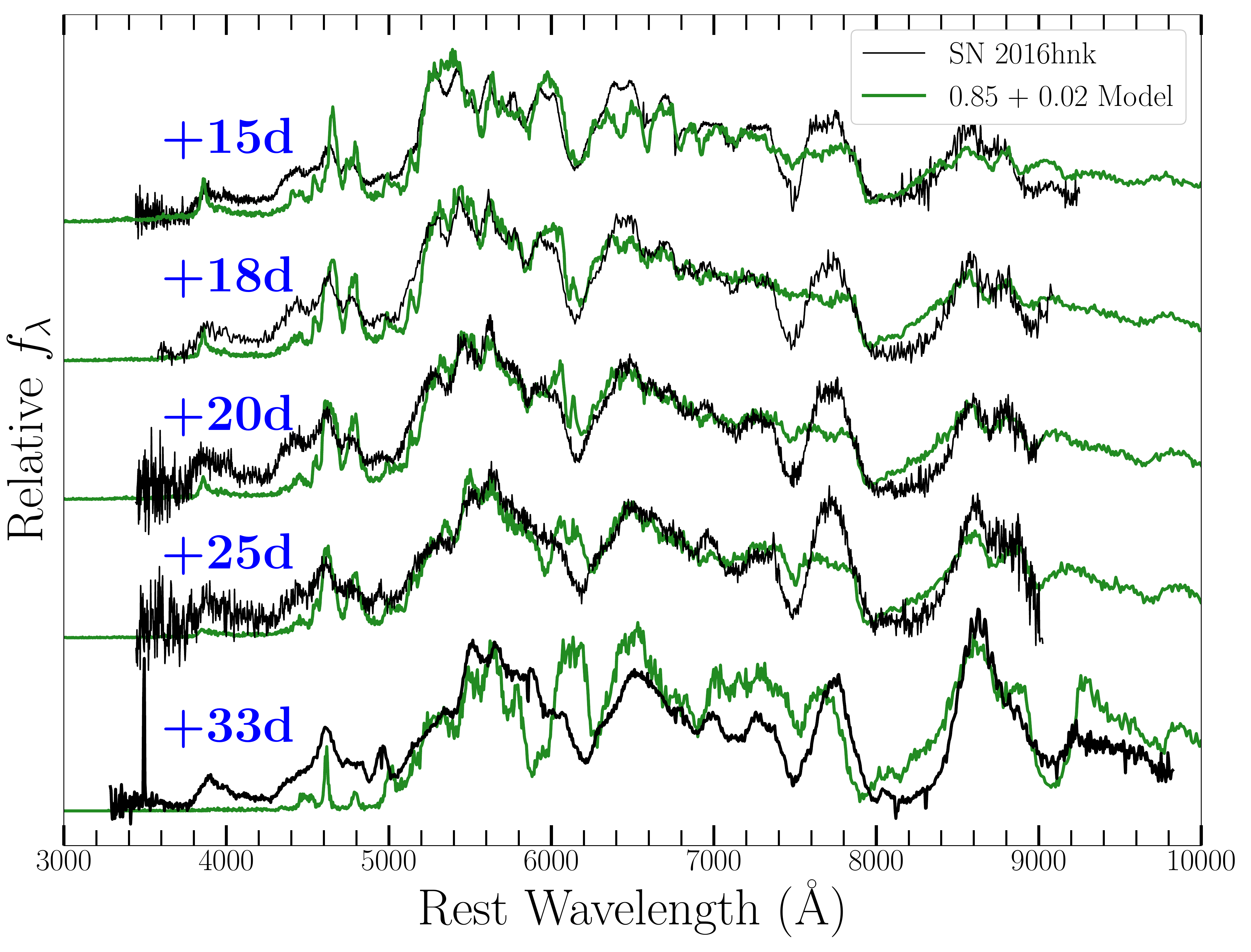}}
\subfigure[]{\includegraphics[width=.47\textwidth]{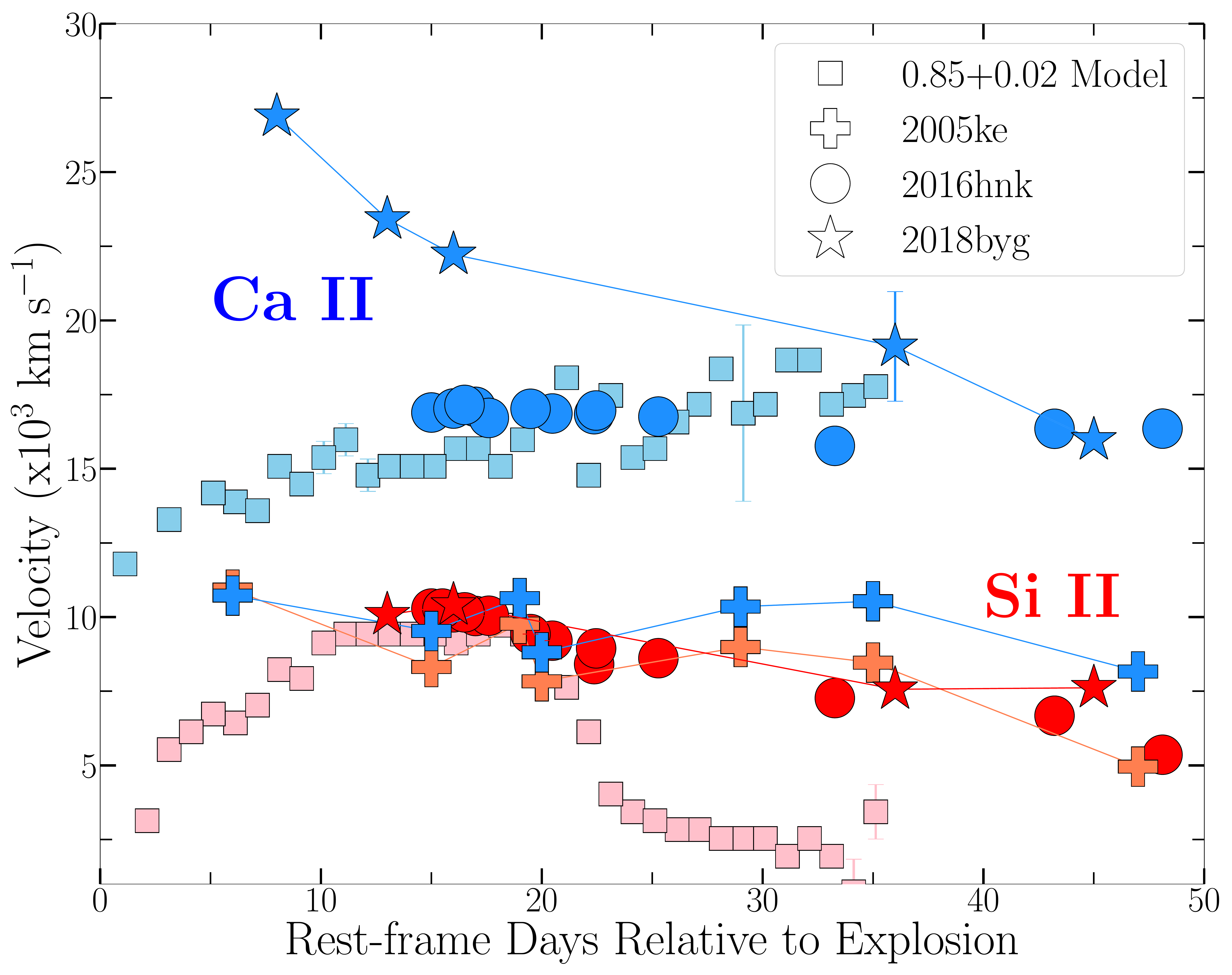}}
\caption{(a) Spectral time-series of SN~2016hnk (black) and the best fitting helium shell detonation model (green) with phases relative to explosion. Spectral models are only shown out to $\sim 30$d, which is where the LTE approximation breaks down and the modeling becomes less reliable. These additional data not shown in Figure \ref{fig:spectral_series} are taken from G19. (b) The evolution of Ca II and Si II velocities, calculated from fitted absorption minima, are presented for SNe~2005ke (plus signs), 2016hnk (circles), 2018byg (stars) with respect to those derived from best-fitting model spectra (squares). } \label{fig:models_spectra_vels}
\end{figure*}

As shown in the Figure \ref{fig:DD_spectra}(a), the photospheric spectra of SN~2016hnk are best reproduced by a detonation model involving a $0.85 \Msun$ WD and $0.02 \Msun$ helium shell. This model has a total synthesized Nickel yield of 0.045$\Msun$. For reference, we compare SN~2016hnk to other model spectra from \citealt{polin19} with varying WD and shell masses in Figures \ref{fig:DD_spectra}(a)/(b). We track the evolution of the best fitting models with respect to observations in a spectral time-series presented in Figure \ref{fig:models_spectra_vels}(a). We include additional spectral data from G19 in order to present comparisons from +15d out to +33d after explosion; the final phase being the time when the LTE model assumption becomes unreliable. In Figure \ref{fig:models_spectra_vels}(b) we present Si II and Ca II velocities for SNe~2005ke, 2016hnk and 2018byg with respect to the same line velocities from the best fitting helium shell model. In both figures we demonstrate a time-dependent consistency between the complete spectral profile, as well as individual ion velocities, observed in SN~2016hnk and the $0.85 \Msun$ WD plus $0.02 \Msun$ helium shell model.

Using the nebular companion to \texttt{Sedona}, \texttt{SedoNeb} \citep{janos2017}, we are able to examine our best fitting $0.85 + 0.02 \Msun$ model in the nebular phase following the methods outlined in \cite{polin2019b}, which investigated the nebular features of double detonations and shows that low mass scenarios would appear Ca-rich in the nebular phase. These methods require the SN ejecta to be fully optically thin in the desired wavelengths in order to produce a nebular spectrum. In this way we are able to examine the observational signatures of our model in the nebular phase (beginning $\sim$150 days after explosion). The resulting spectrum is shown in Figure \ref{fig:DD_spectra}(c) as a green line. This model is  consistent with the nebular spectrum of SN~2016hnk at +277 day after explosion, with [Ca II] emission being the dominant feature in both the simulation and observation. This is the first potential thick helium shell double detonation that we are able to compare to the models in both epochs as the nebular spectrum of SN~2018byg was unable to be obtained.

The thickness of the helium-shell in our preferred double-detonation model ($\sim$0.02 $\Msun$) suggests a explosion scenario involving a larger, non-degenerate companion to the C/O WD. A progenitor system that works well with this detonation mechanism is a He or sdB star + C/O WD binary system, wherein the WD is accreting helium-rich material from the non-degenerate companion. Such a configuration has been explored in simulations for both steady and time-varying dynamical accretion scenarios involving a 0.4-0.6~$\Msun$ He or sdB donor star \citep{nomoto82a, woosley89, woosley11}. Recent modeling has demonstrated that significant build up of a thick helium layer on the WD surface, combined with proper treatment of nuclear reaction networks (e.g., CNO or NCO burning), can trigger a shell detonation \citep{shen14, shen14b, brooks15, bauer17}. Consequently, \cite{de19} discuss how SN~2018byg is consistent with a 0.48~$\Msun$ sdB star model presented by \cite{bauer17} wherein dynamical accretion led to a large He envelope ($> 0.1 \ \Msun$) on the sub-Chandra C/O WD surface, eventually resulting in a shell detonation. While our models do not consider the pre-cursor accretion process to the double-detonation, the success of sdB + C/O WD binary configurations in achieving a He envelope detonation makes these accretion scenarios viable candidates for producing the thick helium shells present within our models.

\section{Discussion} \label{sec:discussion}

\subsection{Reddening} \label{sec:reddening}

The observed color evolution shows that SN~2016hnk is a highly reddened object. Presently, there is an established link between dust reddening and the strength of Na I D absorption for Milky Way stars and extragalactic SNe \citep{phillips13}. The lack of Na I D absorption in SN~2016hnk is a strong indication that the SN is unaffected by host reddening and that its red colors are intrinsic to the explosion. We further demonstrate this fact by de-reddening SN~2016hnk's color evolution. As shown in Figure \ref{fig:colors}(a), we de-redden \textit{B-V} colors to match the bluest SNe~Ia. Assuming a \cite{fitzpatrick99} reddening law, this corresponds to an \textit{E(B-V)} = 0.45 mag, which is the same as the host galaxy extinction reported in G19. While this shift does make the \textit{g-r} colors more consistent with other objects, de-reddening the SN~2016hnk photometry causes the \textit{r-i} color evolution to be even more inconsistent with SNe~Ia. Therefore, if SN~2016hnk has a host-galaxy reddening of \textit{E(B-V)} = 0.45 mag, then it would be $\geq$0.20 mag intrinsically bluer than all other SNe Ia, including those similar to SN~1991bg, in \textit{r-i}. While possible, this scenario requires both a relatively large dust reddening with no Na~D absorption, in contrast to other SNe with similar reddening, and intrinsically peculiar colors.  Our preferred scenario of minimal reddening and intrinsically peculiar colors requires less exceptional circumstances.  Furthermore, a single explosion model can explain the peculiar colors, luminosity, and spectral evolution of SN 2016hnk.


SN~2016hnk's red colors can be explained by the helium shell double detonation model proposed for this object. As shown in Section \ref{sec:models}, the detonation of a helium shell on the surface of a C/O WD will pollute the outer layers of ejecta with Fe-group elements, which will suppress blue-wards flux. The products of the helium shell detonation will produce ashes that will cause the SN to have redder colors throughout its evolution, as observed in SN~2016hnk. Thus we attribute the observed red colors of SN~2016hnk to a helium shell detonation, which we show to be consistent with observations of this event in Figures \ref{fig:DD_LC} and \ref{fig:DD_spectra}. 

\subsection{Modeling Approach} \label{sec:model_approach}

The challenge in modeling SN~2016hnk lies in the lack of early time data. The typical approach is to constrain the model parameters solely with photometry. The magnitude of the early flux excess constrains the mass of the helium shell, and peak magnitude reflects the total mass of the progenitor star. We would then compare the spectra and see if the event is consistent with a double detonation explosion. This method was used for SN 2018byg by \citealt{de19}. However, in the case of 2016hnk, we can only constrain the total mass (or equivalently a $^{56}$Ni mass) from the peak luminosity, and we estimate the mass of the helium shell by examining the synthetic spectra at the time of peak brightness. When we examine the line-blanketed part of the spectrum ($\lambda < 5000 \AA$) we find a good agreement with the 0.02 $\Msun$ helium shell.


Another source of possible concern is the poor fit of our model to the slow decline of 2016hnk (Figure \ref{fig:DD_LC}). This might be the result of the LTE approximation in the radiative transport calculations, which under predict the flux once the ejecta becomes optically thin. This occurs $\sim$30-40 days after explosion for sub-Chandrasekhar mass models. Future NLTE modeling is required to understand if the slow decline of SN~2016hnk can be modeled with a double detonation explosion of this total mass of 0.87$\Msun$ or if the slow decline requires a more massive ejecta to account for the required diffusion time as, suggested in G19. However, an alternative explanation for this light curve plateau is the emergence of bluewards flux that was previous suppressed by Fe-group elements in the outer ejecta. This would occur once the outer ejecta expands and becomes optically thin, thus causing an increase in flux the bands affected by early-time line blanketing (e.g., \textit{u, B, g, V} bands). This scenario seems plausible based on the spectral evolution after $t\sim$20d in both our dataset and in G19, but a more robust analysis of spectral color evolution would need to be done to verify this model. 


\subsection{Comparison to G19} \label{sec:G19_compare}

Our observational inferences on SN~2016hnk are similar to those presented in G19. We both find SN~2016hnk to be a peculiar thermonuclear object with a low overall luminosity and observed red colors. We obtain similar values for $\Delta \rm{m}_{15}(B)$ and $M_B$ at peak, both of which being dissimilar from those found in SNe~Ia sub-classes. Additionally, both of our observations have shown SN~2016hnk to be rich in high-velocity calcium, which dominates the nebular spectrum as [Ca II] emission. We furthermore agree that exotic elements such as Sc II, Cr II and Sr II are most likely not found within the SN~2016hnk photospheric spectra. Despite its spectral similarity to PTF09dav, we find that blue-wards line profiles can be effectively modeled with Fe, Co and Ti ions, similar to those found in G19. We also acknowledge the similarities between SN~2016hnk and 91bg-like peak spectra, as shown in G19. The closest matching spectral features are in the 5000-6300\AA \ wavelength range, which includes similar Si II absorption profiles in both objects (Figure \ref{fig:Ia_spectra_compare}(a)).

However, additional observations and modeling presented in this paper led to a different overall interpretation than G19 on the intrinsic nature of SN~2016hnk. The observed rise-time and velocities, in addition to consistency with shell models, demonstrates that SN~2016hnk produced $~0.9\Msun$ of ejecta, thus making it a sub-Chandrasekhar mass explosion. This SN was extremely Fe-poor, as shown by its low total Ni-mass and lack of Fe-group ions in its nebular spectra (Sections \ref{subsec:LC_bolometric} \& \ref{subsec:spectra_compare}). As discussed in Section \ref{sec:reddening}, the highly reddened colors are intrinsic to SN and require an explosion mechanism to suppress blue-wards flux as shown in photospheric spectra. Consequently, we attribute these observed characteristics to a helium shell double detonation of a sub-Chandrasekhar mass C/O WD.

In this paper we have demonstrated the physical differences between SN~2016hnk and 91bg-like objects. Firstly, SN~2016hnk produced a small amount of Fe-group elements relative to 91bg-like events. This is demonstrated by the low inferred Ni-mass and lack of visible Fe-group elements in late-time spectra.  Furthermore, as illustrated in Figure \ref{fig:MB_dm15}, SN~2016hnk has very different $M_B$ vs. $\Delta \rm{m}_{15}(B)$ measurement relative to 91bg-like events. The intrinsic color evolution of SN~2016hnk is also unlike any sub-luminous SN~Ia; the SN's highly reddened \textit{B-V} colors, in addition to ``bluer'' \textit{r-i}, are inconsistent with other events (Figure \ref{fig:colors}). Additionally, the barred spiral host-galaxy of SN~2016hnk may also indicate a distinct stellar progenitor systems from 91bg-like objects, which are typically found in elliptical galaxies with older stellar populations (\citealt{van05}; however see \citealt{hoflich02,garnavich04}). Conversely, if SN~2016hnk does in fact belong to the 91bg-like sub-class, it is the most extreme example yet observed. 

\subsection{Classification and Origins} \label{sec:classification}

The properties of SN~2016hnk challenge the traditional classification schemes of both sub-luminous SNe~Ia and Ca-rich transients. Upon discovery, SN~2016hnk was classified as a 91bg-like object because of its spectral features and low luminosity. However, its prominent Ca II features and ``gap'' absolute magnitude placed the SN in Ca-rich class. Its spectroscopic similarity to PTF09dav around maximum light also made the Ca-rich classification plausible. The Ca-rich distinction was later confirmed by its pre-nebular and nebular spectra, which were dominated by [Ca II] emission in addition to having an integrated [Ca II]/[O I] flux ratio greater than 2.

The ambiguity of SN~2016hnk's classification points to a need for diligence in understanding the similarities between low luminosity thermonuclear objects. While it is true that SN~2016hnk shows some spectroscopic similarities to 91bg-like objects near peak, there are prominent physical differences that set these objects apart. Like Ca-rich events, SN~2016hnk was significantly more Fe-poor than any 91bg-like object yet observed. This became most apparent in nebular spectra and may be an indication that some 91bg-like or sub-luminous SNe~Ia are mis-classified. There could be a number of sub-luminous SN~Ia that are classified as such using maximum light spectra, but more physically resemble a ``16hnk-like'' or Ca-rich event. This is further supported by the limited number of sub-luminous objects that are tracked out to nebular times.

The color evolution of SN~2016hnk was essential in determining physical distinctions between the SN and sub-luminous SNe~Ia. We emphasize the need for multi-band color information in order to understand how highly reddened, 16hnk-like SNe might compare to the larger thermonuclear sample. Furthermore, high-cadence color evolution could help to identify more than examples of thin shell detonations of sub-Chandrasekhar mass WDs. 

As shown in \cite{de19}, SN~2018byg was thought to be a relatively rare event. Due to the observational similarity and consistent explosion models between both objects, we may initially conclude that SN~2016hnk was also a rare event with respect to thermonuclear SNe. This indicates that there are physical distinctions between the helium shell detonations that can explain SNe~2016hnk and 2018byg, in addition to models that have been shown to reproduce normal and sub-luminous SNe~Ia (e.g., \citealt{shen18, polin19, townsley19}). The thin shell detonation models that best matched these two SNe were distinct in their production of suppressed bluewards flux, Fe-group line blanketing, high-velocity ($>18,000 \kms$) Ca II velocities and an intrinsically red color evolution. These physical properties can thus be applied as tracers for identifying more events like the apparent helium shell detonations that well explain SNe 2016hnk and 2018byg.

Furthermore, it may be possible that many Ca-rich events are caused by variations on the helium shell double detonation model. A similar model was invoked to explain SN~2005E and it may be the case that other Ca-rich events could now be explained with new varieties of helium shell detonations. These events have noticeable similarities to this model such as low luminosities, reddened colors, dominant [Ca II] emission at nebular times and rapidly evolving light curves. Additionally, iPTF16hgs has a double-peaked light curve, which may be matched to the first $^{48}$Cr peak produced in helium shell models. However, this explosion scenario needs to explain the He I observations in many Ca-rich objects; such an observation may require a fine tuning of helium detonation on the WD surface in order to allow for sufficient amounts of un-burned helium to remain in the SN ejecta. Nonetheless, the observed H$\alpha$ emission in PTF09dav poses a serious problem for this model since no hydrogen is produced in such an explosion. 


\section{Conclusion} \label{sec:conclusion}

In this paper we have presented observations and modeling of the Ca-rich transient SN~2016hnk. We summarize our primary observational findings below:

\begin{itemize}
    \item SN~2016hnk is intrinsically red compared to other thermonuclear objects. This is demonstrated by the lack of Na D absorption in photospheric spectra and the even ``bluer'' \textit{r-i} color evolution if any ``de-reddening'' is applied.
    \item Photospheric spectra shows strong, high-velocity Ca II features ($\approx$-18,000 $\kms$) and suppressed bluewards flux from line blanketing of Fe-group elements.
    \item Nebular spectra are O- and Fe-poor, with the most prominent feature being [Ca II] emission.
    \item SN~2016hnk has a rise-time of $t_r = 15 \pm 2$ days and a slow, ``plateau'' decline in B-band relative to SNe~Ia.
    \item Peak absolute magnitude of $M_{B} = -15.40 \pm 0.088$ mag and decline parameter of $\Delta \rm{m}_{15}(B) = 1.31 \pm 0.085$ mag.
    \item Total Nickel and ejecta masses of $M_{\textrm{Ni}} = 0.03 \pm 0.01$ and $M_{\textrm{ej}} = 0.9 \pm 0.3 \ \Msun$, respectively.
\end{itemize}



SN~2016hnk is most similar to SN~2018byg, a thermonuclear SN that is observationally consistent with a helium shell detonation on the surface of a sub-Chandrasekhar mass C/O WD. In their peak spectra, SNe~2016hnk and 2018byg both have high-velocity Ca II absorption features and significant line blanketing amongst Fe-group elements. Both objects are instrinically  reddened and have consistent \textit{g-r} and \textit{r-i} color evolution to within 0.5mag. While the light curve evolution of both objects is similar, SN~2016hnk has a faster rise time and a somewhat slower decline in \textit{r} and \textit{i} bands. 

Given these physical similarities between both objects, we compare SN~2016hnk observations to observables produced in thin helium shell detonations on sub-Chandrasekhar mass WDs. Using the models of \cite{polin19}, we find SN~2016hnk to be consistent to the detonation of a 0.85 $\Msun$ WD with a 0.02 $\Msun$ helium shell. This model is well-matched to photospheric and nebular spectra, but cannot fully reproduce the slow declining light curve evolution observed in SN~2016hnk at $t>30$d post-explosion. However, we attribute this discrepancy to LTE assumptions which break down as the SN becomes optically thin at $>20$ days after peak luminosity. Nonetheless, the ashes from such a helium detonation can effectively explain the red colors of SN~2016hnk. Furthermore, the increased Fe-group elements produced in the outer ejecta by this type of explosion can reproduce the observed Fe-group line blanketing and suppressed bluewards flux in SN~2016hnk's photospheric spectra.

Finally, we have determined multiple observational differences between SN~2016hnk and 91bg-like SNe. Firstly, SN~2016hnk is highly reddened compared to all normal and sub-luminous SNe~Ia. De-reddening SN~2016hnk's \textit{B-V} colors to match those of SNe~Ia results in an even more substanial difference between these obejcts in \textit{r-i}. Additionally, SN~2016hnk has a lower peak $M_B$ and smaller $\Delta \rm{m}_{15}(B)$ than typical 91bg-like events. Lastly, SN~2016hnk is Fe-poor relative to all 91bg-like objects; SN~2016hnk has no nebular spectral signatures of forbidden Fe, Co or Ni lines and its total inferred Ni-mass is lower than all known sub-luminous SNe~Ia. These stark differences may indicate that SN~2016hnk is the most extreme example of a sub-luminous SNe~Ia to date.

\section{Acknowledgements} \label{sec:ack}

We thank R.\ Thomas, P.\ Nugent and S.\ Woosley for helpful comments on this paper. We thank J. Tonry for providing ATLAS photometry used in this paper. 

The UCSC group is supported in part by NSF grant AST-1518052, the Gordon \& Betty Moore Foundation, and by fellowships from the David and Lucile Packard Foundation to R.J.F. This research is supported at Rutgers University through NSF award AST-1615455. 

This research used resources of the National Energy Research Scientific Computing Center (NERSC), a U.S.\ Department of Energy Office of Science User Facility operated under Contract No.DE-AC02-05CH11231.

This work includes data obtained with the Swope Telescope at Las Campanas Observatory, Chile, as part of the Swope Time Domain Key Project (PI Piro, Co-PIs Drout, Foley, Hsiao, Madore, Phillips, and
Shappee). We wish to thank Swope Telescope observers Jorge Anais Vilchez, Abdo Campillay, Nahir Munoz Elgueta and Natalie Ulloa for collecting data presented in this paper. 

Pan-STARRS is supported in part by the National Aeronautics and Space Administration under Grants
NNX12AT65G and NNX14AM74G. The Pan-STARRS1
Surveys (PS1) and the PS1 public science archive have been made possible through contributions by the Institute for Astronomy, the University of Hawaii, the Pan-STARRS Project Office, the Max-Planck Society and its participating institutes, the Max Planck Institute for Astronomy, Heidelberg and the Max Planck Institute for Extraterrestrial Physics, Garching, The Johns Hopkins University, Durham University, the University of Edinburgh, the Queen's University Belfast, the Harvard-Smithsonian Center for Astrophysics, the Las Cumbres Observatory Global Telescope Network Incorporated, the National Central University of Taiwan, the Space Telescope Science Institute, the National Aeronautics and Space Administration under Grant No. NNX08AR22G issued through the Planetary Science Division of the NASA Science Mission Directorate, the National Science Foundation Grant No. AST-1238877, the University of Maryland, Eotvos Lorand University (ELTE), the Los Alamos National Laboratory, and the Gordon and Betty Moore Foundation.

This paper is based on observations obtained at the Southern Astrophysical Research (SOAR) telescope, which is a joint project of the Minist\'{e}rio da Ci\^{e}ncia, Tecnologia, e Inova\c{c}ao (MCTI) da Rep\`{u}blica Federativa do Brasil; the U.S. National Optical Astronomy Observatory (NOAO); the University of North Carolina at Chapel Hill (UNC); and Michigan State University (MSU), and at Kitt Peak National Observatory, NOAO, which is operated by the Association of Universities for Research in Astronomy (AURA) under cooperative agreement with the National Science Foundation. The authors are honored to be permitted to conduct astronomical research on Iolkam Du'ag (Kitt Peak), a mountain with particular significance to the Tohono O'odham.

Some of the data presented herein were obtained at the W. M. Keck Observatory, which is operated as a scientific partnership among the California Institute of Technology, the University of California and the National Aeronautics
and Space Administration. The Observatory was made possible by the generous financial support of the W. M. Keck Foundation. The authors wish to recognize and acknowledge the very significant cultural role and reverence that the
summit of Maunakea has always had within the indigenous Hawaiian community. We are most fortunate to have the opportunity to conduct observations from this mountain. 

Some of the observations reported in this paper were obtained with the Southern African Large Telescope (SALT), and we thank the SALT Astronomers for assistance.

Some observations presented in this paper were obtained with the Las Cumbres Observatory Global Telescope network. 

\facilities{Pan-STARRS-1(GPC1), Swope:1m, SOAR (Goodman spectrograph), Mayall (KOSMOS spectrograph), CTIO:1.3m, CTIO:1.5m, SALT (RSS), Keck I (LRIS), Keck II(DEIMOS)}

\software{emcee \citep{foreman-mackey13}, SNID \citep{Blondin07}, Superfit \citep{Howell05}, IRAF (Tody 1986, Tody 1993), AstroDrizzle \citep{astrodrizzle}, photpipe \citep{Rest+05}, DoPhot \citep{Schechter+93}, HOTPANTS \citep{becker15}, Sedona \citep{kasen06}, SedoNeb \citep{janos2017}, SYN++/SYNAPPS \citep{thomas11}, Castro \citep{almgren10} }

\section{Appendix} \label{sec:appendix}

\begin{deluxetable}{cccccc}[h!]
\tablecaption{Optical Photometry of SN~2016hnk \label{tab:phot_table}}
\tablecolumns{6}
\tablenum{1}
\tablewidth{0.45\textwidth}
\tablehead{
\colhead{MJD} &
\colhead{Phase\tablenotemark{a}} &
\colhead{Filter} & \colhead{Magnitude} & \colhead{Uncertainty} & \colhead{Instrument}
}
\startdata
57693.31 & +3.11 & u & 21.16 & 0.13 & Swope \\
57693.29 & +3.09 & B & 19.11 & 0.08 & Swope \\
57693.30 & +3.10 & B & 19.11 & 0.03 & Swope \\
57695.26 & +5.06 & B & 19.24 & 0.07 & Swope \\
57695.26 & +5.06 & B & 19.27 & 0.02 & Swope \\
57713.30 & +23.10 & B & 20.51 & 0.13 & Swope \\
57719.24 & +29.04 & B & 20.63 & 0.06 & Swope \\
57721.18 & +30.98 & B & 20.65 & 0.05 & Swope \\
57693.30 & +3.10 & V & 17.63 & 0.02 & Swope \\
57695.27 & +5.07 & V & 17.66 & 0.02 & Swope \\
57713.30 & +23.10 & V & 18.86 & 0.06 & Swope \\
57719.25 & +29.05 & V & 19.16 & 0.04 & Swope \\
57721.18 & +30.98 & V & 19.25 & 0.03 & Swope \\
57693.32 & +3.12 & g & 18.36 & 0.02 & Swope \\
57695.28 & +5.08 & g & 18.45 & 0.02 & Swope \\
57695.28 & +5.08 & g & 18.89 & 0.02 & Swope \\
57713.28 & +23.08 & g & 19.90 & 0.07 & Swope \\
57719.26 & +29.06 & g & 20.09 & 0.04 & Swope \\
57721.17 & +30.97 & g & 20.12 & 0.04 & Swope \\
57746.20 & +56.00 & g & 20.33 & 0.04 & Swope \\
57751.15 & +60.95 & g & 20.35 & 0.04 & Swope \\
57752.17 & +61.97 & g & 20.38 & 0.03 & Swope \\
57754.17 & +63.97 & g & 20.42 & 0.04 & Swope \\
57774.11 & +83.91 & g & 20.82 & 0.08 & Swope \\
57778.12 & +87.92 & g & 20.55 & 0.08 & Swope \\
57780.09 & +89.89 & g & 20.75 & 0.12 & Swope \\
57785.07 & +94.87 & g & 20.91 & 0.08 & Swope \\
57801.04 & +110.84 & g & 21.05 & 0.09 & Swope \\
57804.02 & +113.82 & g & 20.97 & 0.12 & Swope \\
57693.32 & +3.12 & r & 17.23 & 0.01 & Swope \\
57713.28 & +23.08 & r & 18.32 & 0.06 & Swope \\
57713.28 & +23.08 & r & 18.29 & 0.03 & Swope \\
57719.26 & +29.06 & r & 18.46 & 0.02 & Swope \\
57719.22 & +29.02 & r & 18.59 & 0.05 & Swope \\
57721.16 & +30.96 & r & 18.42 & 0.06 & Swope \\
57721.16 & +30.96 & r & 18.45 & 0.02 & Swope \\
57744.20 & +54.00 & r & 19.24 & 0.06 & Swope \\
57746.19 & +55.99 & r & 19.37 & 0.12 & Swope \\
57746.19 & +55.99 & r & 19.26 & 0.03 & Swope \\
57751.14 & +60.94 & r & 19.47 & 0.03 & Swope \\
57752.15 & +61.95 & r & 19.48 & 0.09 & Swope \\
57752.15 & +61.95 & r & 19.43 & 0.02 & Swope \\
57754.16 & +63.96 & r & 19.72 & 0.15 & Swope \\
57754.16 & +63.96 & r & 19.56 & 0.03 & Swope \\
57774.10 & +83.90 & r & 20.19 & 0.06 & Swope \\
57778.11 & +87.91 & r & 20.23 & 0.09 & Swope \\
57780.10 & +89.90 & r & 20.42 & 0.11 & Swope \\
57785.08 & +94.88 & r & 20.67 & 0.08 & Swope \\
57801.03 & +110.83 & r & 21.00 & 0.12 & Swope \\
\enddata
\tablenotetext{a}{Relative to B maximum (MJD 57690.18)}
\tablecomments{}
\end{deluxetable}

\begin{deluxetable}{cccccc}[h!]
\tablecaption{Optical Photometry of SN~2016hnk (Cont.) \label{tab:phot_table2}}
\tablecolumns{6}
\tablenum{1}
\tablewidth{0.45\textwidth}
\tablehead{
\colhead{MJD} &
\colhead{Phase\tablenotemark{a}} &
\colhead{Filter} & \colhead{Magnitude} & \colhead{Uncertainty} & \colhead{Instrument}
}
\startdata
57693.32 & +3.12 & i & 17.47 & 0.02 & Swope \\
57695.28 & +5.08 & i & 17.50 & 0.01 & Swope \\
57713.28 & +23.08 & i & 18.19 & 0.03 & Swope \\
57719.26 & +29.06 & i & 18.39 & 0.03 & Swope \\
57721.17 & +30.97 & i & 18.50 & 0.02 & Swope \\
57744.20 & +54.00 & i & 19.35 & 0.09 & Swope \\
57746.19 & +55.99 & i & 19.30 & 0.04 & Swope \\
57751.14 & +60.94 & i & 19.46 & 0.04 & Swope \\
57752.16 & +61.96 & i & 19.45 & 0.03 & Swope \\
57754.17 & +63.97 & i & 19.44 & 0.04 & Swope \\
57774.11 & +83.91 & i & 19.80 & 0.07 & Swope \\
57778.12 & +87.94 & i & 19.93 & 0.09 & Swope \\
57785.08 & +94.88 & i & 20.14 & 0.08 & Swope \\
57780.09 & +89.89 & i & 20.01 & 0.14 & Swope \\
57801.04 & +110.84 & i & 20.18 & 0.09 & Swope \\
57804.02 & +113.82 & i & 20.02 & 0.15 & Swope \\
57690.20 & +0.00 & B & 18.99 & 0.07 & LCOGT \\
57690.81 & +0.61 & B & 19.00 & 0.09 & LCOGT \\
57690.82 & +0.62 & B & 19.01 & 0.09 & LCOGT \\
57692.83 & +2.63 & B & 19.04 & 0.04 & LCOGT \\
57694.92 & +4.72 & B & 19.18 & 0.04 & LCOGT \\
57695.09 & +4.89 & B & 19.29 & 0.02 & LCOGT \\
57697.43 & +7.23 & B & 19.55 & 0.06 & LCOGT \\
57698.91 & +8.71 & B & 19.75 & 0.06 & LCOGT \\
57699.00 & +8.80 & B & 19.83 & 0.06 & LCOGT \\
57701.27 & +11.07 & B & 19.99 & 0.07 & LCOGT \\
57701.62 & +11.42 & B & 20.05 & 0.09 & LCOGT \\
57702.58 & +12.38 & B & 20.17 & 0.14 & LCOGT \\
57707.82 & +17.62 & B & 20.40 & 0.05 & LCOGT \\
57711.25 & +21.05 & B & 20.46 & 0.25 & LCOGT \\
57711.87 & +21.67 & B & 20.47 & 0.08 & LCOGT \\
57713.16 & +22.96 & B & 20.52 & 0.11 & LCOGT \\
57714.51 & +24.31 & B & 20.53 & 0.09 & LCOGT \\
57717.17 & +26.97 & B & 20.58 & 0.07 & LCOGT \\
57717.85 & +27.65 & B & 20.59 & 0.08 & LCOGT \\
57721.06 & +30.86 & B & 20.58 & 0.08 & LCOGT \\
57721.81 & +31.61 & B & 20.62 & 0.09 & LCOGT \\
57726.84 & +36.64 & B & 20.69 & 0.02 & LCOGT \\
57733.90 & +43.70 & B & 20.80 & 0.05 & LCOGT \\
57736.91 & +46.71 & B & 20.81 & 0.26 & LCOGT \\
57741.15 & +50.95 & B & 20.83 & 0.07 & LCOGT \\
57745.04 & +54.84 & B & 20.89 & 0.09 & LCOGT \\
57746.18 & +55.98 & B & 20.93 & 0.09 & LCOGT \\
57747.01 & +56.81 & B & 20.94 & 0.09 & LCOGT \\
57752.87 & +62.67 & B & 21.01 & 0.11 & LCOGT \\
57755.72 & +65.52 & B & 20.95 & 0.10 & LCOGT \\
57757.11 & +66.91 & B & 20.99 & 0.16 & LCOGT \\
\enddata
\tablenotetext{a}{Relative to B maximum (MJD 57690.18)}
\tablecomments{}
\end{deluxetable}

\begin{deluxetable}{cccccc}[h!]
\tablecaption{Optical Photometry of SN~2016hnk (Cont.) \label{tab:phot_table2}}
\tablecolumns{6}
\tablenum{1}
\tablewidth{0.45\textwidth}
\tablehead{
\colhead{MJD} &
\colhead{Phase\tablenotemark{a}} &
\colhead{Filter} & \colhead{Magnitude} & \colhead{Uncertainty} & \colhead{Instrument}
}
\startdata
57689.93 & -0.27 & V & 17.65 & 0.03 & LCOGT \\
57690.82 & +0.62 & V & 17.67 & 0.05 & LCOGT \\
57692.83 & +2.63 & V & 17.63 & 0.03 & LCOGT \\
57695.10 & +4.90 & V & 17.78 & 0.02 & LCOGT \\
57697.44 & +7.24 & V & 17.90 & 0.03 & LCOGT \\
57698.92 & +8.72 & V & 17.93 & 0.03 & LCOGT \\
57701.27 & +11.07 & V & 18.21 & 0.04 & LCOGT \\
57701.63 & +11.43 & V & 18.27 & 0.03 & LCOGT \\
57702.58 & +12.38 & V & 18.33 & 0.07 & LCOGT \\
57707.07 & +16.87 & V & 18.60 & 0.21 & LCOGT \\
57707.82 & +17.62 & V & 18.65 & 0.06 & LCOGT \\
57708.63 & +18.43 & V & 18.70 & 0.14 & LCOGT \\
57711.87 & +21.67 & V & 18.69 & 0.05 & LCOGT \\
57713.16 & +22.96 & V & 18.84 & 0.07 & LCOGT \\
57714.51 & +24.31 & V & 18.94 & 0.08 & LCOGT \\
57717.18 & +26.98 & V & 19.07 & 0.07 & LCOGT \\
57717.85 & +27.65 & V & 19.10 & 0.05 & LCOGT \\
57721.07 & +30.87 & V & 19.26 & 0.06 & LCOGT \\
57721.82 & +31.62 & V & 19.26 & 0.07 & LCOGT \\
57726.85 & +36.65 & V & 19.34 & 0.09 & LCOGT \\
57733.89 & +43.69 & V & 19.51 & 0.20 & LCOGT \\
57736.92 & +46.72 & V & 19.52 & 0.11 & LCOGT \\
57745.05 & +54.85 & V & 19.56 & 0.09 & LCOGT \\
57746.19 & +55.99 & V & 19.57 & 0.08 & LCOGT \\
57747.02 & +56.82 & V & 19.63 & 0.08 & LCOGT \\
57752.88 & +62.68 & V & 19.75 & 0.10 & LCOGT \\
57755.73 & +65.53 & V & 19.80 & 0.09 & LCOGT \\
57761.50 & +71.30 & V & 19.94 & 0.16 & LCOGT \\
57695.28 & +5.08 & g & 18.58 & 0.01 & LCOGT \\
57699.00 & +8.80 & g & 19.26 & 0.03 & LCOGT \\
57701.63 & +11.43 & g & 19.32 & 0.04 & LCOGT \\
57706.89 & +16.69 & g & 19.57 & 0.08 & LCOGT \\
57707.82 & +17.62 & g & 19.62 & 0.08 & LCOGT \\
57711.88 & +21.68 & g & 19.75 & 0.04 & LCOGT \\
57717.86 & +27.66 & g & 20.01 & 0.04 & LCOGT \\
57721.07 & +30.87 & g & 20.07 & 0.04 & LCOGT \\
57721.82 & +31.62 & g & 20.07 & 0.04 & LCOGT \\
57726.84 & +36.64 & g & 20.17 & 0.08 & LCOGT \\
57728.10 & +37.90 & g & 20.18 & 0.05 & LCOGT \\
57729.59 & +39.39 & g & 20.19 & 0.12 & LCOGT \\
57733.88 & +43.68 & g & 20.25 & 0.06 & LCOGT \\
57741.17 & +50.97 & g & 20.25 & 0.05 & LCOGT \\
57745.17 & +54.97 & g & 20.28 & 0.06 & LCOGT \\
57747.16 & +56.96 & g & 20.32 & 0.04 & LCOGT \\
57752.87 & +62.67 & g & 20.40 & 0.05 & LCOGT \\
57761.22 & +71.02 & g & 20.47 & 0.08 & LCOGT \\
\enddata
\tablenotetext{a}{Relative to B maximum (MJD 57690.18)}
\tablecomments{}
\end{deluxetable}

\begin{deluxetable}{cccccc}[h!]
\tablecaption{Optical Photometry of SN~2016hnk (Cont.) \label{tab:phot_table2}}
\tablecolumns{6}
\tablenum{1}
\tablewidth{0.45\textwidth}
\tablehead{
\colhead{MJD} &
\colhead{Phase\tablenotemark{a}} &
\colhead{Filter} & \colhead{Magnitude} & \colhead{Uncertainty} & \colhead{Instrument}
}
\startdata
57690.82 & +0.62 & r & 17.30 & 0.02 & LCOGT \\
57690.81 & +0.61 & r & 17.30 & 0.03 & LCOGT \\
57692.83 & +2.63 & r & 17.28 & 0.02 & LCOGT \\
57694.93 & +4.73 & r & 17.24 & 0.04 & LCOGT \\
57695.27 & +5.07 & r & 17.26 & 0.04 & LCOGT \\
57697.44 & +7.24 & r & 17.34 & 0.02 & LCOGT \\
57698.93 & +8.73 & r & 17.37 & 0.01 & LCOGT \\
57699.10 & +8.90 & r & 17.54 & 0.04 & LCOGT \\
57700.89 & +10.69 & r & 17.62 & 0.12 & LCOGT \\
57701.27 & +11.07 & r & 17.65 & 0.05 & LCOGT \\
57702.58 & +12.38 & r & 17.75 & 0.04 & LCOGT \\
57703.23 & +13.03 & r & 17.80 & 0.14 & LCOGT \\
57704.31 & +14.11 & r & 17.90 & 0.10 & LCOGT \\
57706.92 & +16.72 & r & 17.99 & 0.12 & LCOGT \\
57707.31 & +17.11 & r & 17.99 & 0.14 & LCOGT \\
57708.63 & +18.43 & r & 18.05 & 0.08 & LCOGT \\
57709.87 & +19.67 & r & 18.10 & 0.06 & LCOGT \\
57710.18 & +19.98 & r & 18.16 & 0.11 & LCOGT \\
57711.88 & +21.68 & r & 18.14 & 0.02 & LCOGT \\
57712.85 & +22.65 & r & 18.29 & 0.03 & LCOGT \\
57713.18 & +22.98 & r & 18.27 & 0.06 & LCOGT \\
57714.51 & +24.31 & r & 18.36 & 0.04 & LCOGT \\
57717.18 & +26.98 & r & 18.42 & 0.03 & LCOGT \\
57717.86 & +27.66 & r & 18.42 & 0.02 & LCOGT \\
57721.82 & +31.62 & r & 18.49 & 0.03 & LCOGT \\
57722.22 & +32.02 & r & 18.54 & 0.15 & LCOGT \\
57726.83 & +36.63 & r & 18.60 & 0.12 & LCOGT \\
57728.10 & +37.90 & r & 18.69 & 0.09 & LCOGT \\
57729.55 & +39.35 & r & 18.71 & 0.11 & LCOGT \\
57733.90 & +43.71 & r & 18.80 & 0.08 & LCOGT \\
57736.94 & +46.74 & r & 18.87 & 0.10 & LCOGT \\
57741.21 & +51.01 & r & 18.92 & 0.06 & LCOGT \\
57745.21 & +55.01 & r & 19.03 & 0.11 & LCOGT \\
57747.12 & +56.92 & r & 19.11 & 0.08 & LCOGT \\
57752.87 & +62.67 & r & 19.29 & 0.04 & LCOGT \\
57761.13 & +70.93 & r & 19.50 & 0.20 & LCOGT \\
57762.86 & +72.66 & r & 19.54 & 0.16 & LCOGT \\
\enddata
\tablenotetext{a}{Relative to B maximum (MJD 57690.18)}
\tablecomments{}
\end{deluxetable}

\begin{deluxetable}{cccccc}[h!]
\tablecaption{Optical Photometry of SN~2016hnk (Cont.) \label{tab:phot_table2}}
\tablecolumns{6}
\tablenum{1}
\tablewidth{0.45\textwidth}
\tablehead{
\colhead{MJD} &
\colhead{Phase\tablenotemark{a}} &
\colhead{Filter} & \colhead{Magnitude} & \colhead{Uncertainty} & \colhead{Instrument}
}
\startdata
57690.82 & +0.62 & i & 17.42 & 0.04 & LCOGT \\
57690.20 & +0.00 & i & 17.48 & 0.03 & LCOGT \\
57692.83 & +2.63 & i & 17.46 & 0.04 & LCOGT \\
57694.82 & +4.62 & i & 17.46 & 0.04 & LCOGT \\
57695.29 & +5.09 & i & 17.49 & 0.03 & LCOGT \\
57695.28 & +5.08 & i & 17.50 & 0.04 & LCOGT \\
57695.04 & +4.84 & i & 17.52 & 0.05 & LCOGT \\
57698.76 & +8.56 & i & 17.55 & 0.05 & LCOGT \\
57699.01 & +8.81 & i & 17.60 & 0.06 & LCOGT \\
57700.20 & +10.00 & i & 17.75 & 0.12 & LCOGT \\
57701.64 & +11.44 & i & 17.85 & 0.09 & LCOGT \\
57701.27 & +11.07 & i & 17.89 & 0.04 & LCOGT \\
57702.58 & +12.38 & i & 17.92 & 0.05 & LCOGT \\
57704.33 & +14.13 & i & 17.93 & 0.10 & LCOGT \\
57706.90 & +16.70 & i & 17.96 & 0.09 & LCOGT \\
57707.29 & +17.09 & i & 18.08 & 0.08 & LCOGT \\
57707.83 & +17.63 & i & 18.08 & 0.08 & LCOGT \\
57707.83 & +17.63 & i & 18.11 & 0.08 & LCOGT \\
57708.63 & +18.43 & i & 18.16 & 0.09 & LCOGT \\
57710.85 & +20.65 & i & 18.23 & 0.02 & LCOGT \\
57711.89 & +21.69 & i & 18.28 & 0.07 & LCOGT \\
57711.26 & +21.06 & i & 18.32 & 0.10 & LCOGT \\
57713.16 & +22.96 & i & 18.39 & 0.06 & LCOGT \\
57715.20 & +25.00 & i & 18.45 & 0.10 & LCOGT \\
57717.87 & +27.67 & i & 18.46 & 0.07 & LCOGT \\
57721.83 & +31.63 & i & 18.48 & 0.09 & LCOGT \\
57722.20 & +32.00 & i & 18.59 & 0.15 & LCOGT \\
57726.87 & +36.67 & i & 18.72 & 0.12 & LCOGT \\
57728.12 & +37.92 & i & 18.75 & 0.08 & LCOGT \\
57733.90 & +43.70 & i & 18.81 & 0.14 & LCOGT \\
57736.91 & +46.71 & i & 18.93 & 0.10 & LCOGT \\
57741.19 & +50.99 & i & 19.11 & 0.10 & LCOGT \\
57745.19 & +54.99 & i & 19.18 & 0.13 & LCOGT \\
57747.18 & +56.98 & i & 19.27 & 0.13 & LCOGT \\
57752.88 & +62.68 & i & 19.40 & 0.17 & LCOGT \\
57761.12 & +70.92 & i & 19.66 & 0.15 & LCOGT \\
57762.89 & +72.69 & i & 19.71 & 0.12 & LCOGT \\
57696.97 & +6.77 & z & 17.26 & 0.06 & LCOGT \\
57701.79 & +11.59 & z & 17.71 & 0.19 & LCOGT \\
57710.26 & +20.06 & z & 17.74 & 0.07 & LCOGT \\
57711.79 & +21.59 & z & 17.68 & 0.11 & LCOGT \\
57720.82 & +30.62 & z & 17.87 & 0.21 & LCOGT \\
57726.82 & +36.62 & z & 17.81 & 0.17 & LCOGT \\
57736.95 & +46.75 & z & 18.15 & 0.27 & LCOGT \\
57741.20 & +51.00 & z & 18.65 & 0.16 & LCOGT \\
\enddata
\tablenotetext{a}{Relative to B maximum (MJD 57690.18)}
\tablecomments{}
\end{deluxetable}

\begin{deluxetable}{cccccc}[h!]
\tablecaption{Optical Photometry of SN~2016hnk (Cont.) \label{tab:phot_table2}}
\tablecolumns{6}
\tablenum{1}
\tablewidth{0.45\textwidth}
\tablehead{
\colhead{MJD} &
\colhead{Phase\tablenotemark{a}} &
\colhead{Filter} & \colhead{Magnitude} & \colhead{Uncertainty} & \colhead{Instrument}
}
\startdata
57690.40 & +0.20 & g & 18.15 & 0.02 & PS1 \\
57697.40 & +7.20 & g & 18.51 & 0.04 & PS1 \\
57709.38 & +19.18 & g & 19.48 & 0.13 & PS1 \\
57736.31 & +46.11 & g & 19.77 & 0.14 & PS1 \\
57744.27 & +54.07 & g & 20.01 & 0.13 & PS1 \\
57690.40 & +0.20 & r & 17.29 & 0.01 & PS1 \\
57697.40 & +7.20 & r & 17.31 & 0.01 & PS1 \\
57709.38 & +19.18 & r & 18.11 & 0.03 & PS1 \\
57736.31 & +46.11 & r & 18.98 & 0.05 & PS1 \\
57744.27 & +54.07 & r & 19.23 & 0.05 & PS1 \\
57690.40 & +0.20 & i & 17.40 & 0.01 & PS1 \\
57697.40 & +7.20 & i & 17.44 & 0.02 & PS1 \\
57709.39 & +19.19 & i & 18.10 & 0.03 & PS1 \\
57736.31 & +46.11 & i & 18.85 & 0.05 & PS1 \\
57744.27 & +54.07 & i & 19.26 & 0.07 & PS1 \\
57765.23 & +75.03 & i & 19.61 & 0.06 & PS1 \\
57765.24 & +75.04 & i & 19.72 & 0.07 & PS1 \\
57765.26 & +75.06 & i & 19.74 & 0.06 & PS1 \\
57765.27 & +75.07 & i & 19.75 & 0.06 & PS1 \\
57709.39 & +19.19 & z & 17.68 & 0.03 & PS1 \\
57736.31 & +46.11 & z & 18.28 & 0.04 & PS1 \\
57744.27 & +54.07 & z & 18.57 & 0.05 & PS1 \\
57671.55 & -18.65 & orange & >20.20 & --- & ATLAS \\
57680.53 & -9.67 & orange & 18.90 & 0.34 & ATLAS \\
57700.46 & +10.26 & orange & 17.61 & 0.08 & ATLAS \\
57704.44 & +14.24 & orange & 17.82 & 0.20 & ATLAS \\
57712.46 & +22.26 & orange & 17.99 & 0.25 & ATLAS \\
57736.36 & +46.16 & orange & 18.55 & 0.31 & ATLAS \\
57743.36 & +53.16 & orange & 18.63 & 0.19 & ATLAS \\
57659.57 & -30.63 & cyan & >21.79 & --- & ATLAS \\
57663.56 & -26.64 & cyan & >21.70 & --- & ATLAS \\
57667.56 & -22.64 & cyan & >20.32 & --- & ATLAS \\
57688.51 & -1.69 & cyan & 17.80 & 0.07 & ATLAS \\
57696.46 & +6.26 & cyan & 17.83 & 0.06 & ATLAS \\
57716.43 & +26.23 & cyan & 18.67 & 0.17 & ATLAS \\
57744.34 & +54.14 & cyan & 19.72 & 0.33 & ATLAS \\
57756.34 & +66.14 & cyan & 20.01 & 0.41 & ATLAS \\
57954.56 & +264.36 & B & >23.40 & --- & Keck \\
57981.54 & +291.34 & B & >23.30 & --- & Keck \\
57981.51 & +291.31 & V & >23.50 & --- & Keck \\
57954.57 & +264.37 & R & >23.30 & --- & Keck \\
57981.54 & +291.34 & R & >23.40 & --- & Keck \\
57981.51 & +291.31 & I & 23.57 & 0.09 & Keck \\
\enddata
\tablenotetext{a}{Relative to B maximum (MJD 57690.18)}
\tablecomments{}
\end{deluxetable}

\begin{deluxetable}{cccccc}[h!]
\tablecaption{Optical Spectroscopy of SN~2016hnk \label{tab:spec_table}}
\tablecolumns{5}
\tablenum{2}
\tablewidth{0.45\textwidth}
\tablehead{
\colhead{MJD} &
\colhead{Phase\tablenotemark{a}} &
\colhead{Telescope} & \colhead{Instrument} & \colhead{Wavelength Range}
}
\startdata
57691 & +1 & SOAR & Goodman & 3000-9000\AA \\
57692 & +2 & SALT & RSS & 3000-9000\AA \\
57693 & +3 & NOT & ALFOSC & 3000-9000\AA \\
57694 & +4 & NOT & ALFOSC & 3000-9000\AA \\
57722 & +32 & Mayall & KOSMOS & 3000-9000\AA \\
57749 & +59 & Mayall & KOSMOS & 3000-9000\AA \\
57756 & +66 & SOAR & Goodman & 3000-9000\AA \\
57954 & +264 & Keck I & LRIS & 3000-9000\AA \\
\enddata
\tablenotetext{a}{Relative to B maximum (MJD 57690.18)}
\tablecomments{}
\end{deluxetable}

\bibliographystyle{aasjournal} 
\bibliography{references} 



\end{document}